%% file: main.tex
\documentclass[graybox, secnum]{svmult}

\usepackage{mathptmx}       
\usepackage{helvet}         
\usepackage{courier}        
\usepackage{type1cm}        
\usepackage{makeidx}         
\usepackage[]{graphicx,xcolor}

\usepackage{siunitx}         

\usepackage{multicol}        
\usepackage[bottom]{footmisc}
\usepackage{hyperref}        
\usepackage{soul}            
\usepackage{upgreek}         

\newcommand{\blue}{\textcolor{blue}}

\hypersetup{colorlinks=true,urlcolor=blue}
\usepackage[square,numbers]{natbib}

\newcommand {\urss}[1]{\ensuremath{_{\mathrm{#1}}}}
\newcommand{\Ohm}{\ensuremath{{\mathrm\Omega}}}
\newcommand{\micro}{\ensuremath{\upmu}}

\newcommand{\umux}{\micro mux}
\newcommand{\by}{\ensuremath{\times}\ }
\newcommand{\raiseme}[1]{\raisebox{1.5ex}[0pt]{#1\hspace{6pt}}}  
\newcommand{\phinot}{\ensuremath{\mathrm{\Phi}_{0}}}
\newcommand{\uphinotperrthz}{\ensuremath{\upmu\phinot/\sqrt{\mathrm{Hz}}}}
\mathchardef\mhyphen="2D

\makeindex             

\begin{document}
\title*{Signal readout for Transition-Edge Sensor X-ray imaging spectrometers}
\author{H. Akamatsu
\thanks{corresponding author}, W.B. Doriese, J.A.B. Mates, B.D. Jackson}
\institute{
H. Akamatsu \at 
SRON Netherlands Institute for Space Research, 
Niels Bohrweg 4 2333 CA Leiden, The Netherlands
\email{h.akamatsu@sron.nl}
\and W.B. Doriese \at 
National Institute of Standards and Technology (NIST),
United States Department of Commerce, Boulder, CO, USA
\email{william.doriese@nist.gov}
\and
J.A.B. Mates \at
National Institute of Standards and Technology (NIST),
United States Department of Commerce, Boulder, CO, USA
 \email{john.mates@nist.gov}
 \and
B.D. Jackson \at
SRON Netherlands Institute for Space Research, 
Landleven 12, 9747 AD Groningen, The Netherlands
\email{B.D.Jackson@sron.nl}
}

\maketitle

\abstract{
\input{abstract_2022Aug.tex}
}



\section{The cryogenic X-ray imaging spectrometer}
\input{cryo_sensor_2022Aug.tex}

\section{Basic concepts of signal readout}

\subsection{Impedance matching}\label{sec:impedance}
\input{impedance_matching_2022Aug.tex}

\subsection{dc Superconducting quantum interference device (dc-SQUID)}\label{sec:SQUID}
\input{dc_SQUID_2022Aug.tex}

\section{Principles of multiplexed readout of X-ray TES microcalorimeters}

\subsection{Why is multiplexed readout necessary?}\label{sec:why_mux}
\input{why_mux_2022Aug.tex}

\subsection{General considerations}
\input{general_mux_2022Aug.tex}

\subsection{Time-division multiplexing (TDM)}\label{sec:tdm}
\input{TDM_2022Aug.tex}

\subsection{MHz frequency-domain multiplexing (FDM)}\label{sec:fdm}
\input{FDM_2022Aug.tex}

\subsection{Microwave SQUID multiplexing (\umux)}\label{sec:umux}
\input{umux_2022Aug.tex}

\section{Summary and future prospects}\label{sec:summary}
\input{summary_2022Aug}

\newpage
\bibliographystyle{spbasic}
\bibliography{references_short, references_TDM, ref_Bolometer}

\end{document}

%% file: abstract_2022Aug.tex
Arrays of low-temperature microcalorimeters provide a promising technology for X-ray astrophysics:  the imaging spectrometer.  A camera with at least several thousand pixels, each of which has an energy-resolving power ($E/\Delta E\urss{FWHM}$) of a few thousand across a broad energy range (200~eV to 10~keV or higher), would be a revolutionary instrument for the study of energetic astrophysical objects and phenomena.  Signal readout is a critical enabling technology.  Multiplexed readout, in which signals from multiple pixels are combined into a single amplifier channel, allows a kilopixel-scale microcalorimeter array to meet the stringent requirements for power consumption, mass, volume, and cooling capacity in orbit.  This chapter describes three different multiplexed-readout technologies for transition-edge-sensor microcalorimeters:  time-division multiplexing, frequency-domain multiplexing, and microwave-SQUID multiplexing.    For each multiplexing technique, we present the basic method, discuss some design considerations and parameters, and show the state of the art.  The chapter concludes with a brief discussion of future prospects.

%% file: cryo_sensor_2022Aug.tex
High-energy-resolution X-ray spectroscopy is a powerful probe of elemental and chemical composition that is used across many scientific disciplines, from materials science to the physics of hot plasma in the Universe.  Wavelength-dispersive (WD) X-ray spectrometers, such as gratings, Bragg crystals, and multi-layer materials, can achieve the highest energy-resolving powers ($E/\Delta E \sim 1,000$ to 10,000).  WD spectrometers have been deployed previously for X-ray astronomy, such as the Reflection Grating Spectrometer (RGS) on board the XMM-Newton satellite\cite{rgs} and Chandra's High Energy Transmission Grating Spectrometer\cite{markert94}. However, WD spectrometers have important drawbacks for astronomy, such as a limited simultaneous energy range, a low collecting efficiency (solid-angle coverage times quantum efficiency), and a lack of any inherent imaging capability.  Another common technology, the solid-state energy-dispersive (ED) detector (e.g., the silicon-drift detector, or SDD, and the charged-coupled device, or CCD), which measures the X-ray energy directly during detection, can combine high collecting efficiency with imaging capability across a wide energy band; however the energy resolution is limited ($E / \Delta E \le 50$ @ 6~keV for the best SDDs and somewhat lower for X-ray CCD cameras).

\begin{figure}
\begin{center}
  \includegraphics[width=\textwidth]{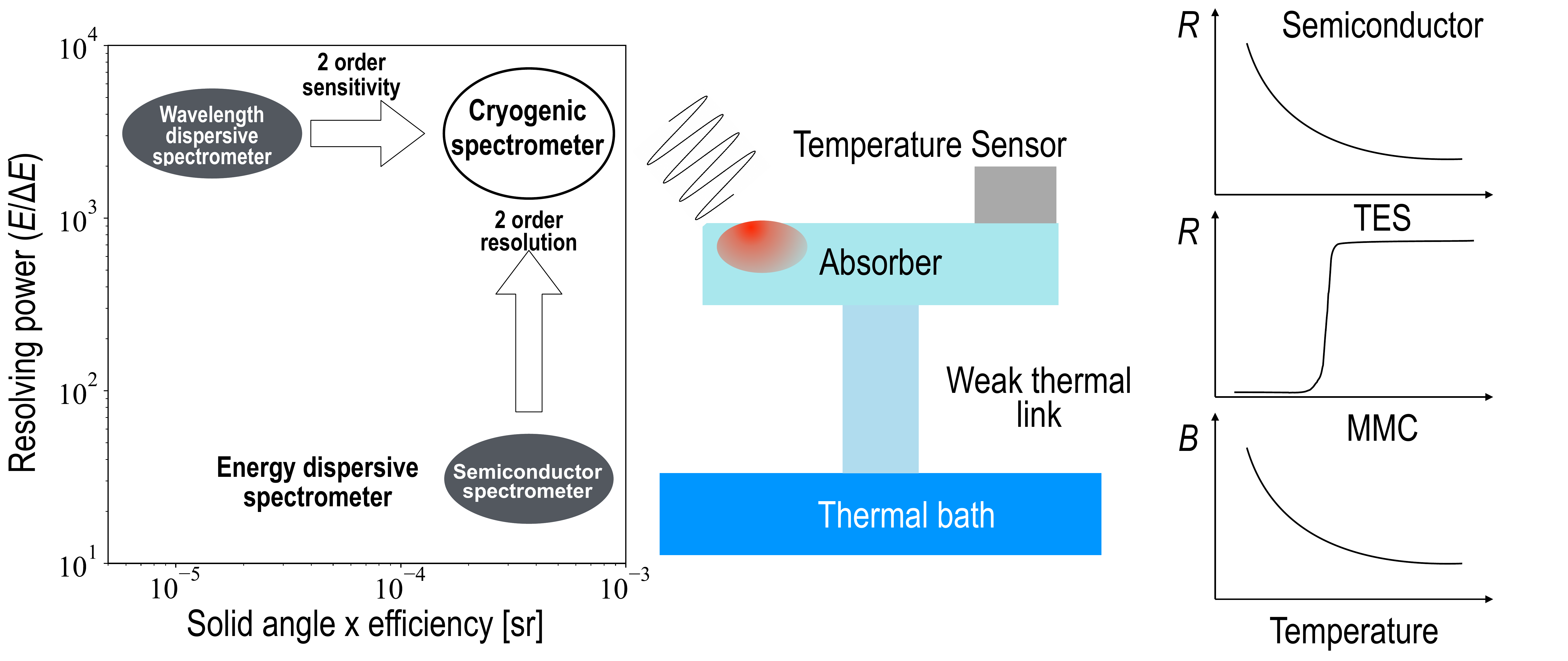}
\end{center}
\caption{The microcalorimeter array as a cryogenic imaging spectrometer.  {\it Left:}  comparison of cryogenic spectrometers to wavelength-dispersive and solid-state energy-dispersive spectrometers on a plane of energy-resolving power vs.\ collecting efficiency\cite{friedrich2006}.  Here, collecting efficiency is the solid angle subtended by the spectrometer multiplied by the cumulative quantum efficiency of all components of the spectrometer.  {\it Middle:}  cartoon diagram of a single cryogenic microcalorimeter pixel.  An X-ray absorber is connected to a thermal bath via a weak thermal link, and also to a thermometer.  Measurement of a pulsed change in the absorber temperature yields the energy of the X-ray.  {\it Right, top to bottom:}  $R$ vs.\ $T$ of a semiconductor calorimeter, $R$ vs.\ $T$ of a transition-edge sensor, and $B$ vs.\ $T$ of a metallic-magnetic calorimeter. 
}
\label{fig:fom}
\end{figure}

A technology that has been developed over the last few decades for a variety of present and future astronomical missions is the microcalorimeter array\cite{mmm, mccammon05} (see Fig.~\ref{fig:fom}).  A microcalorimeter array is an imaging spectrometer that combines some of the best features of WD spectrometers and ED detectors:  $E / \Delta E \ge 2,400$ @ 6~keV and a high collecting efficiency.  A microcalorimeter pixel consists of a sensitive thermometer and an X-ray absorber connected weakly to a cryogenic thermal bath.  When an X-ray strikes the X-ray absorber, the thermometer measures the magnitude of the resulting temperature pulse to determine the energy of the X-ray.

The energy resolution of a cryogenic spectrometer is estimated as follows:
\begin{equation}
\Delta E\sim \sqrt{\frac{k\urss{B}T^2C}{\left|\alpha\right|}}.
\end{equation}
where $k\urss{B}, T, C$, and $\alpha \equiv d\mathrm{ln}R/d\mathrm{ln}T$ are the Boltzmann constant, detector temperature, heat capacity, and thermometer sensitivity, respectively.   The keys to achievement of high energy-resolving power (low $\Delta E$) are a low operating temperature, a small heat capacity, and a high sensitivity.
 
\begin{table}[h]
\begin{center}
\caption{Summary of the three main types of X-ray microcalorimeters:  semiconductor, transition-edge sensor (TES), and metallic-magnetic calorimeter (MMC).  The upper section lists some basic characteristics of each type of microcalorimeter.  Semiconductor microcalorimeters use junction-gate field-effect transistors (JFETs) to read out their voltage, while TESs and MMCs employ superconducting quantum-interference devices (SQUIDs) to read out the device current.  The lower section gives the best energy resolution achieved in a single pixel and the largest operating detector array (as of the publication of this book). 
}
\def\arraystretch{1.5}
\begin{tabular}{l|ccc}\hline
  & Semiconductor   & TES    & MMC   \\ \hline
$T$ of sensor bath [K]    & 0.05           & 0.05         & 0.03   \\
resistance [$\rm \Omega$]  & 10$^6$      & 0.001 to 1        & 0.1       \\  
readout amplifier      & JFET      & SQUID       & SQUID     \\
$T$ of readout amplifier [K] \hspace{0.05in}   & 140           & 0.05         & 0.03       \\
multiplexing      & difficult   & highly developed        & possible      \\
amp. bandwidth	    & NA 		& 10 MHz (SQUID)		& 10 MHz	\\
    &       & \hspace{0.05in}  \raiseme{4 to 8 GHz (HEMT)} \hspace{0.05in}   &   \\ \hline
best $\Delta E$ @ 6 keV  [eV]  & 3.7 \cite{kelley09, porter09, porter10}       & 1.6 \cite{smith12, dewit22_SRON}       & 1.6 \cite{kemph18}       \\
largest array [pixels] & 36 \cite{ishisaki18} & 992 \cite{szypryt_tomography,szypryt_1kpix_operation} & 64 \cite{Mantegazzini22} \\ \hline
\multicolumn{4}{l}{TES: transition-edge sensor} \\
\multicolumn{4}{l}{MMC: metallic-magnetic calorimeter}\\
\multicolumn{4}{l}{JFET: Junction Field Effect Transistor}\\
\multicolumn{4}{l}{SQUID: superconducting quantum interference device (see Sect.\ref{sec:SQUID})}\\
\multicolumn{4}{l}{HEMT: High Electron Mobility Transistor (see Sect.\ref{sec:umux})}\\
\end{tabular}
\end{center}
\label{tab:spec_summary}
\end{table}%

While many types of cryogenic thermometers have been proposed and studied, only a few are developed enough to be considered for space-borne instrumentation:  semiconductor calorimeters, transition-edge sensors (TESs), and metallic-magnetic calorimeters (MMCs).  Table~\ref{tab:spec_summary} summarizes the basic properties of each type of microcalorimeter, while Table~\ref{tab:muxprojlist} in Sec.~\ref{sec:summary} lists present and future missions for each.

The semiconductor calorimeter has a negative thermal coefficient (thermistor $R$ decreases as $T$ increases), a high impedance ($\sim$M$\rm\Omega$), and a moderate sensitivity ($|\alpha| \sim 7$).  Semiconductor calorimeters have been used in many projects, including the X-ray Quantum Calorimeter (XQC) sounding rocket experiment\cite{mccammon02} and the ASTRO-E\footnote{ASTRO-E was the first satellite to be equipped with a cryogenic X-ray spectrometer, but it was lost during launch.}, ASTRO-E2 (Suzaku\cite{mitsuda07}/ XRS\cite{xrs})\footnote{ASTRO-E2/ XRS was the first microcalorimeter instrument to reach orbit on a satellite, but was lost due to rapid evaporation of the cryogens.}, and ASTRO-H (Hitomi\cite{takahashi12}/ SXS\cite{mitsuda14_SXS})\footnote{Hitomi/ SXS was a pioneer in low-temperature X-ray microcalorimetry and successfully observed X-ray emission from astronomical objects\cite{Hitomi16}.  Unfortunately, however, the satellite was lost due to a malfunction in its attitude-control system.} satellites.  They will also be employed for the Resolve\cite{ishisaki18} instrument on XRISM\cite{xrism18}.  McCammon\cite{mcCammon05_semiconductor} gives a detailed review of the technology.

The transition-edge sensor (TES) works in the superconducting transition from the normal to superconducting states, which provides a high temperature sensitivity ($\alpha\sim100$).  The operating impedance of a TES microcalorimeter ranges from $\sim$ 1 to 10 m$\rm\Omega$, depending on the TES design and the material chosen for the superconducting film.  The sensor is voltage biased and its current is read out via a SQUID.  There are several excellent reviews of the TES X-ray microcalorimeter in the literature\cite{irwin05, ullom15, gottardi21}. TES calorimeters are in use in or planned for many present and future X-ray astronomy projects, including Micro-X (the first TES array and readout system operated in space)\cite{adams20_microx}, the X-IFU instrument\cite{xifu18} onboard the ESA Athena mission, HUBS\cite{hubs}, Super-DIOS\cite{sdios20}, and LEM\footnote{https://lem.physics.wisc.edu/}, and the TES is one of the proposed technologies for the Lynx\cite{lynx} imaging spectrometer\cite{bandler19}.

Unlike other resistance calorimeters, the metallic-magnetic calorimeter (MMC) utilizes a non-resistive, paramagnetic temperature sensor; a MMC converts a temperature change into a change in magnetization, which is then measured as a change in magnetic flux in a dc SQUID\cite{fleischmann05}. MMCs have two interesting differences from resistive microcalorimeters: (1) no power is dissipated in the sensor and (2) the readout makes no galvanic contact to the sensor.  MMCs also exhibit excellent energy resolution, high dynamic range, and high linearity in their energy-gain scale.  MMCs will be used in the International Axion Observatory\cite{babyIAXO20} and are also one of the proposed detector technologies for the Lynx imaging spectrometer\cite{stevenson19}.

Because cryogenic X-ray microcalorimeters are formed as small patches of thin films on a silicon substrate, they are naturally fabricated in a two-dimensional array of pixels to create an imaging spectrometer.  An international effort is underway to develop large arrays of low-temperature X-ray microcalorimeters.  The following two technologies are particularly active areas of research and development:
\begin{enumerate}
\item  arrays of the scale of 10$^4$ pixels via microfabrication technology;
\item  signal-multiplexing techniques that read out multiple pixels per readout channel, which is necessary to reduce the volume of electronics on the low-temperature (typically about 50~mK to 100~mK) stage of the spectrometer and the number of wires to that low-temperature stage (see Sec.~\ref{sec:why_mux}).
\end{enumerate}

This chapter presents issues and recent progress in signal-readout technology for arrays of X-ray microcalorimeters.

%% file: impedance_matching_2022Aug.tex
\begin{figure}
\begin{minipage}{.5\textwidth}
\begin{center}
  \includegraphics[width=1.\hsize]{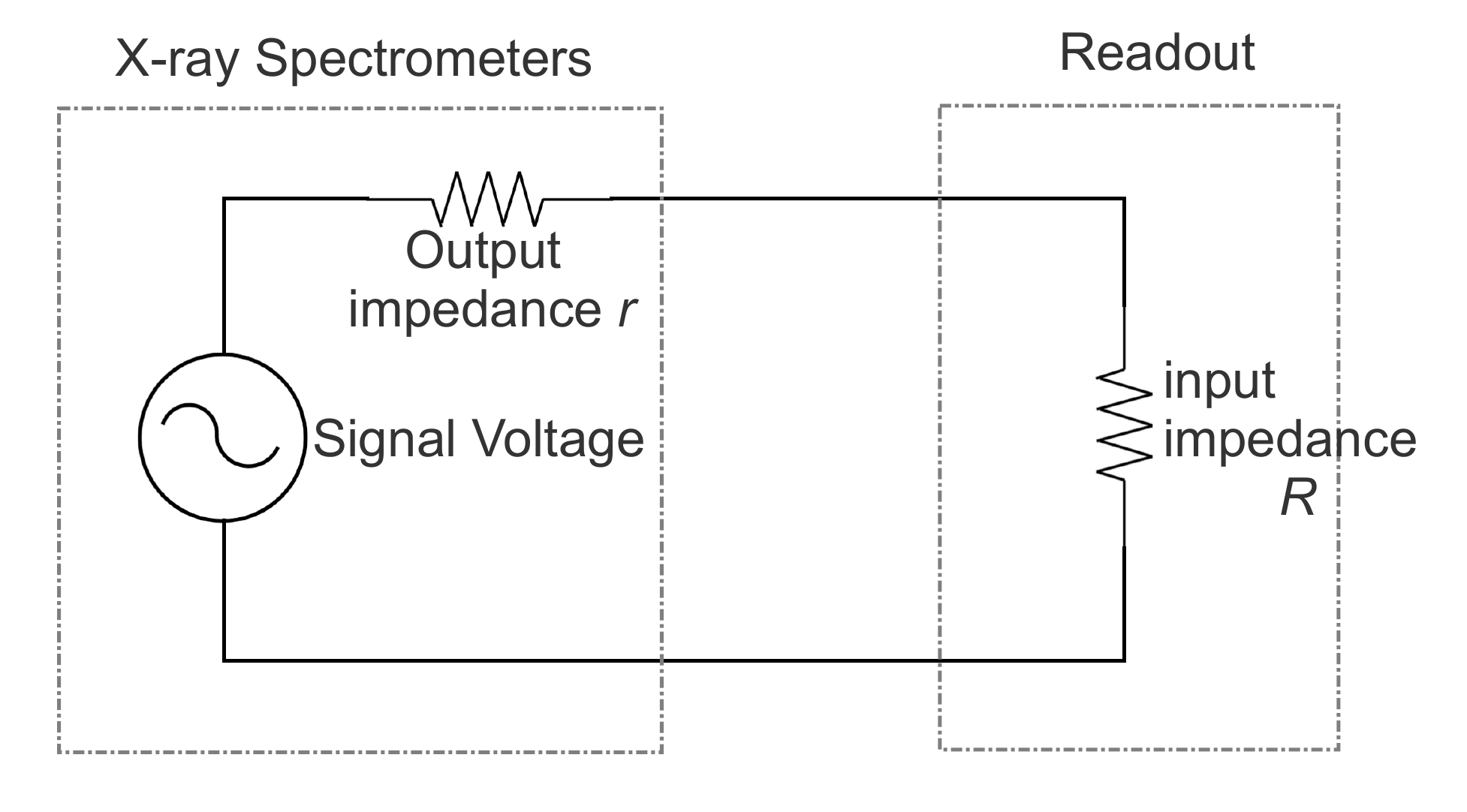}
\end{center}
\end{minipage}
\begin{minipage}{.49\textwidth}
\begin{center}
  \includegraphics[width=1.\hsize]{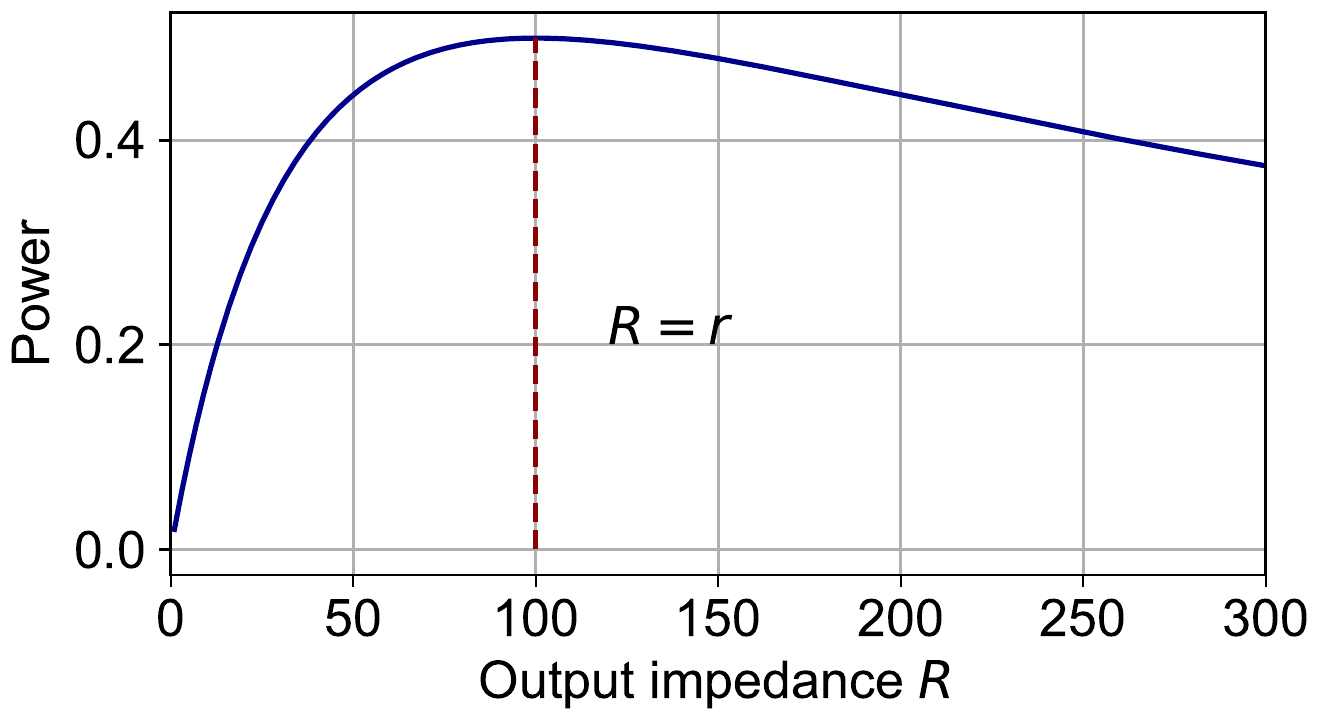}
\end{center}
\end{minipage}
\caption{{\it Left:}  Electrical schematic showing the output impedance, $r$, of an X-ray microcalorimeter and the input impedance, $R$, of its readout.  {\it Right:}  Power transmitted to the readout vs.\ $R$ for $r=100$~\Ohm.  Transmitted power is maximized when the output impedance of the microcalorimeter and the input impedance of the readout electronics are equal.}
\label{fig:circ}
\end{figure}

When two circuit elements are connected, such as a microcalorimeter pixel to its readout, a difference in impedance will cause some signal power to be reflected.  This loss of signal in the readout decreases the signal-to-noise ratio. This can be compared to exchanging water between two hoses of different diameters: different hose diameters result in water loss.

The following simple calculation illustrates the concept of impedance matching. A signal voltage, $V$, drives a current, $I$, through the circuit shown in Fig.~\ref{fig:circ}(left), given by:
\begin{equation}
I=\frac{V}{\textit{R}+\textit{r}},
\end{equation}
where $r$ and $R$ are the output impedance of an X-ray microcalorimeter pixel and input impedance of its readout, respectively. Therefore, the voltage drop across $R$ is:
\begin{equation}
V_R=RI=V\frac{R}{R+r}.
\end{equation}
The power dissipated in $R$, $P_R \equiv IV_R$, and its first derivative with respect to $R$ are:
\begin{equation}
P_R = V^2 \frac{R}{(R+r)^2} \hspace{20pt} {\rm and} \hspace{20pt} \frac{dP_R}{dR} = V^2 \frac{r-R}{(R+r)^3}.
\end{equation}
Thus, power transmission to the readout is maximized ($dP_R/dR = 0$) when the output impedance of the microcalorimeter pixel is equal to the input impedance of the readout (when the impedances are {\it matched}). Fig.~\ref{fig:circ}(right) plots $P_R$ vs.\ $R$ for $r=100$~\Ohm.

As shown in Table~\ref{tab:spec_summary}, the various microcalorimeter types have different output impedances, and so require different readout circuits.

%% file: dc_SQUID_2022Aug.tex
Superconducting quantum interference devices (SQUIDs) are a class of highly sensitive magnetic sensors that use ring-shaped superconductors containing Josephson junctions to measure extremely weak magnetic fields\cite{anderson63, jakelevic64, squid}. When two superconductors are brought close enough together that their wave functions overlap, a tunnel current flows that is proportional to the phase difference between the two superconducting wave functions. A modern Josephson junction is a sandwich of a few-nanometer-thick insulator or normal-conductive metal between two superconducting films. SQUIDs can reach magnetic field sensitivities as low as $\sim 10^{-15}$~T\citep{drung07_SQUID}; by contrast, the strength of a refrigerator magnet is about 0.02~T, while the magnetic fields of the human heart and brain are orders of magnitude lower ($\sim10^{-12}$~T)\cite{cohen72}. There are two main types of SQUIDs: direct current (dc) and radio frequency (rf). The latter are easier to manufacture but are less sensitive because they operate with a single Josephson junction.

\begin{figure}
\begin{center}
  \includegraphics[width=0.9\textwidth]{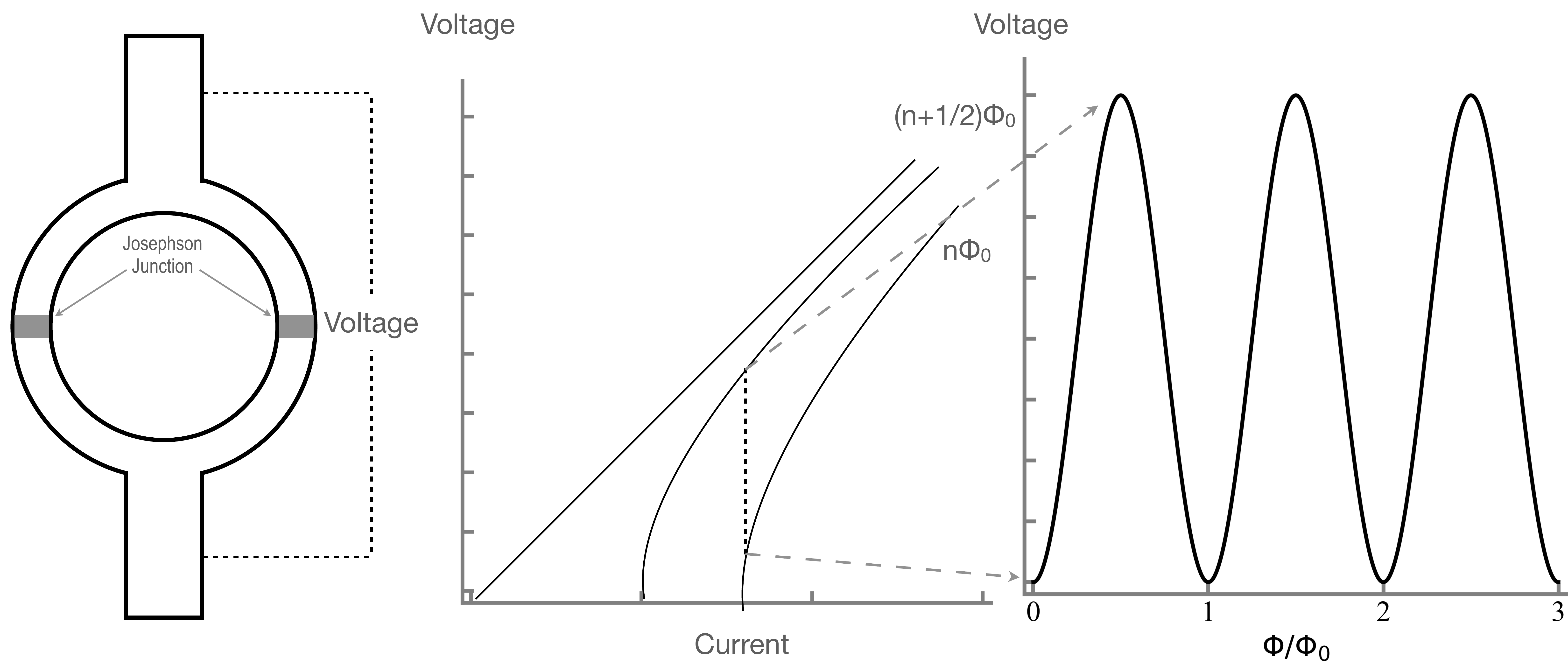}
\end{center}
\caption{{\it Left:}  sketch of a dc-SQUID, consisting of two Josephson junctions in a superconducting loop.  {\it Middle:}  simulated $V$ vs.\ $I$ curves in a dc-SQUID for the cases of an integral ($\Phi = n\Phi_0$) and a half-integral ($\Phi = [n+1/2]\Phi_0$) number of applied flux quanta ($n$ is an integer).  {\it Right:} simulated $V$ vs.\ $\Phi$ curve for a dc-SQUID.}
\label{fig:dcsquid}
\end{figure}

The dc-SQUID is a sensitive magnetometer with a wide inherent frequency bandwidth from dc up to the gigahertz range. It has two Josephson junctions in a washer-shaped superconducting loop (see Fig.~\ref{fig:dcsquid}-left).  An external magnetic flux applied to the loop drives a screening current that maintains the total flux enclosed by the loop as an integral number of flux quanta (the magnetic-flux quantum is $\Phi_0 = h/2e = 2.07\times10^{-15}$~Wb, where $h$ is the Planck constant and $e$ is the charge of the electron).  This modulates the maximum supercurrent that can flow across the loop periodically, with a period of $\Phi_0$.  When the current through either Josephson junction exceeds its critical current, a voltage develops.  The voltage-vs.-current ($V$-$I$) curve (see Fig.~\ref{fig:dcsquid}-middle) is thus modulated as a periodic function of the external magnetic field. When the external flux is zero, the phase difference of the superconducting wave functions in the two sides of the SQUID is small (high critical current and thus lower voltage; $\Phi = n\Phi_0$ case, where $n$ is an integer), while when the external magnetic field is close to $(n+1/2)\Phi_0$, the critical current is suppressed and the generated voltage is higher. When the current bias applied to the device is higher than its highest critical current, the output voltage shows a sinusoidal response to applied magnetic flux.

Series arrays of dc-SQUIDs\cite{welty} are used to increase the voltage gain.  Some designs (typically operating at a temperature of 4~K or lower) can approach quantum-limited energy resolution (the energy resolution per bandwidth approaches $\hbar$).

Parasitic inductances and capacitances in the circuitry of a dc-SQUID can lead to resonances that amplify gigahertz-frequency Josephson oscillations in the SQUID.  These resonances can cause undesirable kinks in the otherwise sinusoidal response of the SQUID and excess noise. Careful design, including the use of resistive structures to damp microwave resonances, can reduce or eliminate such behavior\cite{huber_SSA,huber_SSA_2}.

%% file: why_mux_2022Aug.tex
X-ray astronomy differs significantly from other fields in that X-ray signals from celestial objects do not reach the ground due to atmospheric absorption.  Thus, X-ray observatories (and also, generally, gamma-ray observatories) {\it must} be placed on space-borne platforms such as rockets and satellites.

\begin{figure}
\begin{center}
  \includegraphics[width=0.6\textwidth]{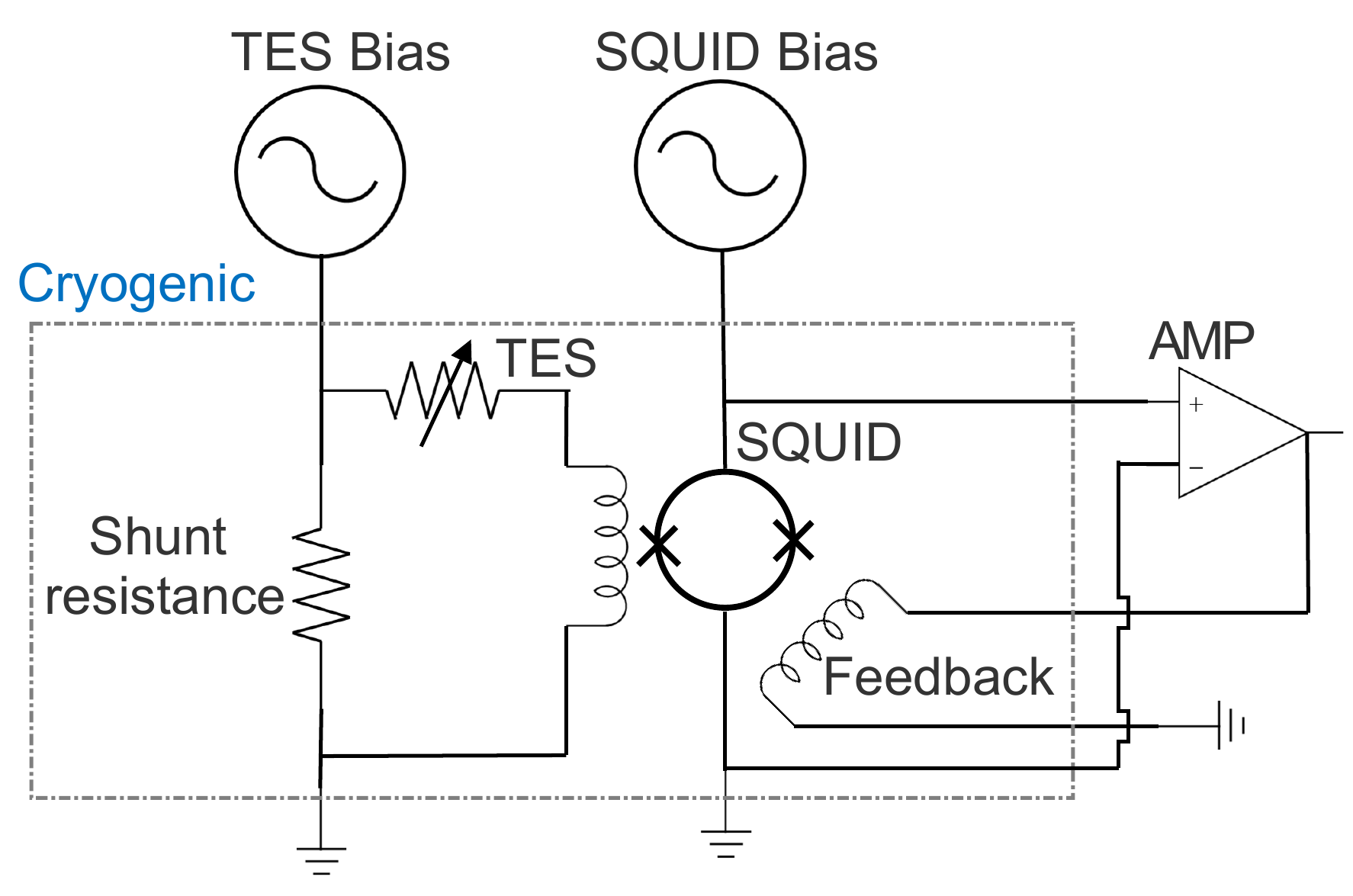}
\end{center}
\caption{Schematic of the readout of a TES via a single-stage dc-SQUID.  Cryogenic components are inside the dashed box.  This is the simplest reasonable implementation of (non-multiplexed) readout of a TES, and requires three cryogenic wire pairs per TES (for TES bias, SQUID bias/output, and SQUID feedback).  However, typically, a two-stage-SQUID architecture is needed to achieve the required gain without dissipation of too much Joule power on the coldest cryogenic stage (see Fig.~\ref{fig:tdm-singlepixel}); in this case, at least five cryogenic wire pairs per TES are needed.}
\label{fig:wires}
\end{figure}

Low-temperature microcalorimeters make excellent X-ray imaging spectrometers, as described by \blue{Gottardi \& Smith in this book}.  However, readout of large arrays of microcalorimeters is nontrivial due to the electrical power and cooling capacity available in a satellite.  For example, the total electrical draw of a typical communication satellite is about 1~kW to 1.5~kW\footnote{https://www.esa.int/Enabling\_Support/Space\_Engineering\_Technology/Power\_Systems} (about the same as a standard household microwave oven), while the Hubble Space Telescope uses about 2.8~kW\footnote{https://www.nasa.gov/content/goddard/hubble-space-telescope-electrical-power-system}.  For X-ray astronomy, the Athena satellite has a planned power draw of $\sim$6~kW, of which the X-IFU instrument will draw about 1.3~kW\cite{xifu14}. 

As discussed in Fig.~\ref{fig:wires}, readout of a single TES microcalorimeter pixel requires a minimum of three cryogenic wire pairs, while five pairs per TES are typical.  Thus, brute-force readout of X-IFU's $\sim$2,400 TESs\cite{xifu18} would require more than 10,000 wire pairs.  The power draw from 2,400 channels of room-temperature electronics plus a cryogenic system capable of overcoming the thermal conductivity of this many wires would overwhelm any feasible satellite platform.  Future arrays of X-ray microcalorimeters will be even larger.  To solve this problem, signal-multiplexing techniques that read signals from multiple sensors with fewer wires are essential.

%% file: general_mux_2022Aug.tex
\begin{figure}
\begin{center}
  \includegraphics[width=\textwidth]{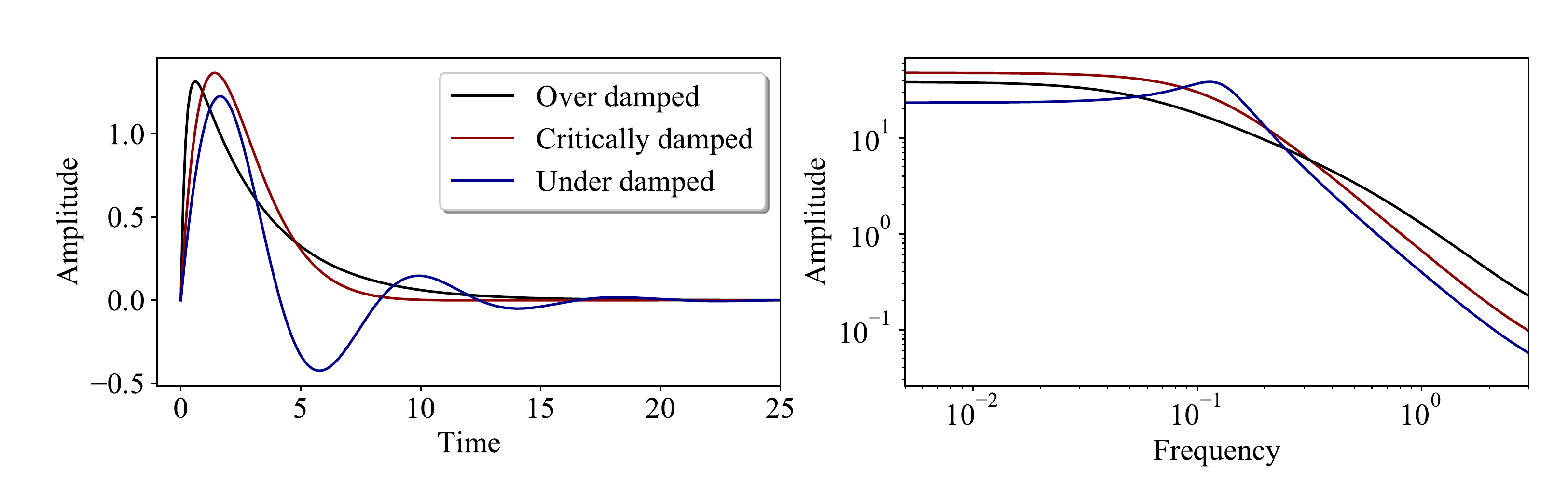}
\end{center}
\caption{Simulated TES-current pulses (left:  current vs.\ time; right:  current$*$time vs.\ frequency) under various damping conditions.  In the overdamped condition, the relation between the TES time constants is $\tau\urss{rise} < \tau\urss{decay}$ (and both are real).  In a critically damped pulse, $\tau\urss{rise} = \tau\urss{decay}$.  In an underdamped pulse, the time constants are complex conjugates. 
}
\label{fig:critical_damp}
\end{figure}

Multiplexed readout should introduce minimal noise and signal degradation.  The noise budget is application-dependent:  e.g. in Athena X-IFU, about 3\% of the energy-resolution budget is allocated to readout.  Successful multiplexing generally entails these three considerations:

{\bf 1.  The signal and noise bandwidths of the sensor should be made as narrow as feasible for the astronomical application.}  The signal and noise currents of a TES X-ray microcalorimeters generally obey two low-pass time constants:  the electrical time constant (determines behavior on the leading edge of an X-ray pulse) and the thermal time constant (determines the pulse's decay behavior).  Irwin and Hilton\cite{irwin05} discuss the theory of TES signals, including how electrothermal feedback (ETF) governs the interaction between the electrical and thermal time constants.  Here, we refer to the TES's ETF-modified time constants on the leading and falling edges of a pulse as $\tau\urss{rise}$ and $\tau\urss{decay}$, respectively.  The achievable X-ray count rate is governed by $\tau\urss{decay}$:  a faster-decaying detector exhibits less pulse pileup for a given input rate of X-rays.  The signal bandwidth of the TES is governed by $\tau\urss{rise}$.  The ratio between $\tau\urss{rise}$ and $\tau\urss{decay}$, is generally adjusted via inductance in the TES-bias loop (see, e.g., $L\urss{Ny}$ in Fig.~\ref{fig:tdm-singlepixel}).  Fig.~\ref{fig:critical_damp} shows simulated TES-current pulses under various damping conditions.  The critically damped (or just a bit overdamped) condition is often the design goal for the TES-bias circuit:  for a given TES model, critical damping provides the largest $\tau\urss{rise}$, and thus the narrowest overall signal bandwidth, and so is often preferred for multiplexed readout.  Also, because $\tau\urss{rise}$ cannot be longer than $\tau\urss{decay}$, per-pixel X-ray counting capability is in natural tension with multiplexing factor (and thus with array size).  Finally, filtering limits sensor noise from being aliased within a multiplexer pixel or from leaking into neighboring multiplexer pixels:  the TES-current noise due to phonon exchange with the bath is low-pass filtered with the time constants $\tau\urss{decay}$ and $\tau\urss{rise}$, while the TES-Johnson noise is low-pass filtered with the time constant $\tau\urss{rise}$.

{\bf 2.  The readout system should have a wide bandwidth.}  Multiplexed readout is essentially an exercise in stuffing many low-bandwidth TES signals into a high-bandwidth readout system.  Higher readout bandwidth generally allows more TESs per readout channel.

\begin{table}[t]
\def\arraystretch{1.5}
\caption{Summary information for four multiplexed-readout technologies for TESs X-ray calorimeters:  time-division (TDM), code-division (CDM), MHz-frequency-division (FDM), and microwave-SQUID (\umux) multiplexing.  The top section lists the type of TES bias and the type of modulation.  The second section discusses the front-end SQUID.  The third section discusses the 2nd-stage amplifier; in \umux\ readout, a high-electron-mobility transistor (HEMT) amplifier is traditionally used, but the recently developed kinetic-inductance traveling-wave parametric amplifier (KITWPA) offers an attractive lower-power alternative.  Finally, the fourth section lists the state of the art in the readout of X-ray-TES arrays.}
\begin{center} 
\scriptsize
\begin{tabular}{l|cccc} \hline
	&	TDM	\cite{chervenak99, durkin2021}	& 	CDM\cite{irwin10, morgan16}			& MHz FDM\cite{yoon01, kiviranta02}  		& \umux\cite{irwin04, mates2008demonstration}		\\
\hline
TES Bias    &	dc	&	dc	&	ac	& 	dc	\\ 
how modulated?	&   SQUIDs on/off	&	Walsh codes \cite{walsh}	&   resonating TESs 	&   resonating SQUIDs  \\
\hline
front-end SQUID		&	dc	&	dc	& 	dc	&	rf		\\ 
FE SQ:  $P\urss{Joule}$&	450 pW\cite{durkin20}	&   450 pW\cite{doriese_ASC2019}		&    450 pW\cite{kiviranta21}	&    20 pW \\ 	\\
\hline
2nd-stage amplifier & 	SQUID array		&	SQUID array	&	SQUID array	& 	HEMT\cite{duh1988ultra} \\
\raiseme{2nd stage:  $P$ @ $T$}	&	\raiseme{$\sim$300~nW @ 2~K}	&	\raiseme{$\sim$300~nW @ 2~K}	& \raiseme{$\sim$300~nW @ 2~K}	& \raiseme{$\sim$5~mW @ 4~K} \\
  &  &  &  &  KITWPA \cite{malnou}\\
  &  &  &  &  \raiseme{{(0.2 to 20)~$\mu$W @ 4~K}$^{*}$}\\ 
\hline
demonstrated $\Delta E\urss{FWHM}$ @ $E$ &  2.23~eV @ 6~keV  &  2.77~eV @ 6~keV &  2.23~eV @ 6~keV &  3.3~eV @ 6~keV  \\
\raiseme{@ channels $\times$ mux factor}  &  \raiseme{@ $1 \times 40$ \cite{durkin19}}  &  \raiseme{@ $1 \times 32$ \cite{morgan16}}  &  \raiseme{@ $1 \times 37$ \cite{akamatsu21}}  &  \raiseme{@ $1\times37$ \cite{nakashima20}}  \\
  &  1.98~eV @ 6~keV &  &  2.14~eV @ 6~keV  &  3.4~eV @ 6~keV   \\
  &  \raiseme{@ $8 \times 32$ \cite{smith2021}}  &  &  \raiseme{@ $1 \times 31$ \cite{akamatsu21}} &  \raiseme{@ $1 \times 28$ \cite{yoon18}} \\
  &  &  &  & 2.04 eV @ 1.25 keV   \\
  &  &  &  & \raiseme{@ $1 \times 88$ \cite{bennett2019microwave}} \\
  &  &  &  & 10 eV @ 6$^{\dagger}$ keV   \\
  &  &  &  & \raiseme{@ $1 \times 116$ [pre-pub. comm.]} \\
\hline
\multicolumn{5}{l}{$*$ KITWPA power dissipation depends on pump power and therefore scales with compression point of the amplifier.} \\
\multicolumn{5}{l}{$\dagger$ TOMCAT\cite{szypryt2021design} TESs were optimized for very high count rates and not for energy resolution.} \\
\end{tabular}
\end{center}
\label{tab:summary}
\end{table}

\begin{figure}
\begin{center}
  \includegraphics[width=\textwidth]{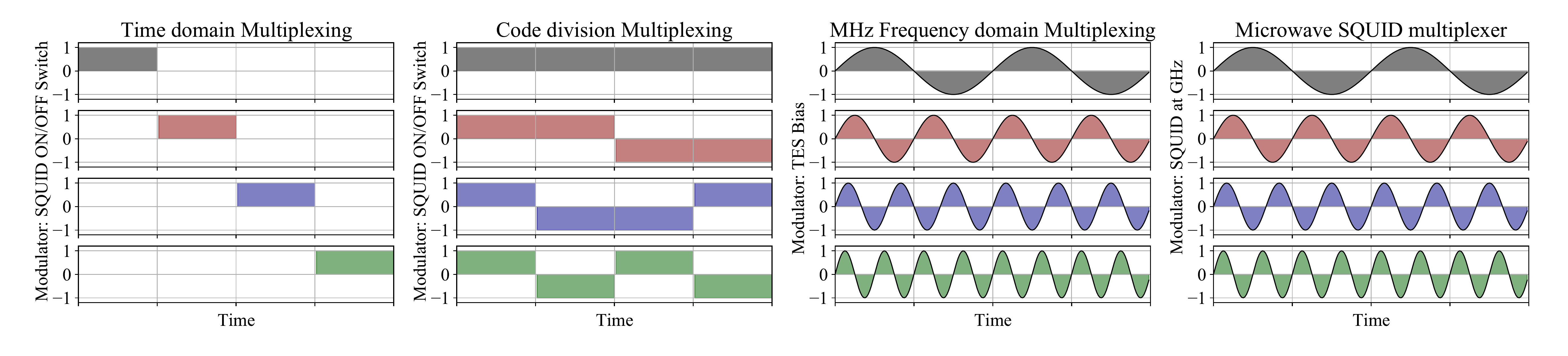}
\end{center}
\caption{Multiplexer modulation functions vs.\ time for the basis vectors of four-channel versions of time-division (TDM), code-division (CDM), MHz-band frequency-domain (FDM), and microwave-SQUID multiplexing (\umux).  Colors indicate the different pixels.  Modulation in TDM is via switching SQUIDs on and off, so the basis set is the identity matrix.  In CDM, TESs are coupled to different SQUIDs with different polarities to create orthogonal Walsh codes.  In FDM, different TESs are biased at different megahertz frequencies.  In \umux, different rf SQUIDs are biased at different gigahertz frequencies.  Both FDM and \umux\ thus use sine waves to form their orthogonal basis sets.  
}
\label{fig:readout_example}
\end{figure}

{\bf 3.  The multiplexer should be built from an orthogonal basis set to prevent signal leakage from pixel to pixel.}  Fig.~\ref{fig:readout_example} shows the basis sets used by the main multiplexing techniques for TESs to be discussed in Secs.~\ref{sec:tdm} to \ref{sec:umux}:  time-division multiplexing (TDM), frequency-domain multiplexing (FDM), and microwave-SQUID multiplexing (\umux).  In addition, this figure shows the basis set of code-division multiplexing (CDM)\cite{irwin10, irwin_FAS, morgan16, doriese_ASC2019, yu20}.  Irwin, et al.\cite{irwin09} discuss, in the context of information theory, the number of pixels that can be multiplexed via the various techniques.  A summary of these techniques is given in Table~\ref{tab:summary}.

TDM and \umux\ for TESs are under development by the Quantum Sensors Group at NIST's Boulder Labs.  FDM for TESs was initially demonstrated\cite{cunningham02} by a team from Lawrence Livermore National Lab, UC Berkeley, and Lawrence Berkeley National Lab, and is under development by UC Berkeley, McGill University, and SRON.  CDM for TESs was pursued by NIST as an early option for Athena X-IFU and Lynx, but is not presently under development for any mission and so is not discussed further in this chapter.  MMCs can be read out via \umux\cite{wegner18}.

%% file: TDM_2022Aug.tex
In the time-division-multiplexing (TDM) technique, each TES has its own first-stage dc-SQUID (SQ1).  The TESs are dc-biased and are always on, while the SQ1s are turned on and off such that one SQ1 is on at a time per readout column\footnote{In TDM, an amplifier chain is usually referred to as a readout ``column.''  Throughout this chapter, we use the terms ``amplifier chain,'' ``readout channel,'' ``multiplexer channel,'' and ``readout column'' interchangeably to refer to the logical readout unit that is divided into detector pixels via multiplexing.}.  Only the signal current from the TES whose SQ1 is on is read out, instantaneously, by the column.

The TDM scheme was proposed in 1999\cite{chervenak1999} for the readout of arrays of TES bolometers and microcalorimeters for various astronomical applications.  In the 20+ years since, several TDM-based bolometric arrays\cite{scuba2, act_swetz, hawc_plus, piper}, each of the kilopixel to multi-kilopixel scale, have been deployed for far-IR, millimeter-wave, and submillimeter-wave astronomy.  In addition, about 15 TDM-based X-ray spectrometers\cite{doriese2017, spring8, ssrl}, each of the few-hundred-pixel scale (to be discussed further in Sec.~\ref{sec:tdm_lab}), have been deployed to X-ray-science facilities around the world.  A 128-pixel TDM X-ray array flew on the Micro-X sounding rocket in 2018\cite{microx} and again in August 2022.  Presently, TDM is undergoing refinement\cite{smith2021, durkin2021} for the multi-kilopixel X-IFU imaging spectrometer for ESA's Athena mission (see Sec.~\ref{sec:tdm_XIFU}).

How does TDM compare to the other main multiplexing technologies for TES microcalorimeters, philosophically and at the systems level?  Philosophically, in TDM, the SQ1s are the modulated elements, while the TESs themselves are not modulated; this is the same as in the microwave-multiplexing (\umux; Sec.~\ref{sec:umux}) scheme, while the frequency-domain scheme (FDM; Sec.~\ref{sec:fdm}) takes the opposite approach.  At the systems level, in TDM the cryogenic electronics are the most complicated, followed by those in \umux\ and then FDM, while the TDM room-temperature electronics are simpler than those of either FDM or \umux.

\subsubsection{Principles of TDM operation}\label{sec:tdm_oper}

\begin{figure}
\begin{center}
  \includegraphics[width=0.8\textwidth]{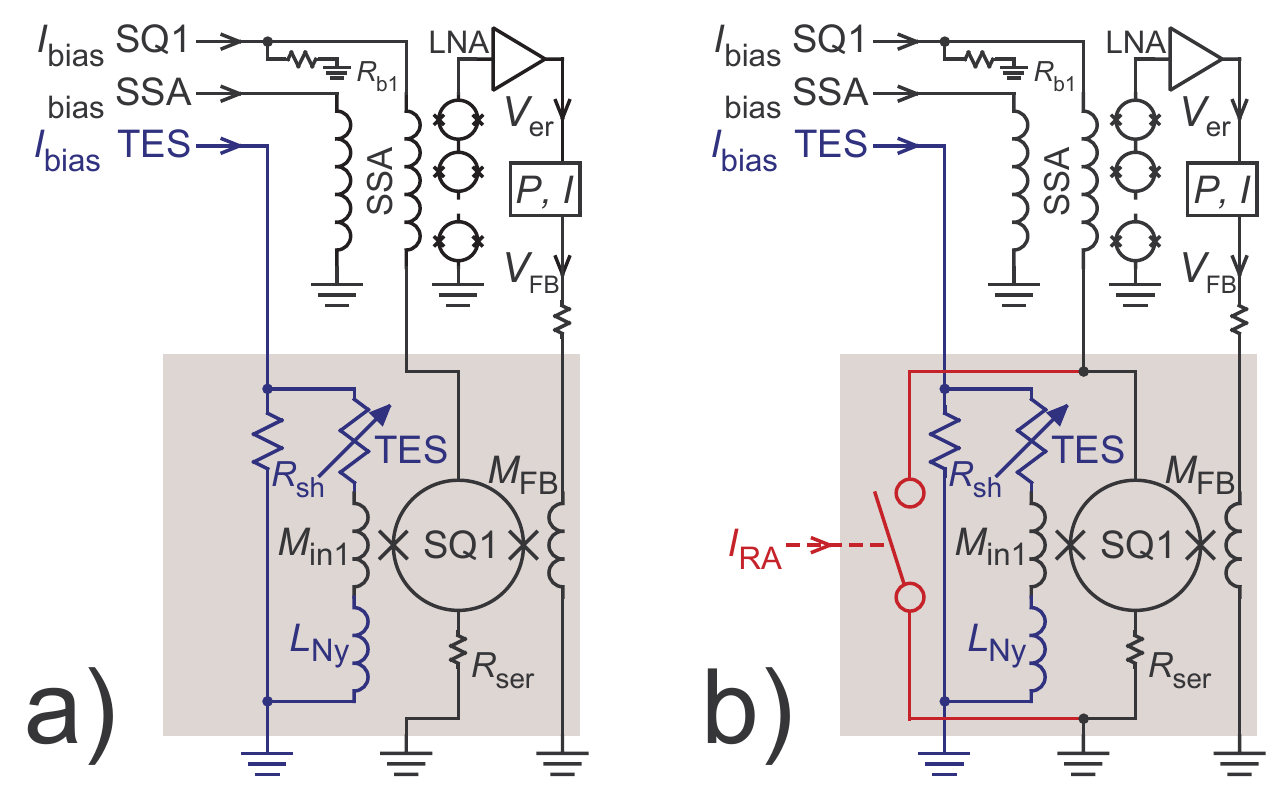}
\end{center}
\caption{Readout of a dc-biased TES by a dc-SQUID chain.  (a) A single TES under dc bias is read out by a first-stage dc SQUID (SQ1) and a series array of dc SQUIDs (SSA).  The signal of interest, the TES current, is converted to flux in SQ1 via its input coil ($M\urss{in1}$).  The SSA reads out the SQ1 current in the same manner.  The elements in the grey box are located on the same temperature stage as the TES, which for X-ray TESs is typically around 50~mK.  TES-related circuit elements (the TES shunt resistor, or $R\urss{sh}$, and the bandwidth-limiting Nyquist inductor, or $L\urss{Ny}$) are blue, while the SQUID-related circuit elements are in black.  The SSA is usually located at a higher-temperature cryogenic stage to accommodate its Joule-power dissipation.  (b) The same circuit diagram as (a), but with the addition of a superconducting switch (red) to turn on and off the SQ1.  When the switch is closed, the SQ1-bias current bypasses SQ1, so the TES signal current is not transmitted up the readout chain.  When the switch is open, circuit (b) behaves like circuit (a).}
\label{fig:tdm-singlepixel}
\end{figure}

Fig.~\ref{fig:tdm-singlepixel} shows circuit diagrams that contain the building blocks of a TDM readout system.  Fig.~\ref{fig:tdm-singlepixel}(a) shows readout of a single dc-biased TES by two stages of dc SQUIDs (see Fig.~\ref{fig:dcsquid}).  The signal of interest in a TES X-ray microcalorimeter is its current, which is modulated as a pulsed decrement by each incident X-ray; the amplitude of the TES-current pulse is roughly proportional to the energy of the X-ray.  A change in the TES current modulates the flux in the first-stage dc-SQUID (SQ1), which in turn modulates the SQ1 bias current.  The current-biased SQUID-series-array\cite{huber_SSA,huber_SSA_2} (SSA) amplifier transduces the SQ1 current to a voltage at the SSA output, which is amplified further by a room-temperature low-noise amplifier (LNA).

Because the (quasi-sinusoidal) transfer function of the combined $V$-$\Phi$ curve of the SQ1 and the SSA is highly nonlinear, the readout is run as a flux-locked loop (FLL) to linearize the response.  Thus, the ``output'' signal of this readout system is the feedback voltage, $V\urss{FB}$, that is applied to null the TES-signal current to maintain a constant flux in its SQ1.

Fig.~\ref{fig:tdm-singlepixel}(b) shows the addition of a superconducting switch to turn on and off the SQ1.  The switch chosen for modern TDM implementations\cite{beyer_drung_2008, doriese2016} is based on the dc-SQUID-like Zappe interferometer\cite{zappe}.  An interferometer element consists of several Josephson junctions in parallel (with the number of junctions varying across implementations).  The ``row-address'' (RA) current ($I\urss{RA}$) is coupled to the SQUID loops such that with $I\urss{RA}=0$ the interferometer acts as a superconducting wire (the switch is ``closed''), while with $I\urss{RA}=\Phi\urss{0}/2$ the junctions are perfectly out of phase and the nominal current-carrying capacity of the interferometer is zero (the switch is ``open'').  In the NIST implementation\cite{irwin_FAS}, each interferometer element contains four junctions with equal critical currents for ease of fabrication and wide operating margins.  Tens of these interferometric elements are wired in series to create an operating dynamic resistance of this ``flux-actuated switch'' (FAS) that is much larger than that of the signal SQ1.

\begin{figure}
\begin{minipage}{.57\hsize}
\begin{center}
  \includegraphics[width=1.\hsize]{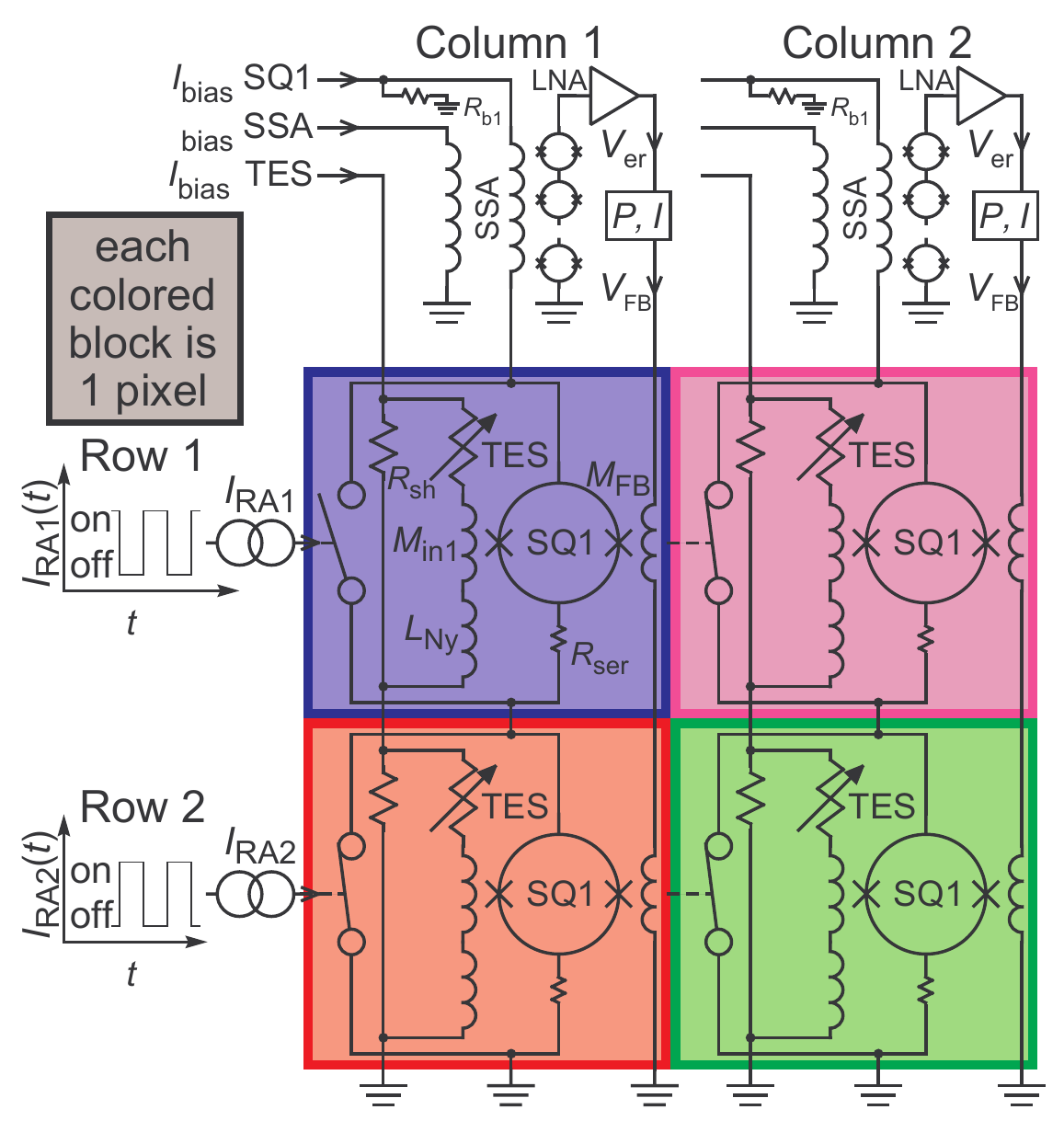}
\end{center}
\end{minipage}
\begin{minipage}{.41\hsize}
\begin{center}
  \includegraphics[width=1.\hsize]{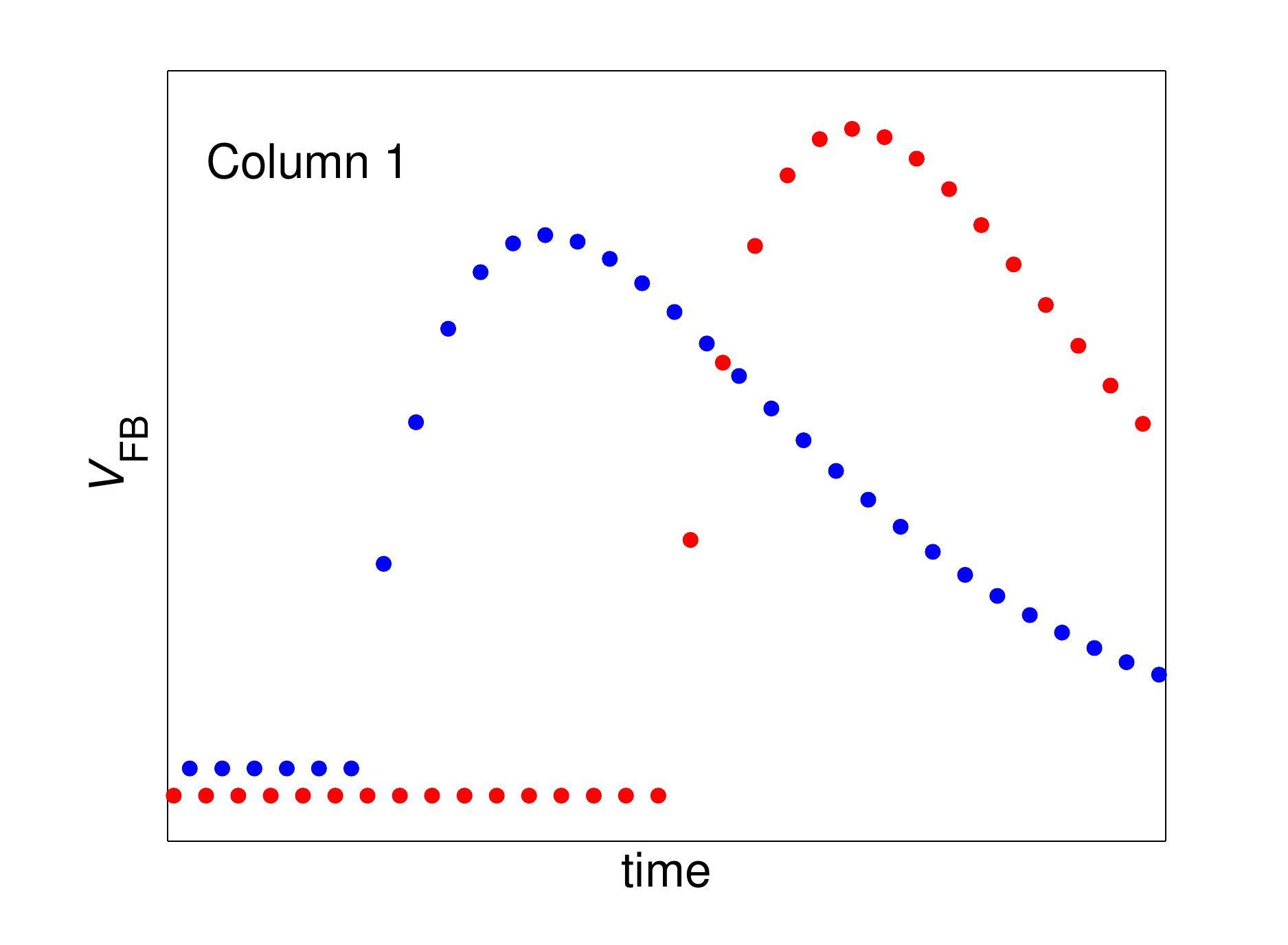}
  \includegraphics[width=1.\hsize]{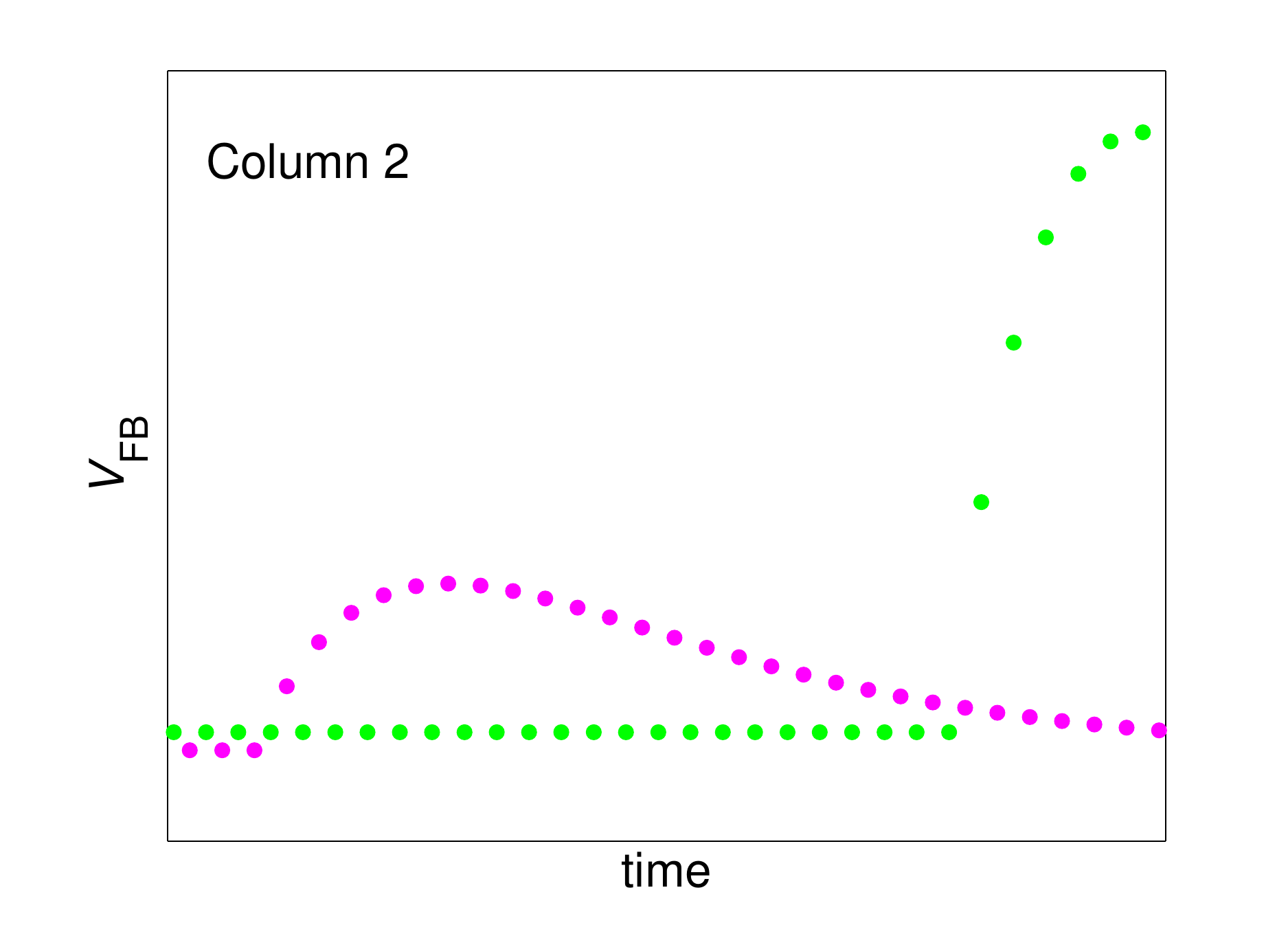}
\end{center}
\end{minipage}
\caption{(left)  Schematic of a two-column by two-row TDM.  Each colored block is a unit pixel and contains the same cryogenic components as shown in the grey box in Fig.~\ref{fig:tdm-singlepixel}(b).  Rows of SQ1s are sequentially turned on via application of a row-address current ($I\urss{RA}$) to their flux-actuated switches.  Thus, only one TES at a time per column is read out.  To keep the non-linear, two-stage SQUID amplifier in a quasi-linear range, each column is run as a set of digitally interleaved flux-locked loops.  (right)  Color-coded, simulated data streams showing the interleaved $V\urss{FB}$ signals for Columns 1 and 2 as the four TESs respond in current pulses to X-ray events.}
\label{fig:tdm-2x2}
\end{figure}

Fig.~\ref{fig:tdm-2x2}(left) shows a cartoon circuit schematic of a 2-column \by 2-row TDM system.  In a general $M$-column \by $N$-row array of TESs, both the count of cryogenic wires and the number of amplifier chains would scale as $MN$ under brute-force (non-multiplexed) readout; with TDM readout these scale only as $\sim(M+N)$ and $M$, respectively.  These savings in channels and wires are the main reason to use a multiplexed-readout scheme.

Rows of SQ1s are sequentially turned on via application of a RA current ($I\urss{RA}$) to their FASs.  TDM columns are read out in parallel.  The signal from one TES at a time per column is passed to that column's SSA.  During a row, the LNA's output voltage, or error signal ($V\urss{er}$), is digitally sampled, and then the proportional-integral $(P,I)$ flux-feedback signal ($V\urss{FB}$) that would servo the row's $V\urss{er}$ to a constant value is digitally calculated and stored to be applied inductively (via $M\urss{FB}$) on the digital FLL's next visit to the row.  The digital-readout electronics (DRE; to be discussed further in Sec.~\ref{sec:tdm_electronics}) synchronize the RA and D-FLL signals and stream the TES-current data to the back-end processor.  The TDM system samples each TES once per frame time ($t\urss{fr} = t\urss{row}N\urss{rows}$); if the frame time is short compared to the time constants of the TES-current signals, then the system can reconstruct each TES's current with high fidelity. Fig.~\ref{fig:tdm-2x2}(right) shows the multiplexed time streams of the flux-feedback signals of the two two-pixel TDM columns.

\subsubsection{Circuit parameters, multiplexing factor, and noise scaling}\label{sec:tdm_params}

For the dimensioning of a TDM system, an important characteristic of the TES signal-current ($I\urss{TES}$) is its slew rate.  The highest required TES-current slew-rate, $\dot{I}\urss{max}$, will occur at the leading edge of an X-ray pulse and is a function of the TES design and its bias-circuit inductance and the highest X-ray energy of interest.  If the D-FLL system is to function correctly, the flux difference in any SQ1 from one sample to the next, injected into $M\urss{in1}$ by $I\urss{TES}$, must result in an error signal that remains within a quasi-linear region of the combined SQ1-SSA $V$-$\Phi$ curve (see Fig.~\ref{fig:dcsquid} for an example curve).  Thus, the flux difference must obey $\Delta\Phi \le F\Phi\urss{0}$.  The fraction, $F$, of the flux quantum over which the SQUID curve is roughly linear depends on the details of the SQUID designs and is usually between 0.2 and 0.5.  The flux difference between frames is $\Delta\Phi = \dot{I}M\urss{in1}t\urss{fr}$.  Thus\cite{doriese_2006},
\begin{equation}
\dot{I}\urss{max}M\urss{in1}t\urss{row}N\urss{rows} \le F\Phi\urss{0}
\label{eq:tdm-dPhiMax}
\end{equation}
sets a joint condition on $M\urss{in1}$ and $N\urss{rows}$.  If all else is equal, a larger multiplexing factor requires a proportionally smaller input coupling of TES current to SQ1 flux.

The second important consideration in dimensioning a TDM system is readout noise.  A given TES X-ray microcalorimeter design will have a value of the spectral density of (white) multiplexed readout noise ($\sqrt{S_{I\mhyphen\mathrm{mux}}}$; referred to the TES current; units of A/$\surd$Hz) above which it is unable to meet the mission/experimental specification on energy resolution.  A goal of the system design, then, is to keep $\sqrt{S_{I\mhyphen\mathrm{mux}}}$ below this maximum-allowed value.  The SQUID-amplifier system's noise (including contributions from both stages of dc-SQUIDs and the room-temperature electronics) is approximated as being white (with spectral density $\sqrt{S_{\Phi}}$; referred to SQ1 flux; units of $\Phi\urss{0}/\surd$Hz) out to a high-frequency, with a single-pole, low-pass rolloff with time constant $\tau\urss{OL}$.  In a modern TDM system\cite{doriese2016} values of $\sqrt{S_{\Phi}}= 0.2 \mu\Phi\urss{0}/\surd$Hz and $f\urss{OL} \sim 8$~MHz to 10~MHz are reasonable.  The large open-loop bandwidth, $f\urss{OL}$, needed to switch quickly from row to row in TDM, combined with the readout strategy that samples each row only once per frame, means the Nyquist anti-aliasing criterion, $1/t\urss{fr} \ge 2f\urss{OL}$, is not met for the readout noise.  As a result, high-frequency amplifier noise is aliased into the TES-signal band.  This noise aliasing is an inherent condition of the TDM method and is generally what limits the multiplexing factor for a given application.  The TES-referred, multiplexed readout-noise level is given by\cite{doriese_2006}:
\begin{equation}
\sqrt{S_{I\mhyphen\mathrm{mux}}} = \sqrt{A N\urss{rows}S_{\Phi}}/M\urss{in1}.
\label{eq:tdm-noise}
\end{equation}
Here, $A \ge 1$ is an alias-scaling factor that depends on the digital-sampling strategy and the ratio $t\urss{row}/\tau\urss{OL}$.  For the common operational case in which $t\urss{row} \sim 2\pi\tau\urss{OL}$ and the digital sampling occupies a small fraction of $t\urss{row}$ at the end of the row period\cite{doriese_2006}, $A\sim\pi$.  In the limit of $f\urss{OL} \rightarrow \infty $, in which the digital sampling can occupy the full row period, $A \rightarrow 1$.  In both the modern lab systems (see Sec.~\ref{sec:tdm_lab}) and the proposed design of Athena X-IFU (see Sec.~\ref{sec:tdm_XIFU}), the row time is chosen to be $t\urss{row}=160$~ns; this is well matched to the achievable open-loop bandwidth in standard dc-coupled twisted-pair and flexible-microstrip circuits of $f\urss{OL} \sim 8$~MHz to 10~MHz.

Eq.~\ref{eq:tdm-noise} shows that the TES-referred readout noise scales as $\sqrt{N\urss{rows}}/M\urss{in1}$, while Eq.~\ref{eq:tdm-dPhiMax} shows that $M\urss{in1}$ scales as $1/N\urss{rows}$.  Thus, in a TDM system, TES-referred readout noise scales with the multiplexing factor as $\sqrt{S_{I\mhyphen\mathrm{mux}}} \propto N\urss{rows}^{3/2}$.

\subsubsection{Room-temperature electronics}\label{sec:tdm_electronics}

The room-temperature electronics for TDM readout separate neatly into two functional pieces:  the warm-front-end electronics (WFEE) and the digital-readout electronics (DRE).

The WFEE contains the LNA and generates various low-frequency analog-bias signals for the SQUIDs and TESs.  Because the pre-LNA SSA-output signals are small (a few millivolts) and thus EMI-sensitive, the WFEE often shares a Faraday cage with the cryostat.  For high-performance TDM readout of TES X-ray microcalorimeters, the LNA is challenging to implement.  Its needed specifications are $f\urss{OL} \ge 10$~MHz and input-voltage noise below 1~nV/$\surd$Hz; this combination is on the edge of what is available in commercial op-amps, so the front end of the LNA is often assembled from discrete transistors.  Various groups have produced WFEE modules that have been used for TDM readout over the last 20+ years\cite{MCE, drung_XXF,reintsema_2003, doriese2016, kazu_COTS, APC_WFEE}.

The DRE generates the fast RA signals to drive the FASs and runs the D-FLL for each column.  In most ground-based applications, raw D-FLL data are streamed to a host computer for demultiplexing, triggering, filtering, and other signal processing; however, on Athena X-IFU, these higher-level functions will be performed within the DRE.  Various versions of TDM-DRE have been produced and used over the last 20+ years\cite{MCE, reintsema_2003, reintsema_2009, kazu_COTS}.

\subsubsection{Laboratory TDM systems}\label{sec:tdm_lab}

Between 2010 and 2022, the NIST Quantum Sensors Group (Boulder, CO, USA) has deployed about 15 TDM-based TES-microcalorimeter arrays to various X-ray-spectroscopy experiments and facilities around the world\cite{doriese2017}.  This development effort has allowed TDM technology to be tested in real-world conditions beyond the traditional environment of the detector lab.  Application areas have included time-resolved absorption and emission spectroscopy with a tabletop, broad-band source\cite{uhlig_2013, miaja_2016, oneil_2017}, synchrotron-based X-ray-emission and absorption spectroscopy\cite{spring8, ssrl}, synchrotron-based energy-resolved scattering\cite{lee_2022}, particle-induced X-ray-emission spectroscopy\cite{palosaari_2016}, spectroscopy of hadronic and muonic atoms\cite{hashimoto_kaonic, okumura_muonic}, the metrology of X-ray-line energies \cite{fowler_2021, szypryt_EBIT}, and X-ray tomography of integrated circuits\cite{szypryt_tomography}.  Similar arrays of gamma-ray TESs have been used for the assay of special-nuclear materials\cite{bennett_gamma}.

\begin{figure}
\begin{center}
  \includegraphics[width=0.6\textwidth]{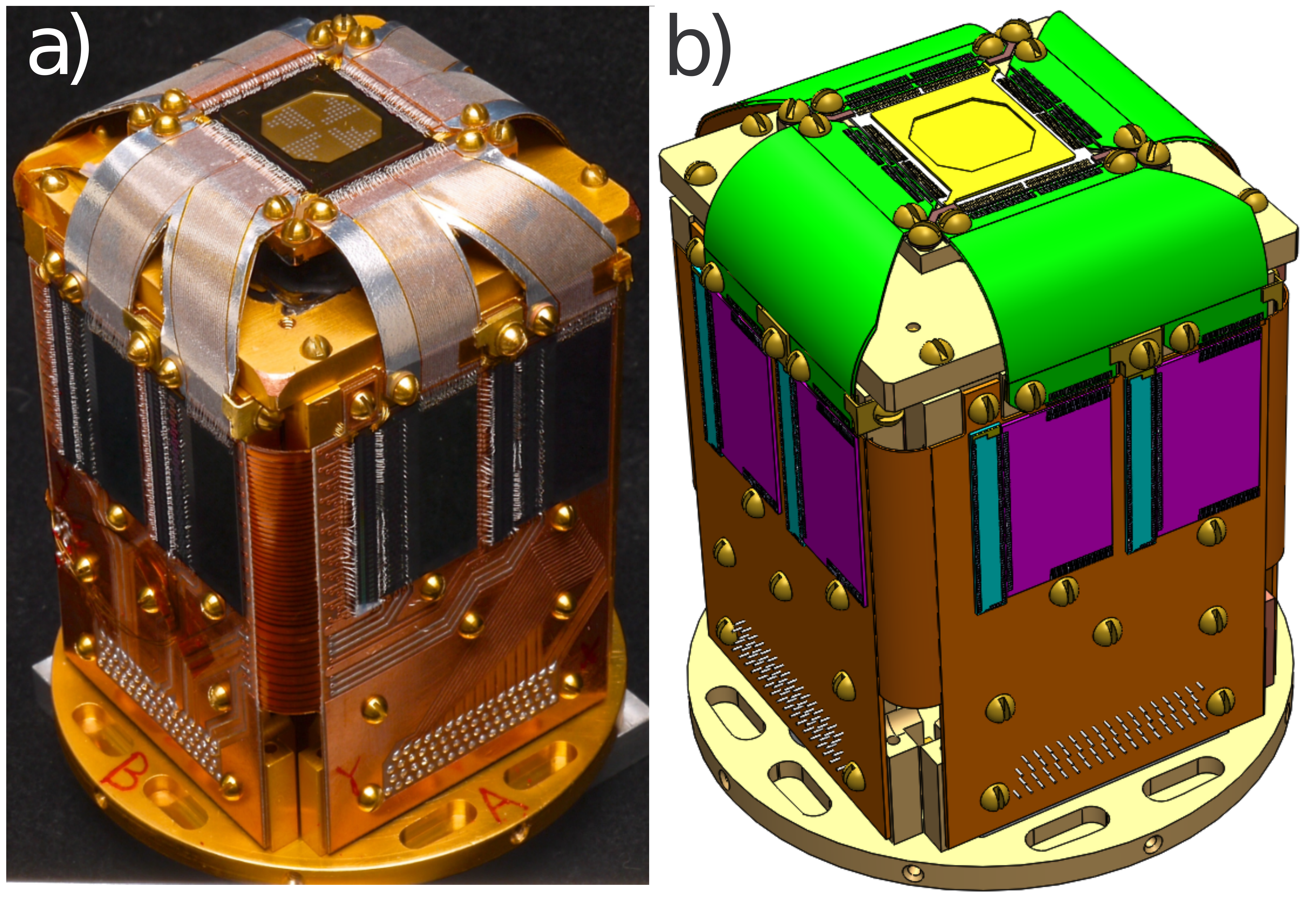}
\end{center}
\caption{The 50~mK snout package.  (a)  Photograph of the snout.  The circular pedestal at the bottom of the image has a diameter of 58.4~mm and the snout has a height of 67~mm.  (b)  CAD image of the snout, with various parts highlighted in different colors.  {\it Yellow:} The TES chip and X-ray aperture sit on top.  {\it Green:} on flexible circuits, a total of 512 Al traces (256 pairs) run from the detector plane to the four side panels.  {\it Purple:} The ``interface chips'' contain the TES-shunt resistors and the Nyquist inductors; there is one of these chips per TDM column, or 8 chips total.  {\it Cyan:} The TDM multiplexer chips contain the SQ1s and FASs.  {\it Orange:} A rigid-flexible printed-circuit board (PCB) connects to the interface and TDM chips via Al wirebonds and wraps around the snout to carry signal lines from panel to panel.  A trio of 65-lead ``Nano-D'' connectors on the inside of the rigid-flex PCB connect to twisted-pair cables that carry signals to and from higher-temperature cryogenic circuitry.  \blue{From Doriese, et al., {\it Rev. Sci. Instrum.} (2017)\cite{smith2021}; re-printed with permission.}}
\label{fig:tdm-snout}
\end{figure}

The NIST lab TDM system for X-ray TESs accommodates up to 8 TDM columns and up to 32 TDM rows (up to 256 TES pixels).  The heart of each spectrometer is the 50~mK ``snout'' package (see Fig.~\ref{fig:tdm-snout}).  This package contains the TES array, the biasing circuitry for the TESs (blue components in Fig.~\ref{fig:tdm-singlepixel}), and the SQ1s and FASs.

\begin{figure}
\begin{center}
  \includegraphics[width=\textwidth]{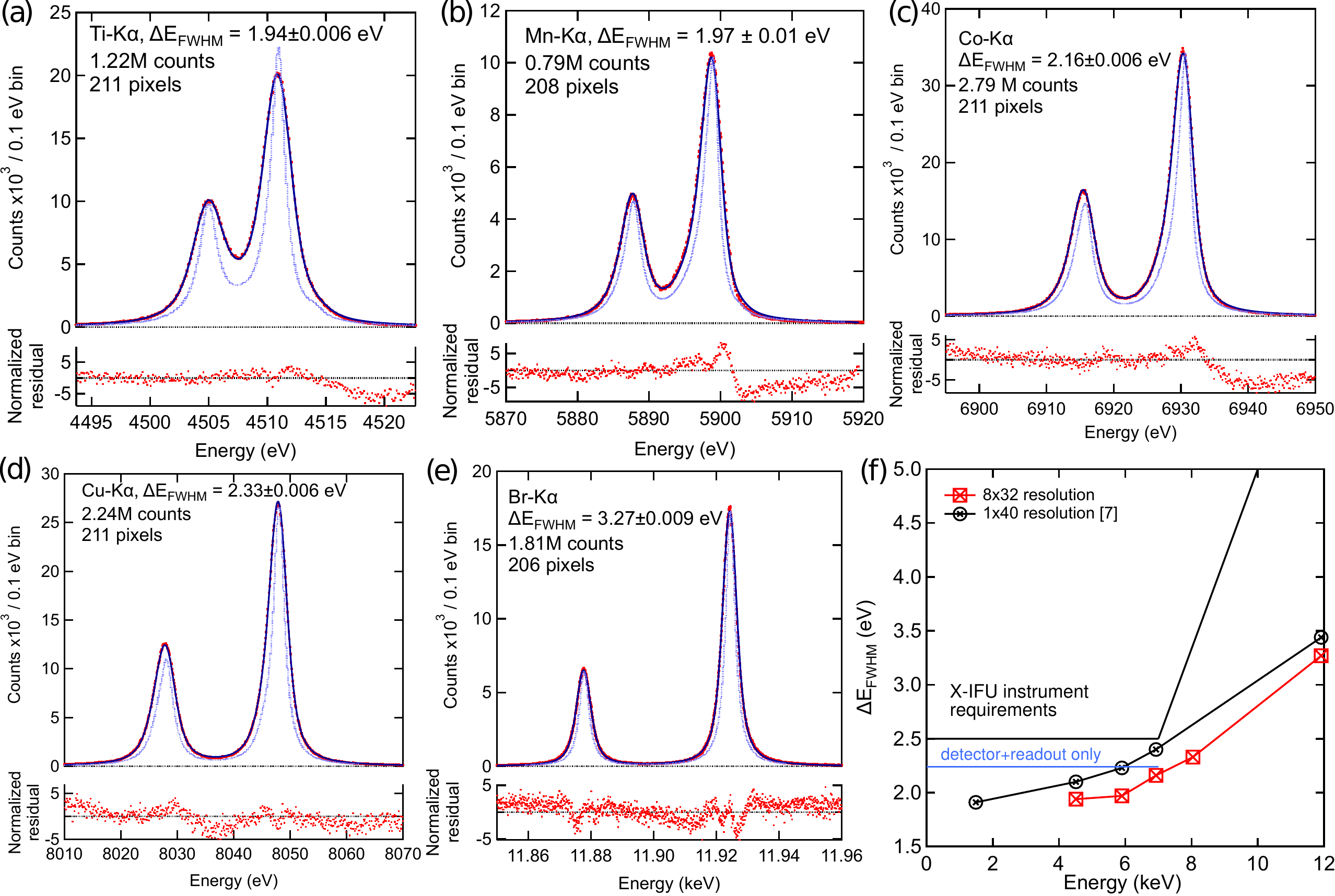}
\end{center}
\caption{Coadded 8-column by 32-row TDM spectra for (a) Ti-K$\alpha$, (b) Mn-K$\alpha$, (c) Co-K$\alpha$, (d) Cu-K$\alpha$, and (e) Br-K$\alpha$; all are measured from pure-metal foils except Br-K$\alpha$, which is measured from a KBr salt. The red dots in the main plots are the histogrammed spectra data, the light-blue lines are the natural line shapes (not broadened by the finite resolution of the spectrometer), and the dark blue line is the best fit to the data.  Below each main plot, the red dots show residuals (data minus fit).  (f) $\Delta E\urss{FWHM}$ vs. X-ray energy for these 2019 measurements (red squares) and a previous data set acquired with a less mature version of the SQ1s (black circles).  The black line shows the X-IFU instrument-level requirements on energy resolution, while the blue line shows the requirements considering only the detector and readout subsystem and excluding margin.  \blue{From Smith, et al., {\it IEEE Trans. Appl. Supercond.} (2021)\cite{smith2021}; re-printed with permission.}}
\label{fig:tdm-8x32}
\end{figure}

The 8x32 TDM snout system has also been used to develop TDM readout to meet the stringent performance requirements for Athena's X-IFU.  As Fig.~\ref{fig:tdm-8x32} illustrates, by late 2019 TDM had achieved the needed performance for X-IFU.

\subsubsection{Optimizations for space flight:  Athena X-IFU}\label{sec:tdm_XIFU}

TDM has been improved in three significant ways from its 8x32-lab configuration\cite{doriese2017, doriese2016} to optimize it for Athena X-IFU.

The first optimization was to increase the open-loop bandwidth in the link between the SQ1s and the SSA.  This bandwidth is $R\urss{dyn}/(2\pi L\urss{tot})$, and for X-IFU, the desired value is 15~MHz (12~MHz specification plus a 25\% margin).  Here, $R\urss{dyn} \equiv dV/dI$ is the dynamic resistance of the components in the SQ1-bias loop (SQ1, $R\urss{ser}$, $R\urss{b1}$; see Figs.~\ref{fig:tdm-singlepixel} and \ref{fig:tdm-2x2}), while $L\urss{tot}$ is the series inductance in the SQ1-bias loop (dominated by the self inductance of the SSA input coil and the wiring that connects the TES/SQ1 stage to the SSA stage).  In X-IFU, it is desired to allow $L\urss{wiring} \le 350$~nH, which is higher than in the 8x32 TDM-snout systems.  In early 2022, $R\urss{dyn} \ge 35$~\Ohm\ was demonstrated (Durkin, et al.\cite{durkin2021} give an intermediate report of progress toward this milestone).

The second set of optimizations was to the packaging of the TDM componentry.  Cryogenic components, such as the TES-shunt resistors\cite{doriese_ASC2019} and SQ1s, were shrunk in area.  The various wire coils (SQ1-input transformer and Nyquist inductors) were designed in ``even'' and ``odd'' versions to allow close-packing of TDM cells in a 2-dimensional grid without an increase in cell-to-cell inductive crosstalk.  Most importantly, a cold-indium bump-bonding process\cite{lucas2022indium} was developed to allow attachment of the X-IFU TDM chips to the FPA side panels without the need for a challenging number of wirebonds.

The third optimization was a change in the architecture of the flux-actuated switches.  X-IFU has baselined a ``two-level-switching'' scheme in which each TDM pixel retains its own ``pixel-select'' switch and a second layer of switch, a ``cluster-select'' switch, shorts out a larger cluster of pixels.  This scheme's chief advantage is a significant reduction in the number of wires needed for row addressing.  Dawson, et al.\cite{Dawson_TLS} describe the idea further and provide some preliminary demonstrations.

%% file: FDM_2022Aug.tex
\begin{figure}
\begin{tabular}{c}
\begin{minipage}{\hsize}
\begin{center}
  \includegraphics[width=1.\hsize]{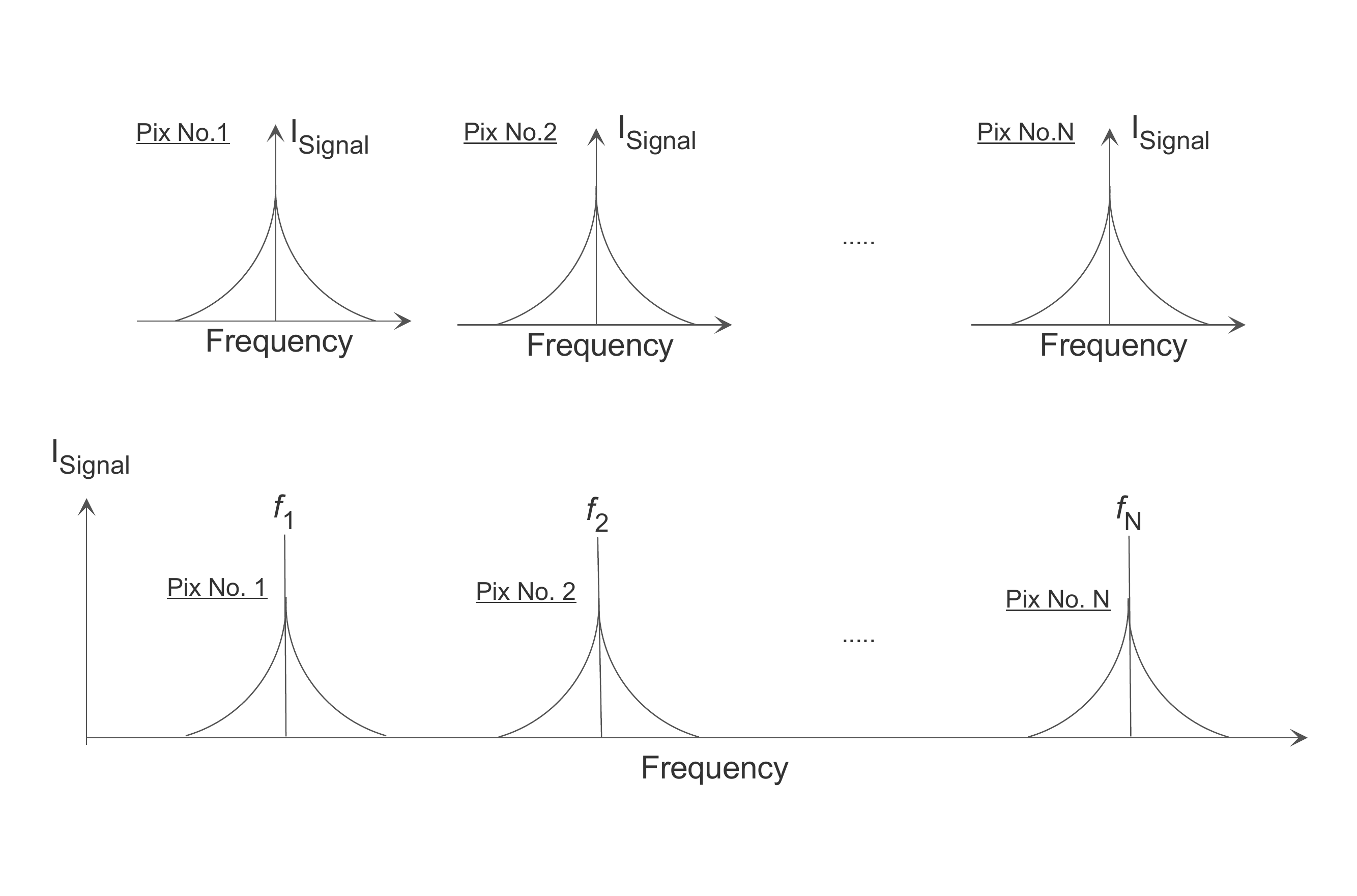}
\end{center}
\end{minipage}
\end{tabular}
\caption{Cartoon of the distribution of signals in frequency space in FDM. (top panel) In the TDM and \umux\ readout schemes, the TESs are biased via a dc current so the signals are distributed around zero frequency. (bottom panel)  In the FDM scheme, each TES is biased by its own MHz ac current and therefore the signals are distributed in frequency space. The bias frequencies of the TESs should be separated enough to avoid cross-talk between TESs. The signals are demodulated by the room temperature electronics.}
\label{fig:fdm}
\end{figure}

In the MHz-FDM scheme, signals from the TESs are multiplexed in frequency space and the TESs are biased with megahertz-frequency alternating currents (ac bias)\cite{yoon01,cunningham02,kiviranta02,ullom03,dobbs08,takei09_FDM}. Each TES is connected to a passive $LC$ resonator. The resonators are sufficiently separated in frequency space to avoid interferences (e.g. Fig.~\ref{fig:fdm}). Signals from the different TESs are summed in the front-end SQUID amplifier. There are three major advantages of the MHz-FDM architecture.  First, the cold electronics are relatively simple (there is one dc-SQUID front-end amplifier per multiplexer chain).  Second, MHz-FDM is relatively robust to electromagnetic interference (EMI) because the signals are transported at frequencies well above the electromagnetic disturbances due to the cryocoolers and the frequency dependence of the skin depth makes shielding easier and lighter.  Third, MHz-FDM has lower environmental susceptibility\cite{vaccaro22_sus} (see also \blue{Gottardi\&Smith in this book}).  The X-IFU demonstration model (Sec.~\ref{sec:fdm-DM}), on which we base this presentation of MHz-FDM, will use a second-stage series-array SQUID\cite{kiviranta21} located on the 2~K stage.  Multiplexed signals are demodulated per TES (per assigned frequency) in the room temperature electronics.

The sum of signals can be achieved in several ways in SQUID readout: current summing\cite{mikko04}, flux summing\cite{mitsuda99, kimura08}, and voltage summing\cite{yoon01,cunningham02}. Traditionally, voltage summing has been performed via summing loops; however, because SQUIDs are inherently sensitive to changes in magnetic flux, and current is straightforwardly transduced to flux via a wire coil, it is more straightforward to sum the signals via current or magnetic flux. The flux-summing approach requires a separate input coil for each TES. This complicates the design of the SQUID and may cause electrical crosstalk between input coils via mutual inductance. On the other hand, the current-summing method does not require a complicated SQUID design, and crosstalk can be avoided via careful implementation of the coils. This section focuses mainly on the current-summing method.

The multiplexed signal is demodulated (at the assigned frequency) for each TES in the room-temperature electronics. The modulation (bias) frequency must be much higher than the signal bandwidth of the TES response ($\sim$50~kHz). Therefore, a readout bandwidth of several megahertz is necessary to ensure a reasonable multiplexing factor. As mentioned in the previous section, dc-SQUID amplifiers combine sufficient bandwidth ($\sim10$~MHz) with a high flux dynamic range ($\sim10^{8} \rm or \sim 0.01$\uphinotperrthz). This allows multiplexing of tens of TES X-ray microcalorimeter signals, or more than 100 lower-bandwidth bolometric TESs, in a single readout chain.

In the FDM method, the signal-to-noise ratio is not degraded by an increased multiplexing factor as it is in the TDM method (see Eq.~\ref{eq:tdm-noise} in Sec.~\ref{sec:tdm_params}). However, in FDM each TES must be activated by a megahertz-frequency ac bias, which introduces additional physical phenomena into the TES, such as the ac Josephson effect due to the lateral proximity effect of the superconducting leads\cite{sadleir10, sadleir11, gottardi14a, gottardi18} (see also the chapter by \blue{Gottardi \& Smith} in this book for details). Subsections \ref{sec:FDM_roomelec} and \ref{sec:FDM_LC} below detail some important technologies needed to realize FDM readout, while subsections \ref{FDM_demo} and \ref{sec:fdm-DM} describe some experimental demonstrations of FDM readout of X-ray-TES arrays.

\subsubsection{Room temperature electronics}\label{sec:FDM_roomelec}

The FDM room-temperature electronics consist of two main components:  the analog electronics (Front-End Electronics, or FEE) and the digital electronics (DEMUX board).

\begin{figure}
\begin{tabular}{cc}
\begin{minipage}{.5\hsize}
\begin{center}
  \includegraphics[width=.65\hsize, angle=-90]{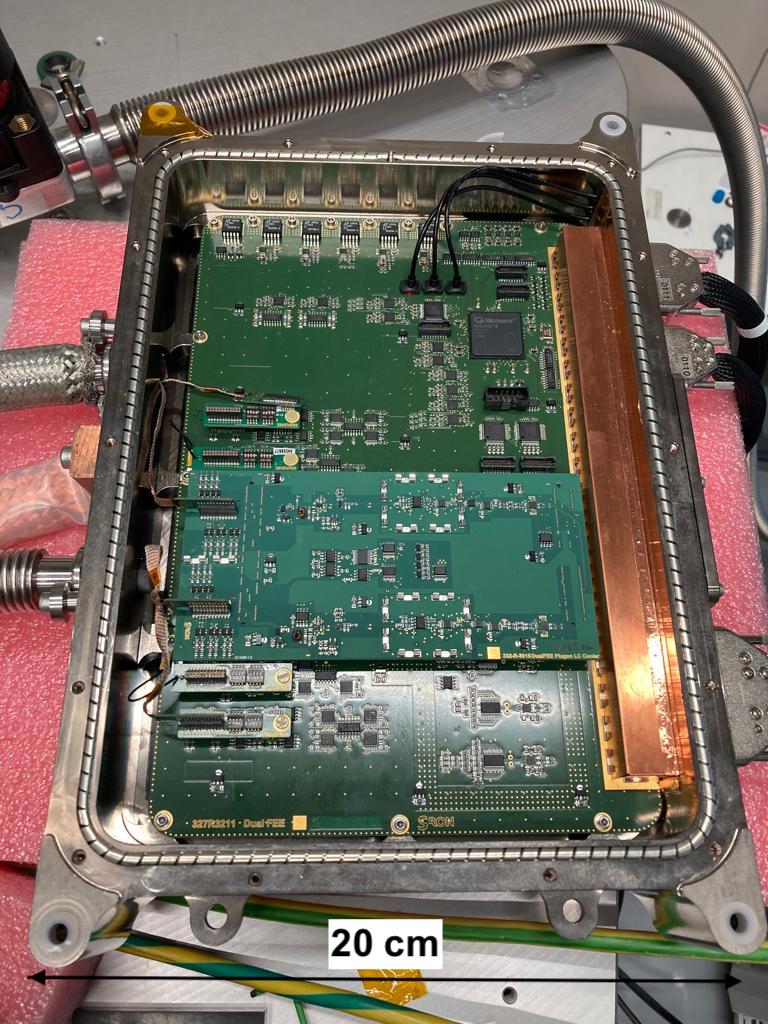}
\end{center}
\end{minipage}
\begin{minipage}{.5\hsize}
\begin{center}
  \includegraphics[width=1.\hsize]{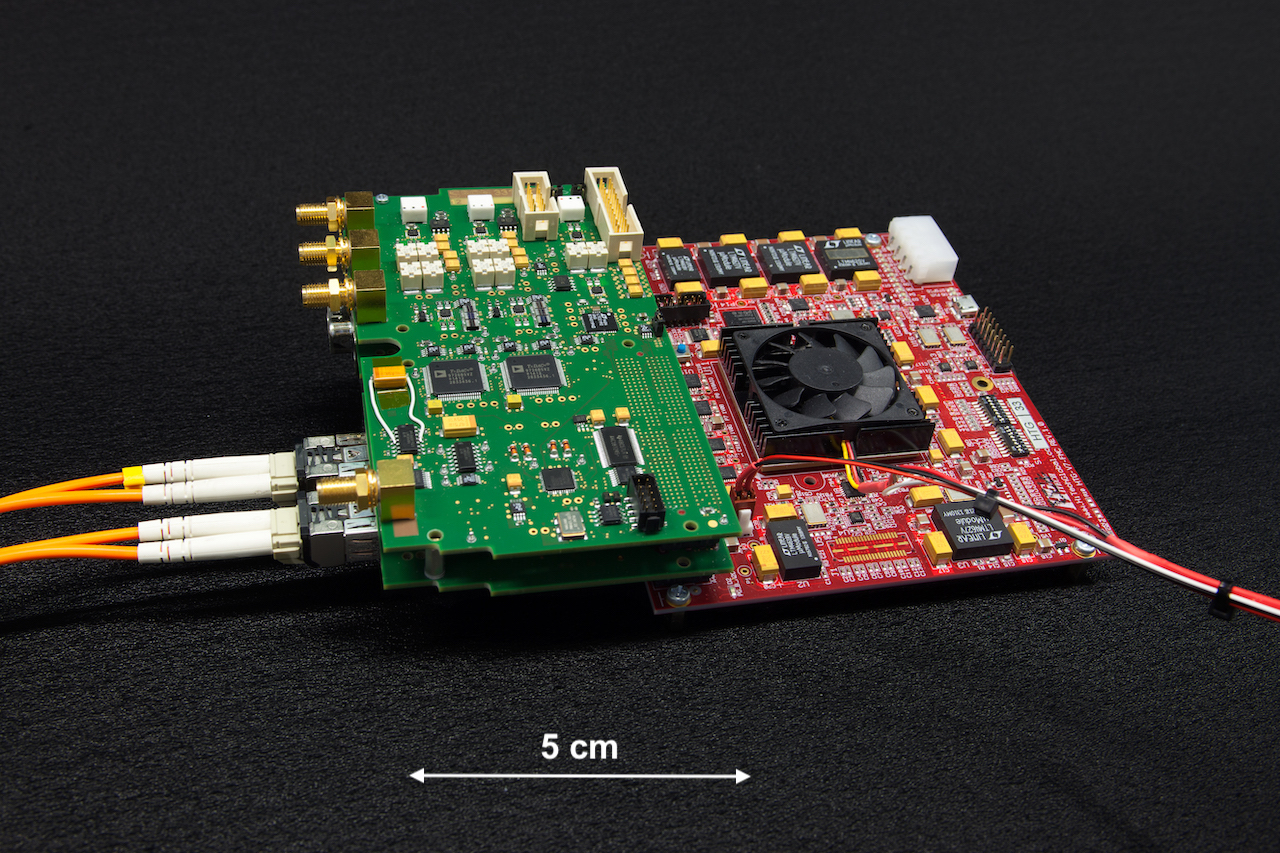}
\end{center}
\end{minipage}
\end{tabular}
\caption{{\it Left:} photograph of the dual FEE with 2-channel LNA and SQUID controller. {\it Right:} photograph of the DEMUX board for FDM readout. Both of these electronics were developed at SRON. 
}
\label{fig:elec}
\end{figure}

\texttt{FEE:} The primary role of the FEE is to amplify the signal from the cryogenic SQUID array so its signal size and output-voltage noise are compatible with the room-temperature ADC. A low-noise amplifier (LNA) is thus needed.  The FEE developed at SRON, pictured in Fig.~\ref{fig:elec}, has an input-voltage noise of $\sim 300~ \rm pV/\sqrt{Hz}$.
Information about the SRON FEE can be found in the literature\cite{wang20}.

\begin{figure}
\begin{center}
  \includegraphics[width=\textwidth]{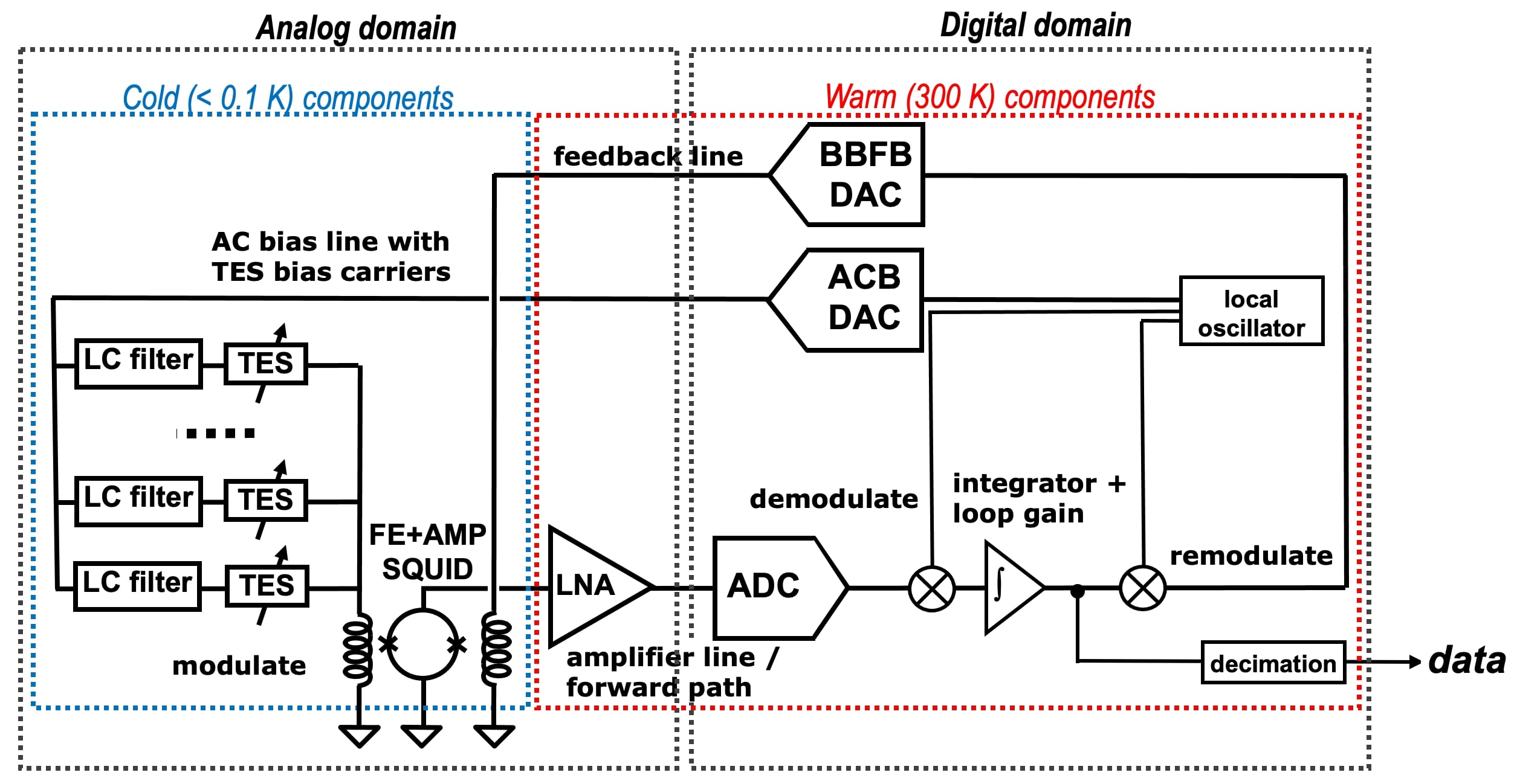}
\end{center}
\caption{Block diagram of the FDM readout architecture.
The ac-bias carrier is generated in the FPGA and converted to an analog signal by a DAC in the digital electronics (DEMUX board). Each TES in the readout channel is equipped with an $LC$ filter with a unique frequency; the two-stage SQUID is read out by an LNA (low-noise amplifier) at room temperature. Signals are converted from analog to digital by an ADC and demodulated by an FPGA. The demodulated signal is decimated by a 4-stage filter to reduce the data volume. The final sampling rate is 156.3~kSample/s. The feedback signal is calculated and fed back to the system after compensating for phase delays caused by electronics and the length of the wiring harness in the system. \blue{From den Hartog, et al., (2009)\cite{BBFB}; re-printed with permission.}
}
\label{fig:BBFB}
\end{figure}

\texttt{DEMUX board:}
The DEMUX board handles many functions, including demodulation of signals and generation of the TESs' ac-bias carrier combs and the feedback signal applied to the front-end SQUID. In addition, the DEMUX board must be capable of handling delays and phase shifts introduced by the environment.  In the presence of the long wiring harnesses that are expected in satellites,\footnote{For Athena X-IFU instrument, the length of the harness between the 50~mK stage and the room-temperature digital electronics may be as long as 5~m. For other similar cryogenic instruments such as SPICA/SAFARI\cite{roelfsema12}, even longer harnesses ($\sim$12~m) have been proposed.} standard feedback methods (e.g., the conventional flux-locked-loop, or FLL) do not properly compensate the input signal to the SQUIDs amplifier due to the phase difference between the input signal and the feedback signal. The lack of phase margin in a standard FLL would thus make the feedback loop unstable above some maximum frequency, $f\urss{max}$. The gain-bandwidth product, $f\urss{max}*G\urss{FB}$, of a standard FLL controller is limited by the phase shift due to signal travel time through a total travel length in the cable, $l$, as\cite{BBFB}

\begin{equation}
    f\urss{max}*G\urss{FB}=\displaystyle{\frac{c}{8l\sqrt{\eta}}},
\end{equation}
where $G\urss{FB}$, $c$, and $\eta$ are the loop gain of the feedback, the speed of light, and the dielectric constant of the cable's dielectric material, respectively. Assuming $G\urss{FB}=10$ (which is required to keep the total flux in the linear regime of SQUID response) and $\eta$=3, for a 1~m cable harness ($l=2$~m for the round-trip travel distance of signal and feedback) the maximum frequency will be about 1~MHz. With even longer cables and FDM's higher bias frequencies, standard FLLs cannot work.

Baseband feedback (BBFB\cite{BBFB}; Fig.~\ref{fig:BBFB}) is one way to overcome this limitation\footnote{Another method is digital-active nulling (DAN\cite{dehaan12}). The main difference is that in BBFB there is feedback to the summing junction shared by the carrier voltage. DAN is or will be implemented in the SPT-3G instrument\cite{montgomery22} and the LiteBird satellite\cite{hazumi20, dehaan19}}.  The BBFB method sends the signal from the TES back to the SQUID after correcting for delay and phase rotation at each carrier frequency. BBFB significantly improves the bandwidth, linearity, and dynamic range of the SQUID amplifier. At SRON, the DEMUX board consists of an AD9726 16-bit DAC and a Xilinx XC7V585T Virtex 7 Field Programmable Gate Array (FPGA). Signals are measured at 20~MSample/s and decimated to 156.3~kSample/s by a 4-stage filter. The DAC performance is described by den Hartog, et al.\cite{denhartog18_DAC}. The firmware of the DEMUX board will allow up to 64 X-ray microcalorimeters or up to 172 bolometers.

\subsubsection{Lithographic $LC$ filter}\label{sec:FDM_LC}

\begin{figure}
\begin{center}
  \includegraphics[width=\textwidth]{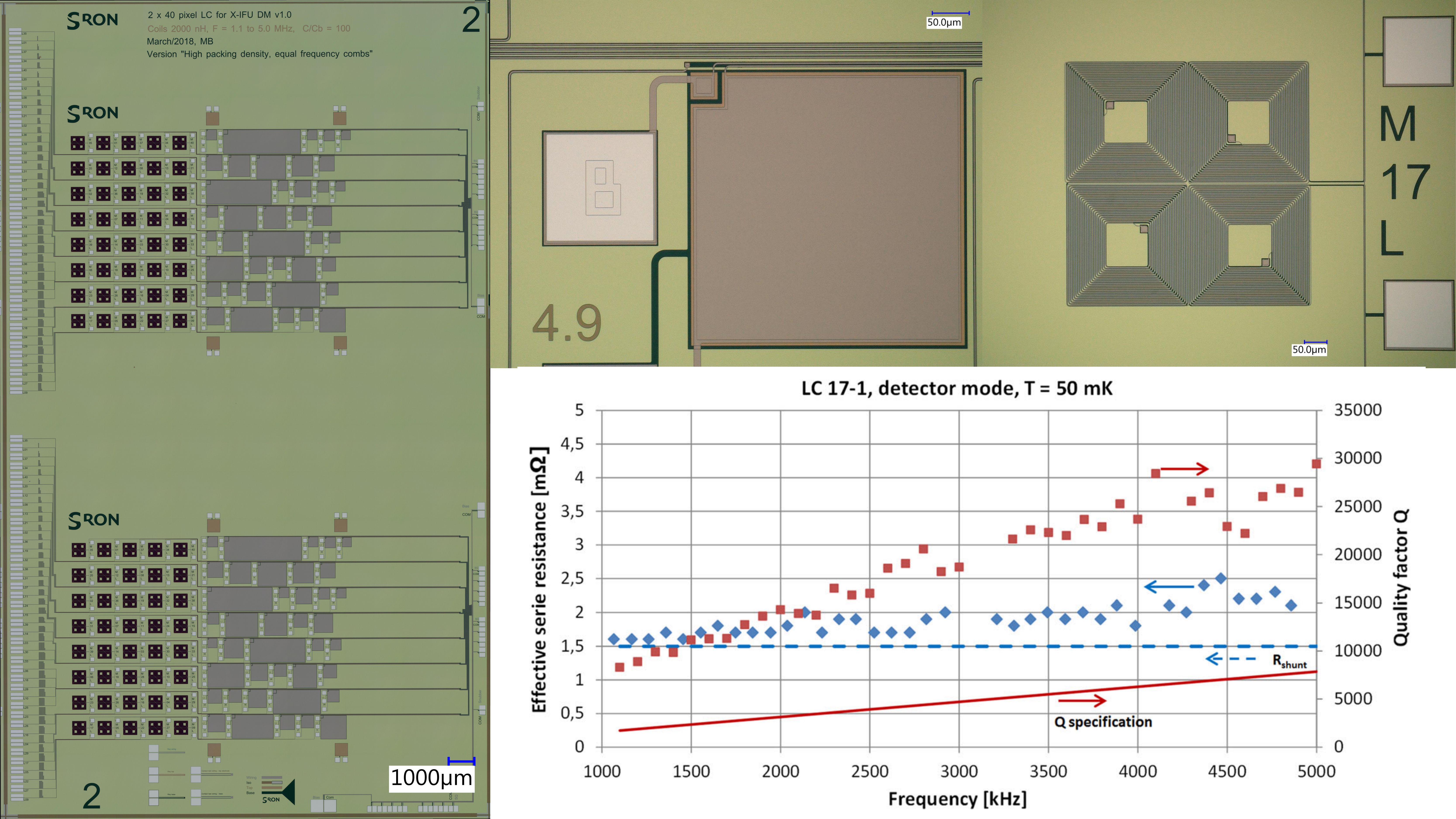}
\end{center}
\caption{{\it Left:} photograph of a lithographic $LC$ filter chip containing 2~$\times$~40 resonators. The chip size is 36~mm $\times$ 60~mm$^2$.  {\it Upper right:} an enlarged photograph of one capacitor (left) and one 2~\micro H spiral coil with back-spiraling return line (right). {\it Bottom Right:} Effective series resistance (blue diamonds) and quality factor (red squares) of the $LC$ resonators vs.\ frequency\cite{bruijn14, bruijn18}. The blue-dashed and red-solid lines indicate the contribution of the shunt resistance and the $Q$ specification of this design.}
\label{fig:lc}
\end{figure}

High-quality $LC$ resonators/filters are one of the key cryogenic components to achieve high multiplexing factors in FDM readout; a higher density of $LC$ filters allows more efficient use of the available bandwidth. However, there is a trade-off between fabrication accuracy due to technology limitations and performance metrics such as crosstalk and stability. For a single-pole $LC$ filter, the resonator bandwidth ($ \Delta f_{\mathrm{BW}}$), resonant frequency ($f\urss{c}$), and quality ($Q$) factor are described as follows:
\begin{equation}
{\Delta f_{\mathrm{BW}}}=\frac{R}{2\pi L}, \,\,\,\,\,\,\,\,\,\,
f\urss{c}=\frac{1}{2\pi\sqrt{LC}},\,\,\,\,\,\,\,\,\,\, 
\rm and
\,\,\,\,\,\,\,\,\,\,
Q=\frac{1}{\textit{R}}\sqrt{\frac{\textit{L}}{\textit{C}}},
\end{equation}
where $R$ and $L$ represent the total resistance and inductance, respectively. Typically, the same inductance value is used for different $LC$ filters and the resonant frequencies are tuned via the capacitance. In this way, the design and uniformity of the $LC$ filter can be well controlled. It is common for proposed satellite-borne FDM systems to have a minimum bias frequency of about 1~MHz. This is because at lower frequencies, the area required for the capacitors begins to dominate the overall area of the cryogenic components. At the higher-frequency end of the readout band, challenges arise due to the equivalent series resistance (ESR) generated in the dielectric medium within the capacitor, the analog bandwidth in the full chain, and power dissipation in the warm electronics.

SRON uses an $LC$ filter with a 2~\micro H coil. A superconducting transformer is also used to adjust the detector bandwidth via the damping inductance. SRON's $LC$ filter is fabricated via lithographic technology\cite{bruijn14, bruijn18}.

\begin{figure}
\begin{tabular}{cc}
\begin{minipage}{.5\hsize}
\begin{center}
  \includegraphics[width=1.\hsize]{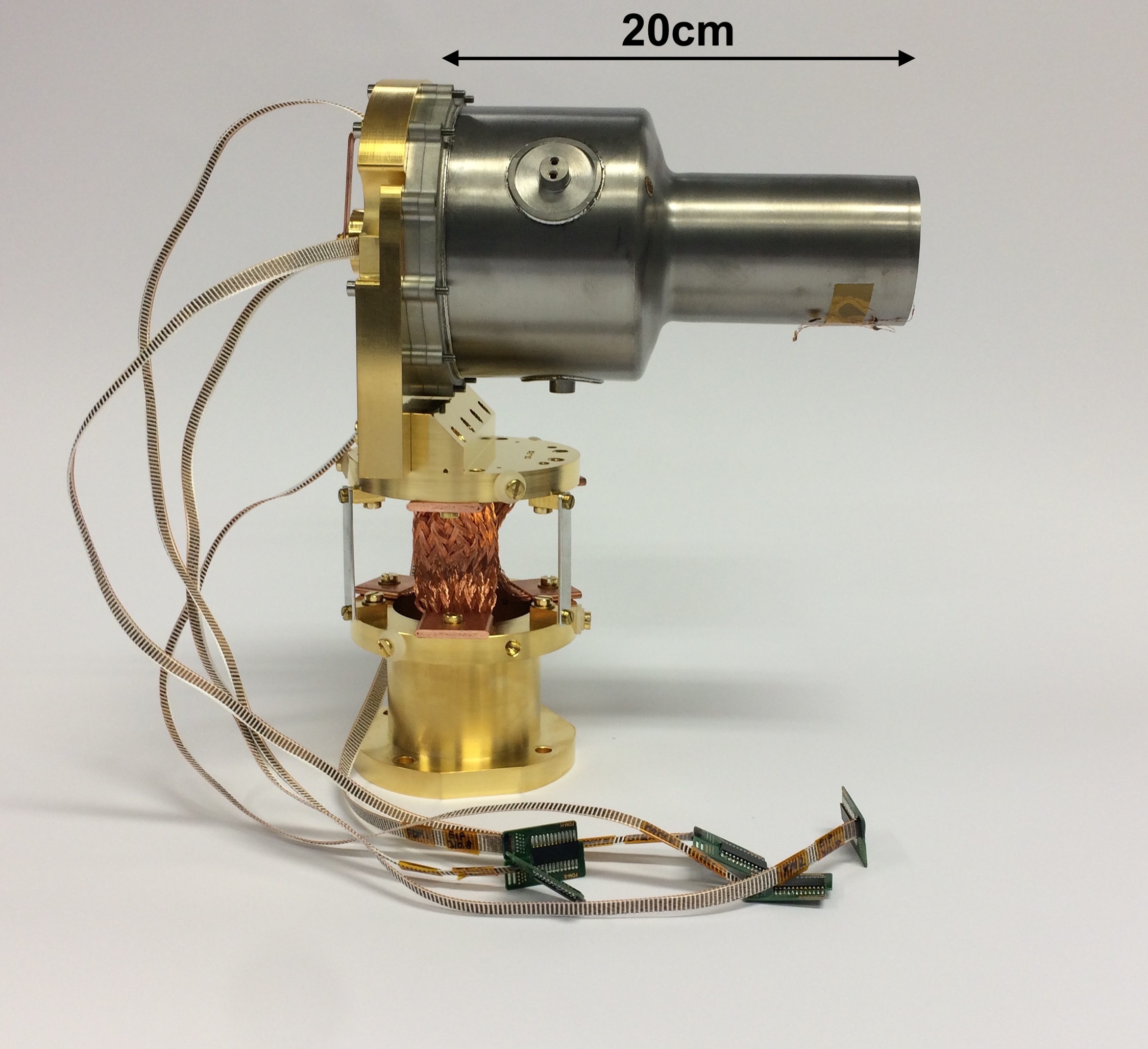}
\end{center}
\end{minipage}
\begin{minipage}{.5\hsize}
\begin{center}
  \includegraphics[width=.8\hsize]{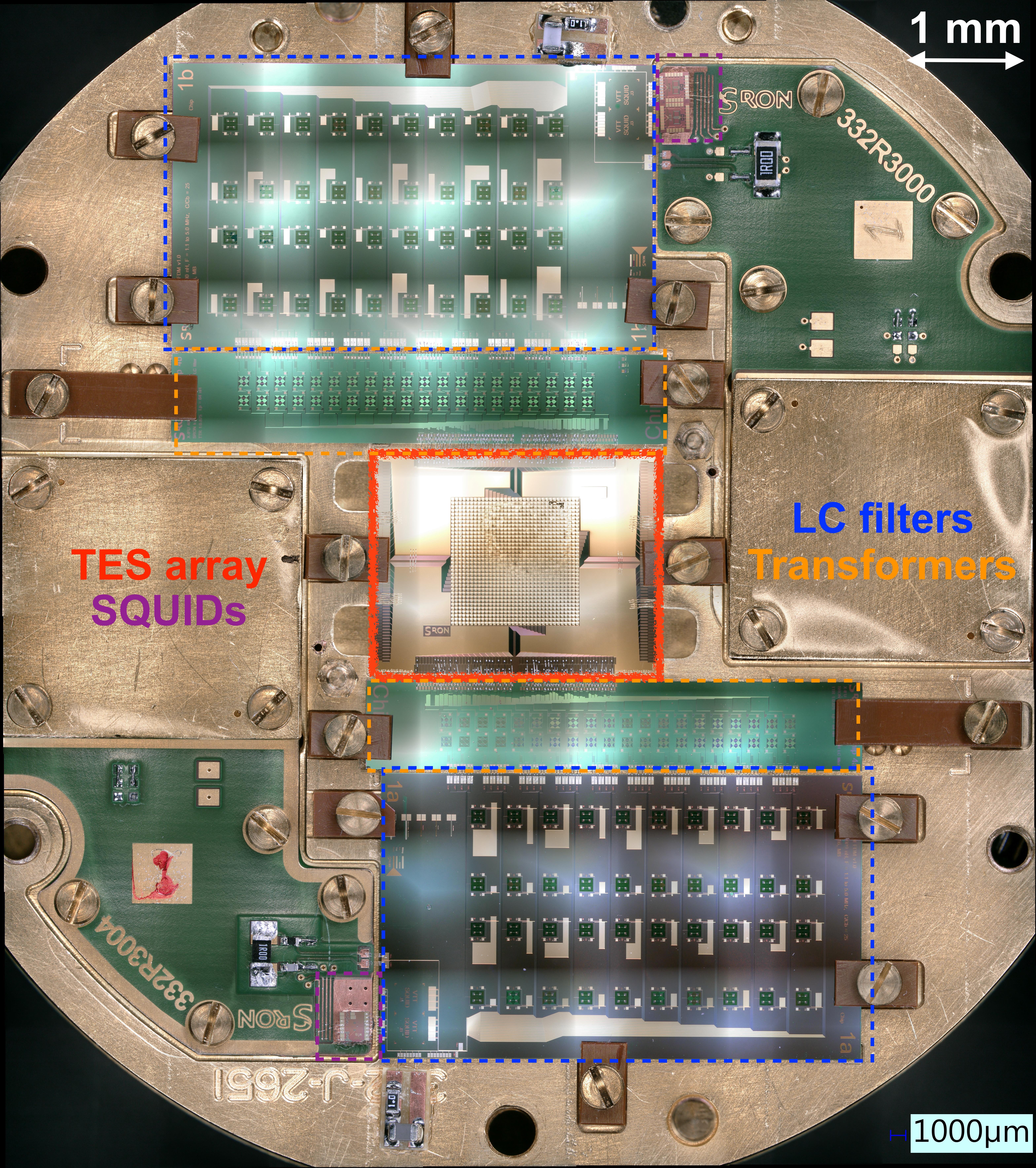}
\end{center}
\end{minipage} \\
\begin{minipage}{.5\hsize}
\begin{center}
  \includegraphics[width=1.\hsize]{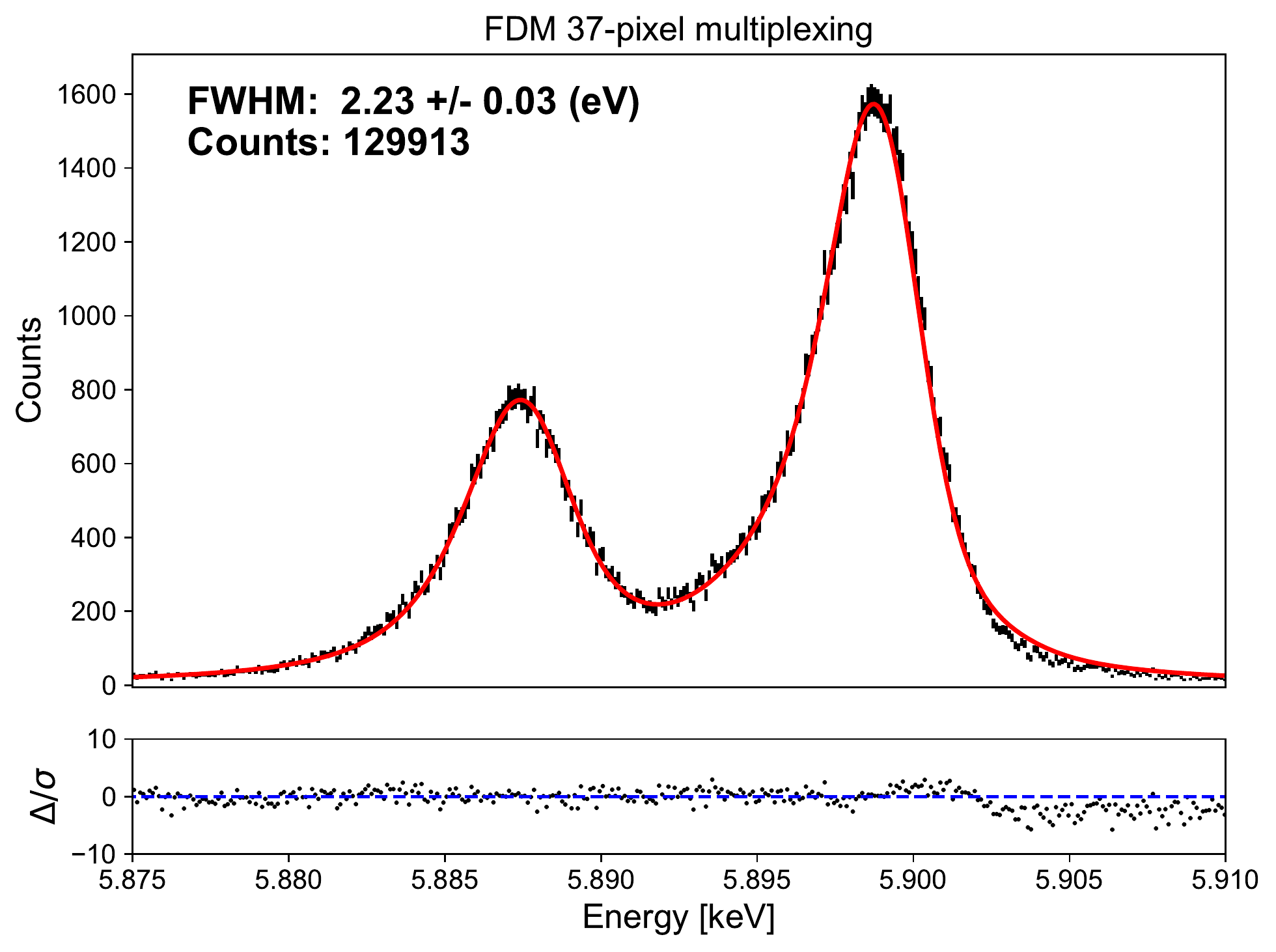}
\end{center}
\end{minipage}
\begin{minipage}{.5\hsize}
\begin{center}
  \includegraphics[width=1.0\hsize]{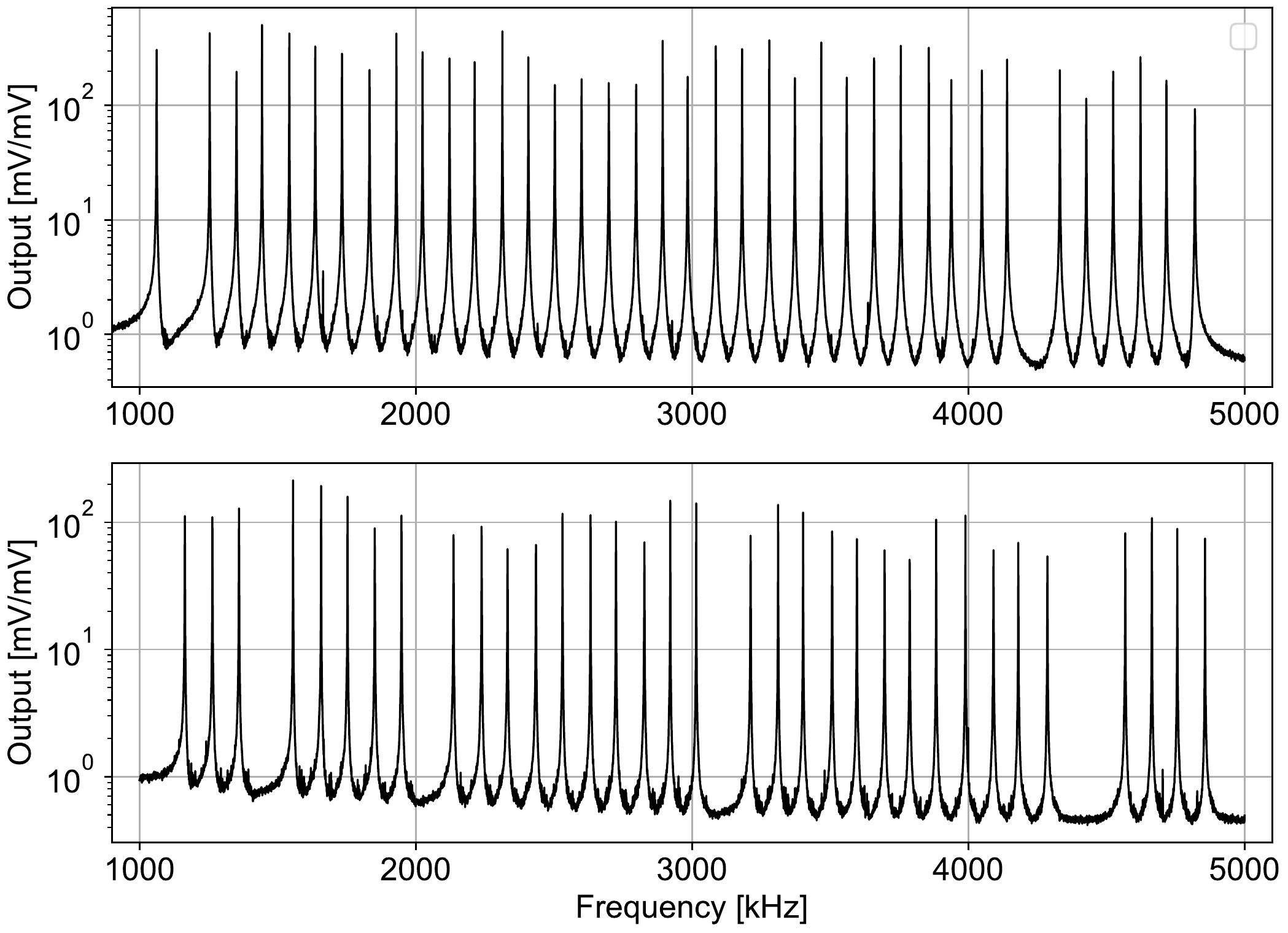}
\end{center}
\end{minipage}
\end{tabular}
\caption{{\it Upper left:}  
photograph of the 2-column FDM-readout demonstrator. The gray cylinder is the Nb superconducting magnetic shield. {\it Upper right:} Photograph of the focal plane showing the SRON kilo-pixel TES array\cite{nagayoshi20}, superconducting transformers, $LC$ filters, and FE SQUIDs. Two FDM columns are connected. {\it Lower left:} co-added energy spectrum of the Mn-K$\alpha$ complex from 37 TES pixels under FDM readout with a combined energy resolution of 2.2~eV at 6~keV\cite{akamatsu21}. The spectral data were fitted to the Mn-K$\alpha$ line model of H\"{o}lzer, et al.\cite{holzer} convolved with the Gaussian detector response. Cash statistics~\cite{cash79, kaastra16} are employed to avoid fitting bias. {\it Lower right:} resonator distribution vs.\ frequency for column-1 (top) and column-2 (bottom) of the 2-column FDM system.}
\label{fig:FDM_demo}
\end{figure}

\subsubsection{Demonstrations}\label{FDM_demo}
Achievement of the best energy resolution in TES X-ray microcalorimeters under FDM readout has required resolution of several deep and difficult issues related to the fundamental physics of superconductivity and SQUID readout\cite{akamatsu20}.  The three main areas of research over the last 10+ years have been 1) the physics of superconducting thin films under ac bias\cite{sadleir10, sadleir11, akamatsu14_ACBias,gottardi14, gottardi18}, 2) performance degradation due to carrier leakage\cite{vanderkuur04, akamatsu20}, and 3) performance degradation due to intermodulation products\cite{vanderkuur16, vaccaro21_FSA, FSA}.  As of late 2021, all issues were resolved; Akamatsu, et al.\cite{akamatsu21} give more information on these issues and their resolution.

Fig.~\ref{fig:FDM_demo} (upper) shows a 2-column FDM demonstrator developed at SRON. It employs magnetic shields of both high-\textmu\ and superconducting (Nb) materials against the earth's magnetic field\cite{bergen16}. The system is installed in a dry dilution-refrigerator with a vibration-reduction mechanism\cite{gottardi19}. Forty $LC$ resonators, each consisting of a 2~\micro H coil\cite{bruijn14, bruijn18}, a capacitor, and a superconducting transformer, were implemented to tune the electrical circuits seen by selected TESs in a kilopixel array of SRON X-ray TESs\cite{nagayoshi20, taralli20, dandrea21}. The effective inductance seen by the TES (equivalent to $L\urss{Ny}$ in TDM) is tuned to be 60\% of the critical inductance at a TES resistance of 15\% of $R$\urss{n}. The resonator centers span the 1~MHz to 5~MHz readout bandwidth with 100~kHz separation, meaning there are 40 resonators per column. The rms scatter of the frequency centers is about 4~kHz and is dominated by lithographic accuracy.  A 2-stage SQUID is employed, which is provided by VTT/Finland\cite{kiviranta18, kiviranta21}.  Fig.~\ref{fig:FDM_demo} (lower left) shows a co-added X-ray spectrum of the Mn-K$\alpha$ complex from 37 multiplexed pixels (two resonators did not yield and one pixel was turned off); the achieved energy resolution was $\Delta E\urss{FWHM}=2.23$~eV at 6~keV. In a different experiment performed in a different cryo-module\cite{akamatsu14_ACBias}, 31 multiplexed TESs achieved a summed spectral resolution of $\Delta E\urss{FWHM}=2.14$~eV.  The level of the thermal cross-talk is evaluated to be $\sim10^{-4}$\cite{vaccaro22} for pixels that are physically nearest neighbors, which surpasses the Athena X-IFU instrument requirement of $<10^{-3}$.

The performance degradation due to readout is estimated to be 0.9~eV in quadrature (see Fig. 3 in Akamatsu et al.\cite{akamatsu21}) and is dominated by known and non-fundamental problems such as thermal gradients in the cryogenic stage. In other words, there is ample room for further improvement and increasing multiplexing factor in future experiments. A two-column FDM demonstration is currently underway (see Fig.\ref{fig:FDM_demo} bottom right).

\subsubsection{Demonstration Model of Focal Plane Assembly of Athena X-IFU}\label{sec:fdm-DM}

\begin{figure}
\begin{tabular}{cc}
\begin{minipage}{.5\hsize}
\begin{center}
  \includegraphics[width=1.\hsize]{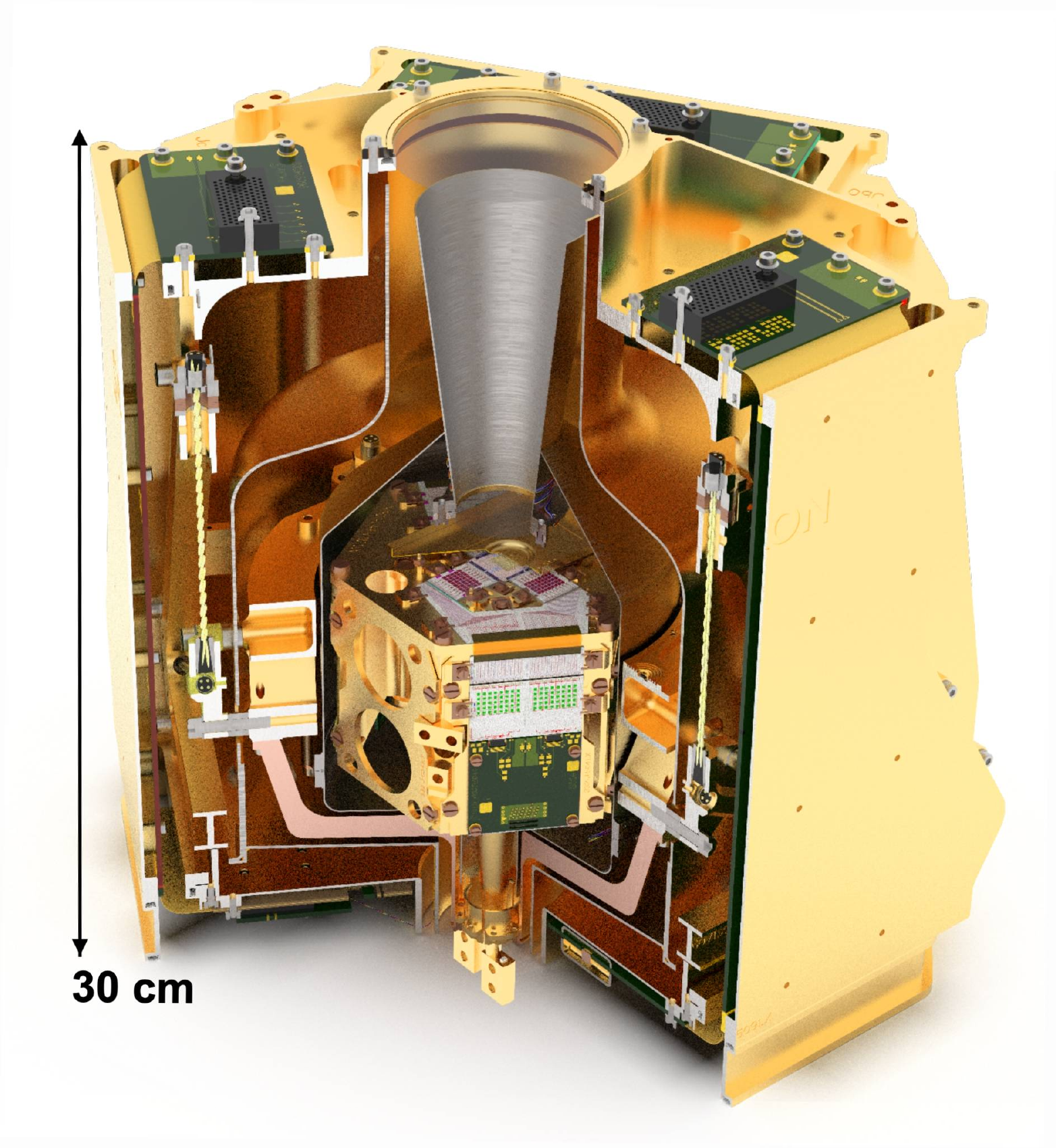}
\end{center}
\end{minipage}
\begin{minipage}{.5\hsize}
\begin{center}
\includegraphics[width=1.\hsize]{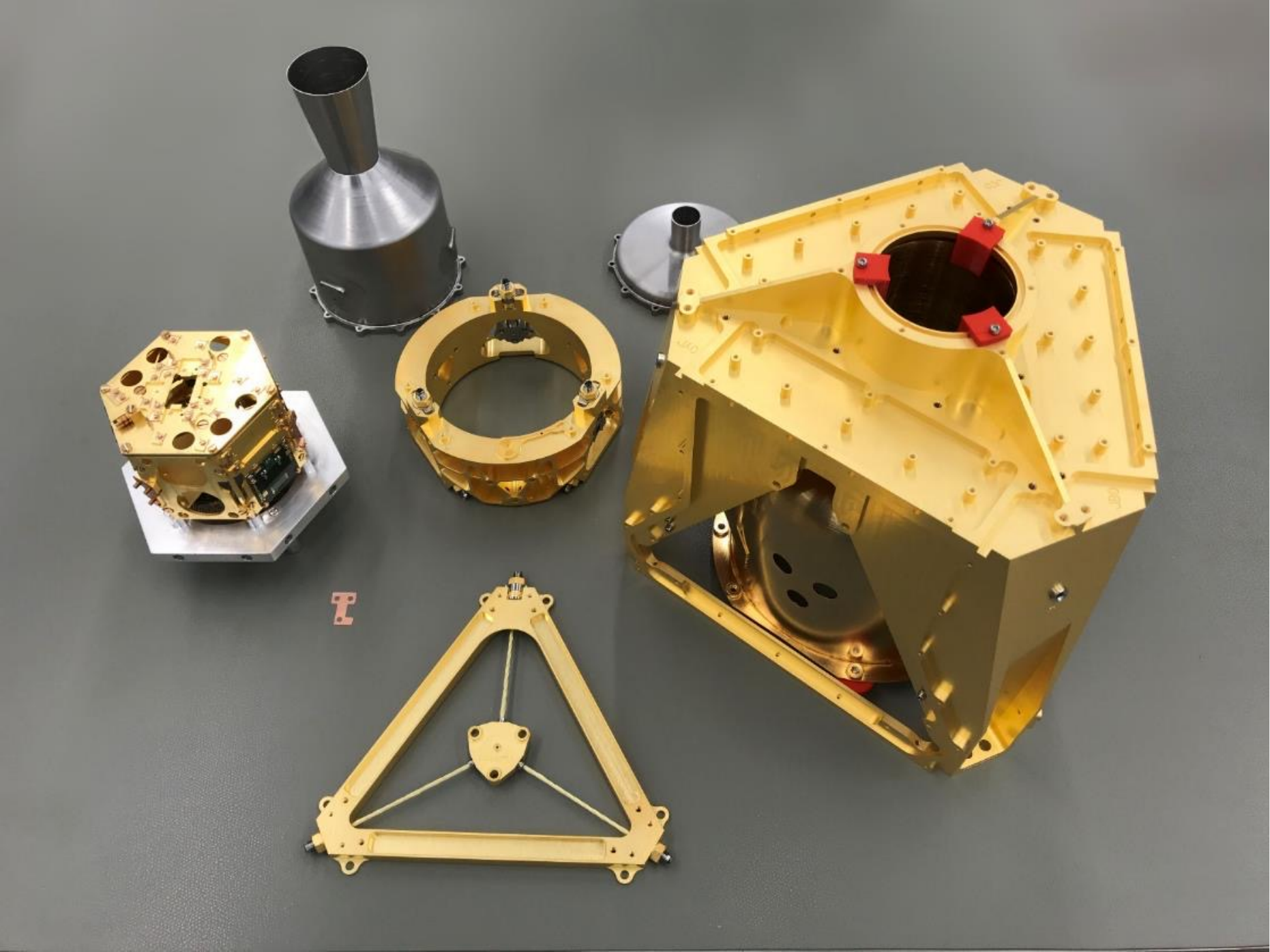}
\end{center}
\end{minipage} \\
\begin{minipage}{.5\hsize}
\begin{center}
\includegraphics[width=1.\hsize]{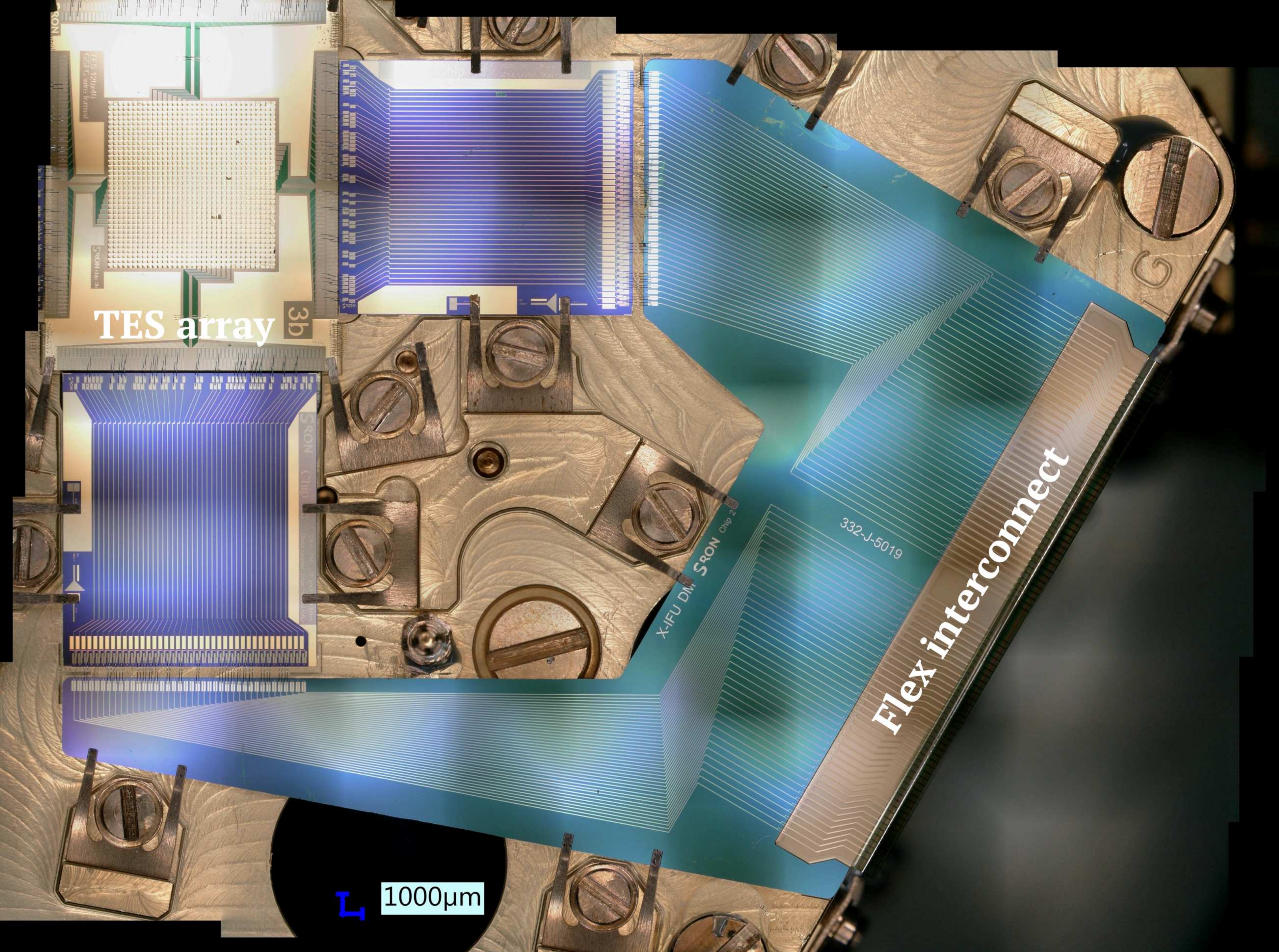}
\end{center}
\end{minipage}
\begin{minipage}{.5\hsize}
\begin{center}
  \includegraphics[width=1.\hsize]{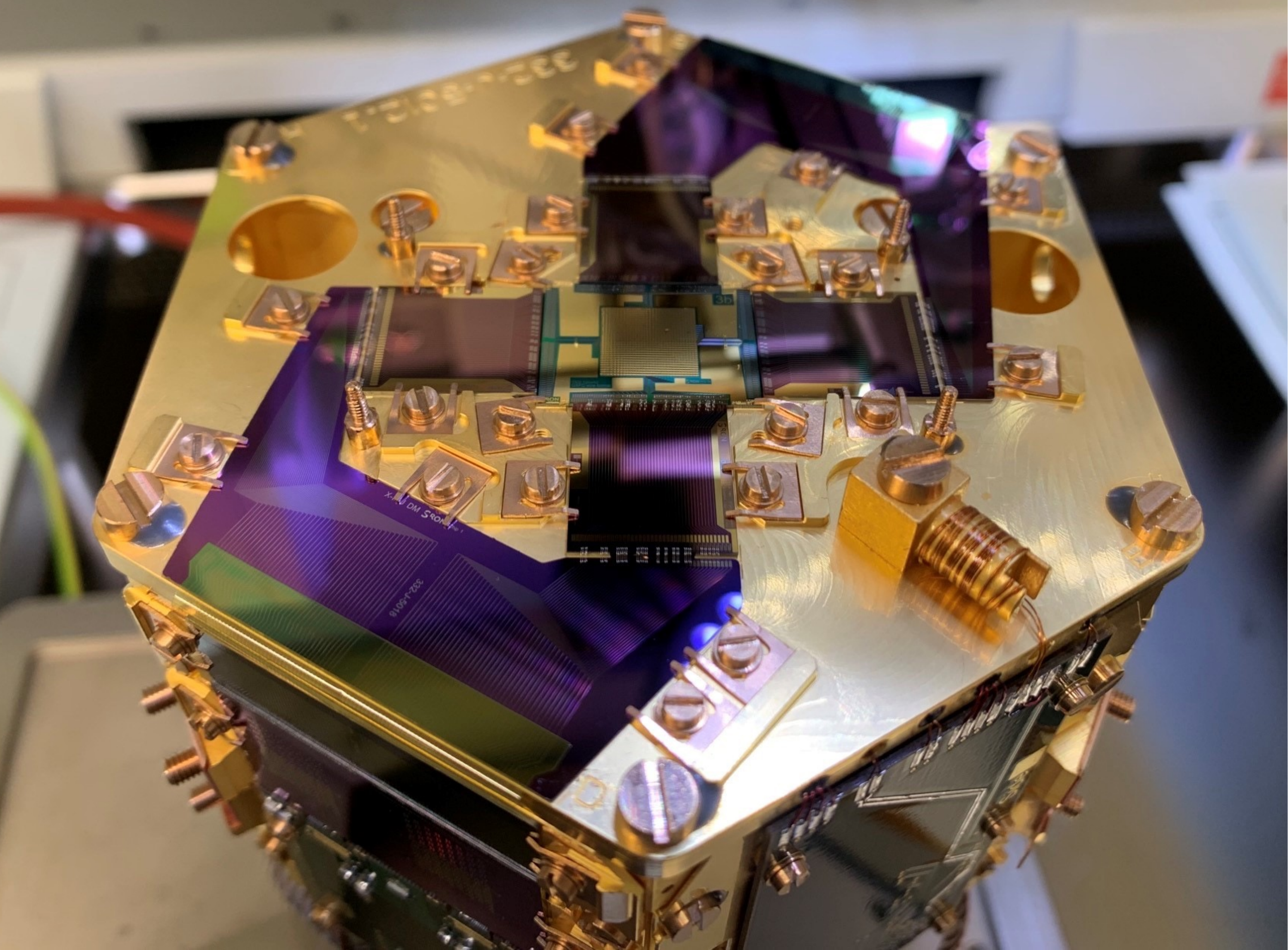}
\end{center}
\end{minipage}
\end{tabular}
\caption{{\it Top left:}  
CAD-rendered image of the X-IFU FPA DM. Inside the gold-plated-copper housing are two magnetic shields (high-\textmu~ metal and superconducting Nb) and a prototype of the 50~mK detector assembly. The FPA is suspended by Kevlar wires to damp micro-vibrations from the mechanical cryocoolers and to thermally isolate the 50~mK stage from the 2~K environment. {\it Top right:}  Photograph of some disassembled components of the X-IFU FPA DM. The triangular component is the mechanical supporting structure based on Kevlar wires. {\it Bottom left:}  photograph of the flexible-interconnect chip\cite{vanweers13}. {\it Bottom right:}  photograph of the 50~mK detector assembly. The central kilopixel TES array is connected to 4 FDM columns with 20 pixels per column.}
\label{fig:XIFU_FPA}
\end{figure}

An important application of FDM readout of TES X-ray microcalorimeters is verification of the the Demonstration model (DM) of the Athena X-IFU Focal Plane Assembly (FPA\cite{jackson16}; see Fig.~\ref{fig:XIFU_FPA}). The X-IFU FPA DM is designed to investigate various technological challenges.
The X-IFU FPA DM hosts a kilo-pixel TES array that was extensively characterized under ac bias\cite{taralli20, dandrea21,taralli21}, an anti-coincidence detector\cite{macculi16, dandrea17, dandrea20} and 4 FDM columns with 20 pixels per column connected (i.e., 80 TESs in total). The X-IFU FPA DM has three main purposes:
\begin{enumerate}
  \setlength{\parskip}{0cm} 
  \setlength{\itemsep}{.15cm}
\item A test-bed for key technology items that have not yet been demonstrated at all, or have not been demonstrated in combination with other parts of the FPA. Examples are the dual-stage SQUID amplifiers which are divided over two temperature stages, the magnetic shields, and the Kevlar suspension (see Fig.~\ref{fig:XIFU_FPA} top left).\footnote{Micro-vibration can degrade the performance of the instrument significantly. It is of importance to isolate the instrument from the source of the vibration. See Takei et al.\cite{takei18} for more details of a similar investigation for Hitomi/SXS.}
\item A physical reference for thermal, magnetic, and mechanical simulations that will guide the design of the engineering model and eventually the flight model. Measurements obtained from the FPA DM allow the simulations to be checked and, when necessary, can guide improvements to the simulations.
\item A demonstrator for the environments in which the TES sensor array needs to perform. As the main task of the FPA is to shield the TES array from adverse thermal, magnetic, and mechanical influences from outside, the FPA DM will provide a test in the environment of a cryostat. In particular, it will serve as a demonstrator of the 2K core in the CEA Cryostat XIII, where the FPA will be operated in proximity to an adiabatic demagnetization refrigerator and flight-representative Joule-Thomson coolers.
\end{enumerate}

As of August 2022, following milestones had been met:
\begin{enumerate}
  \setlength{\parskip}{0cm} 
  \setlength{\itemsep}{.15cm} 
\item the thermal loads on the various cryogenic stages were measured;
\item temperature stability of the detector stage was 0.9~\micro K with 1~s sampling repetition over the required period of more than one hour; and
\item on the first cool-down of the system,  energy resolution better than 2.5~eV at 6~keV was demonstrated.
\end{enumerate}
These results verify that the thermal and mechanical environment provided by the FPA is suitable for the needed performance in X-IFU.

%% file: umux_2022Aug.tex
\begin{figure}
\begin{center}
  \includegraphics[width=\textwidth]{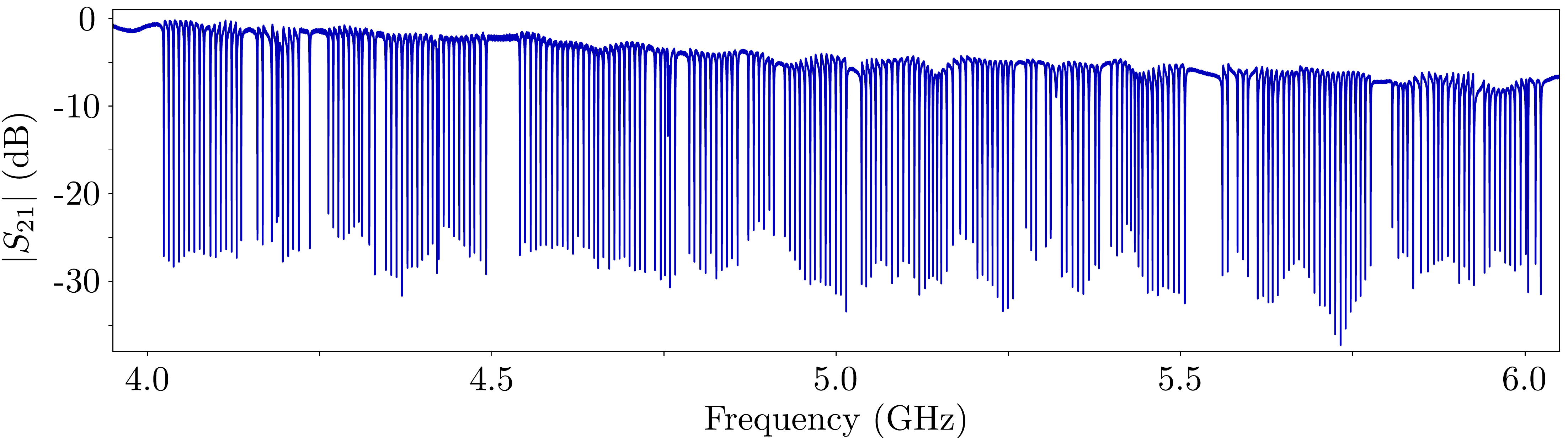}
\end{center}
\caption{Survey of microwave transmission $\left|S_{21}\right|$ through a microwave-SQUID multiplexer operating in the 4~GHz to 6~GHz band, with 256 channels distributed across 4 discrete \umux\ chips.}
\label{fig:umux-widesurvey}
\end{figure}

Multiplexing techniques utilizing gigahertz of bandwidth became possible with the development of low-thermal-conductance cryogenic microwave cables, in the form of narrow gauge Cu-Ni and Nb-Ti coaxial cables, and a low-noise cryogenic microwave amplifier, in the form of the High Electron-Mobility Transistor (HEMT) amplifier\cite{duh1988ultra}.

\begin{figure}
\includegraphics[width=0.54\textwidth]{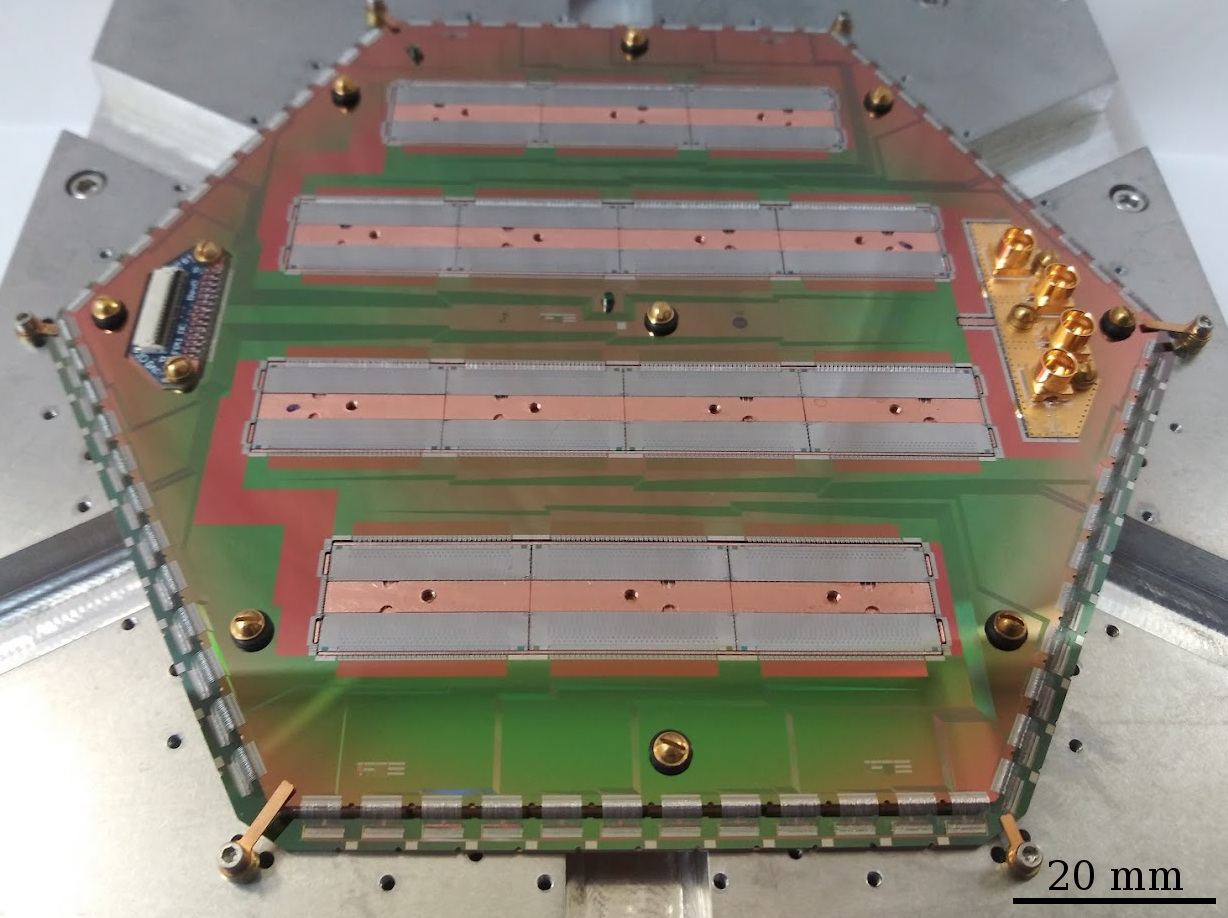} \includegraphics[width=0.45\textwidth]{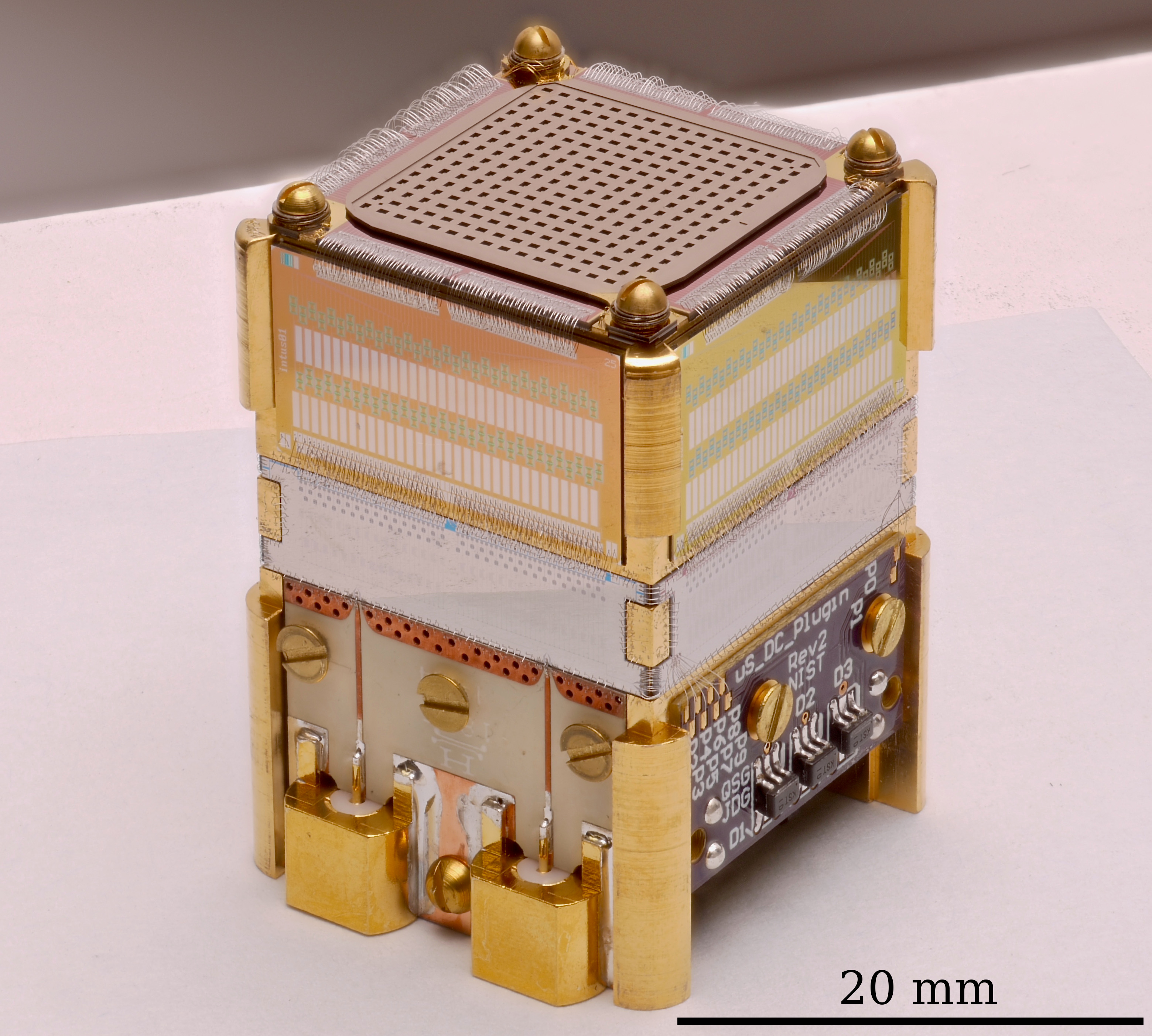}
\caption{\textit{Left} \umux\ readout circuitry on the backside of a 1,820 pixel CMB polarimeter array, read out by 28 \umux\ chips on two pairs of coaxial cables. \textit{Right} ``microsnout'' readout package for 256 X-ray microcalorimeters. The microsnout is modular: microsnouts can be tiled within a focal plane and electrically connected in series within a multiplexer channel.}
\label{fig:umux-examples}
\end{figure}

Microwave SQUID multiplexing (\umux)\cite{mates2008demonstration,hirayama2013microwave,mates2017simultaneous,ahrens2019superconducting} is a form of frequency-domain multiplexing that allocates this bandwidth between input channels by the use of distinct, high-$Q$ microwave resonances, each coupled to its own rf-SQUID and reading out the current signal from its own detector. As a superposition of microwave tones passes through the circuit, each tone is modulated by its own SQUID/resonator circuit before being amplified by the HEMT and brought to room temperature on a single coaxial cable.

\begin{figure}
\begin{center}
  \includegraphics[width=\textwidth]{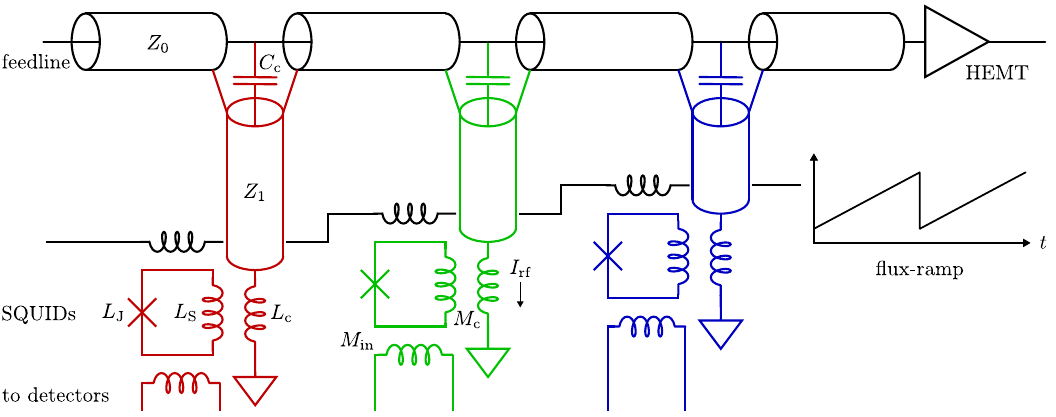}
\end{center}
\caption{Schematic of a 3-pixel microwave-SQUID multiplexer, consisting of quarter-wave resonators inductively coupled to rf-SQUIDs. Resonance frequencies are set by the lengths of the transmission lines, resonance bandwidths by $C\urss{c}$, and resonance frequency shifts by $M\urss{c}$, $L\urss{\mathrm{S}}$, and $L\urss{J}$.}
\label{fig:umux-schematic}
\end{figure}

An example circuit is shown in Figure~\ref{fig:umux-schematic}. In this circuit, distributed quarter-wave resonators of different lengths are capacitively coupled to a common microwave ``feedline'' at one end and inductively coupled to their individual SQUIDs at the other end. Two notable variations utilize lumped-element resonators\cite{ahrens2019superconducting} rather than quarter-wave resonators and direct incorporation of the SQUID into the resonator\cite{hirayama2013microwave} rather than inductive coupling to the SQUID.

In any implementation, the multiplexer works by loading the resonator with the flux-variable inductance of a Josephson junction:
\begin{equation}
L_{\mathrm{J}}(\phi) = \frac{\Phi_{0}}{2\pi I\urss{c} \mathrm{cos}(\phi)}
\end{equation} 
where $I\urss{c}$ is the critical current of the Josephson junction, $\Phi_{0}$ is the magnetic flux quantum, and $\phi \equiv 2 \pi (\Phi / \Phi_{0})$ is the difference in phase of the superconducting wave function across the junction.

\begin{figure}
\begin{center}
  \includegraphics[width=\textwidth]{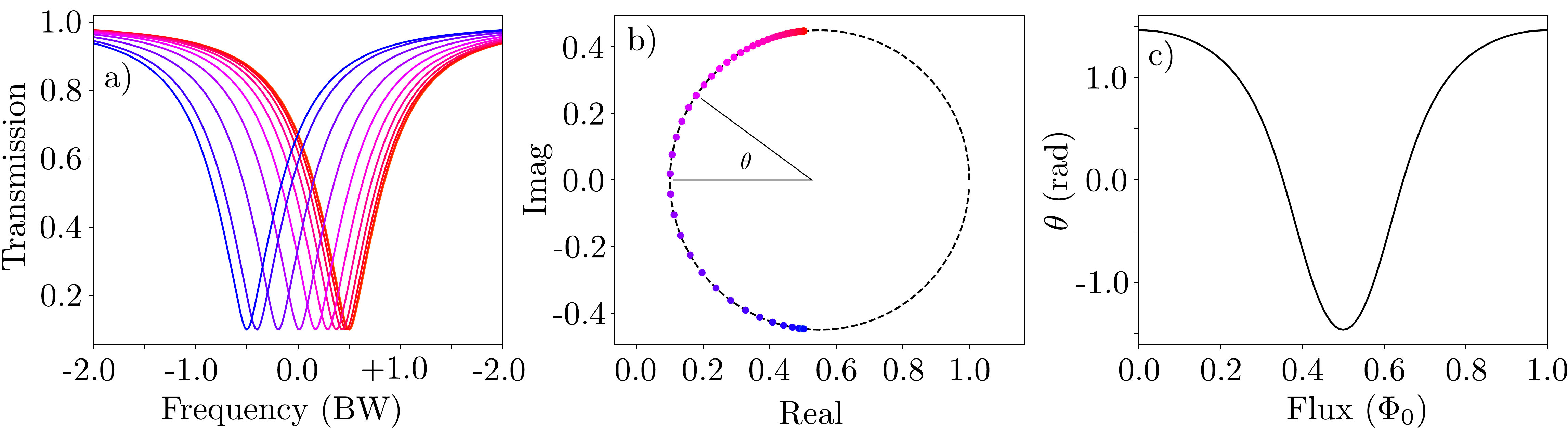}
\end{center}
\caption{Illustration of the principle of operation of a microwave SQUID multiplexer channel. a) The resonance frequency shifts in response to magnetic flux. b) For a fixed microwave frequency, the complex transmission parameter $S_{21}$ describes an arc on a circle. c) The SQUID response is periodic with a magnetic flux quantum $\Phi_{0}$.
}
\label{fig:umux-s21vsflux}
\end{figure}

As the flux in the rf-SQUID varies, the inductive load it applies to the resonator varies, both positively and negatively, causing the resonance frequency $f_{0}$ to change without significant change in microwave loss (Figure \ref{fig:umux-s21vsflux}). For a fixed microwave probe tone the complex transmission traverses an arc on a circle with the angular position on this arc being a periodic function of magnetic flux.

\begin{table}[!ht]
\centering
\renewcommand{\arraystretch}{1}
\renewcommand{\arraystretch}{2}
\begin{tabular}{|c|c|c|}
\hline 
SQUID coupling: & Inductive & Direct \\ 
\hline 
$\Phi_{\mathrm{rf}}$ & $M_{\mathrm{c}}I_{\mathrm{rf}}$ & $L_{\mathrm{S}}I_{\mathrm{rf}}$ \\ 
\hline 
$\Delta L\urss{pp}$ & $\frac{2{M_{\mathrm{c}}}^2}{L_{\mathrm{J}}\left(1-\lambda^{2}\right)}$ & $\frac{2\lambda^2 L_{\mathrm{J}}}{1-\lambda^{2}}$ \\ 
\hline
Resonator type: & Quarter-wave & Lumped-element \\ 
\hline 
$\frac{\mathrm{d}f_{0}}{\mathrm{d}L}$ & $-\frac{4{f_{0}}^{2}}{Z_{1}}$ & $-\frac{\pi}{\sqrt{L/\left(C+C_{\mathrm{c}}\right)}}$ \\ 
\hline 
$\Delta f\urss{BW}$ & $8\pi{f_{0}}^{3}{C_{\mathrm{c}}}^{2}Z_{0}Z_{1}$ & $2\pi^{2}{f_{0}}^{3}{C_{\mathrm{c}}}^{2}Z_{0}\sqrt{L/\left(C+C_{\mathrm{c}}\right)}$ \\ 
\hline 
$P_{\mathrm{feed}}$ & $\frac{\pi Z_{1} \Delta f\urss{BW}}{16 f_{0}}{I_{\mathrm{rf}}}^{2}$ & $\frac{\Delta f\urss{BW}}{4 f_{0}}\sqrt{L/\left(C+C_{\mathrm{c}}\right)}$ \\ 
\hline 
\end{tabular}
\caption{Table of equations describing the behavior of a SQUID-resonator pair. Most variables are defined in Figure \ref{fig:umux-schematic}. $\Phi_{\mathrm{rf}}$ is the amplitude of microwave flux excitation in the SQUID, $\Delta L_{\mathrm{pp}}$ is the peak-to-peak variation in load inductance with flux, $\Delta f\urss{BW}$ is the resonator bandwidth, and $P\urss{feed}$ is the power per tone on the feedline. For a lumped-element design, $L$ and $C$ are the lumped inductance and capacitance of the resonator, respectively.}
\label{tab:umux-equations}
\end{table}
One key parameter of the rf-SQUIDs is the ``screening-parameter''\cite{likharev2022dynamics}, sometimes denoted $\beta_{L}$, which gives the ratio of SQUID self-inductance to Josephson inductance:
\begin{equation}
\lambda \equiv \frac{L\urss{S}}{L\urss{J}} = \frac{2\pi I\urss{c} L\urss{S}}{\Phi_{0}}.
\end{equation}
For $\lambda < 1$ the SQUID response is single-valued, while for $\lambda > 1$ the SQUID may switch hysteretically between multiple states\cite{likharev2022dynamics}. An rf-SQUID may be operated in the hysteretic regime, but it requires damping to make its behavior predictable. Because this damping would limit resonator $Q$-factors, microwave-SQUID multiplexers operate in the non-hysteretic regime, typically targeting $\lambda \approx 1/3$.

Table~\ref{tab:umux-equations} lays out some of the equations\cite{mates2011microwave} describing the behavior of an rf-SQUID and resonator pair. Although these equations are only strictly valid in the regime of both low-power slow-flux-ramp, they provide a useful guide to device design.

For example, the equations show how $M\urss{c}$ and $C\urss{c}$ may be adjusted in tandem to achieve designs with different pixel bandwidths while holding the ratio of peak-to-peak frequency shift to bandwidth ($\eta \equiv \Delta f\urss{pp}/\Delta f\urss{BW}$) constant. They further show that the microwave-flux excitation (on-resonance and in the inductive coupling, quarter-wave-resonator configuration of Figure \ref{fig:umux-schematic}) in the SQUID is related to the power-per-tone on the feedline as:
\begin{equation}
P_{\mathrm{feed}} = {\Phi_{\mathrm{rf}}}^{2} \frac{\pi^{2} f_{0} I\urss{c}}{\eta \Phi_{0} \left(1 - \lambda^{2}\right)}.
\end{equation}

In Section~\ref{sec:umuxreadoutnoise} we will use this information to derive expected readout noise and discuss the trade-off between noise and maximum signal slew-rate. In Section~\ref{sec:umuxcrosstalk} we will discuss the trade-off between frequency packing and crosstalk. This analysis will show what capabilities we can expect of a \umux\ system designed for any particular application.

\subsubsection{Flux-ramp modulation}\label{sec:umuxfluxrampmodulation}

\begin{figure}
\begin{center}
  \includegraphics[width=\textwidth]{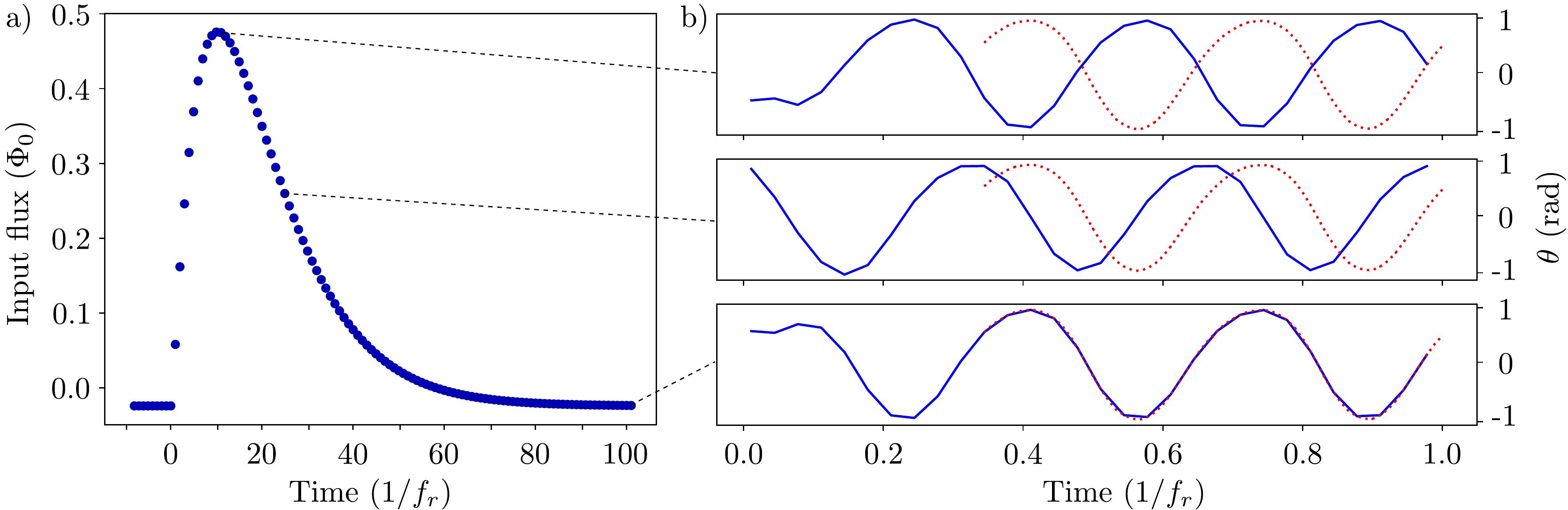}
\end{center}
\caption{Illustration of the principle of flux-ramp modulation, which transforms (a) the input flux signal into (b) a phase-shift of the SQUID response, shown as blue lines with the baseline response as dashed-red lines. The flux-ramp reset transient is visible at the start of each ramp response window.
}
\label{fig:umux-frdm}
\end{figure}

With microwave-SQUID multiplexing there is no practical way to apply feedback without the addition of a large number of wires. Instead, a flux-ramp modulation scheme\cite{mates2012flux} is typically employed that sweeps rapidly across the SQUID response such that the input signal is transduced into a phase shift of the response function (Figure \ref{fig:umux-frdm}). This phase may then be extracted at room temperature either with a fit or with Fourier techniques over an integral number of oscillations. The initial section of the flux-ramp response is usually contaminated with a transient from the reset and must be discarded, with analysis only over a useful fraction $\alpha$.

This both linearizes the SQUID readout and upconverts it above the significant low-frequency noise that is due to two-level systems, while reducing the readout signal power by approximately $\frac{\alpha}{2}$, where the factor of $\frac{1}{2}$ arises from averaging across the roughly sinusoidal SQUID response curve and the factor of $\alpha$ arises from discarding transient-contaminated data.

The flux-ramp repetition rate $f\urss{r}$ thus becomes the sampling rate of the input signal. To fit the SQUID modulation within the resonator bandwidth it must obey $f\urss{r} n_{\Phi_{0}} < \Delta f\urss{BW}/2$, where $n_{\Phi_{0}}$ is the number of flux quanta per ramp. It must also substantially exceed the frequency content of the input signals in order to read them out accurately, as the fidelity of the modulation scheme begins to degrade at an input signal slew rate of order $f\urss{r}\Phi_{0}/2$. These considerations determine the resonator bandwidth necessary for readout of a detector system.

\subsubsection{\umux\ readout noise}\label{sec:umuxreadoutnoise}

There are multiple sources of noise in a microwave-SQUID multiplexer: HEMT-amplifier noise is generally dominant, two-level system noise becomes significant for narrow resonator bandwidths, and digital noise limits the number of tones we can cleanly generate and digitize in a set of room temperature electronics.

The HEMT amplifier produces broadband white noise with typical noise temperatures on the order of $T\urss{N} \approx 4$ K ($k\urss{B}T\urss{N} \approx -193$ dBm/Hz). To convert this noise to units of magnetic flux we must find the power in the microwave probe tone, which should be as large as possible, up to the limits of SQUID non-linearity.

\begin{figure}
\begin{center}
  \includegraphics[width=\textwidth]{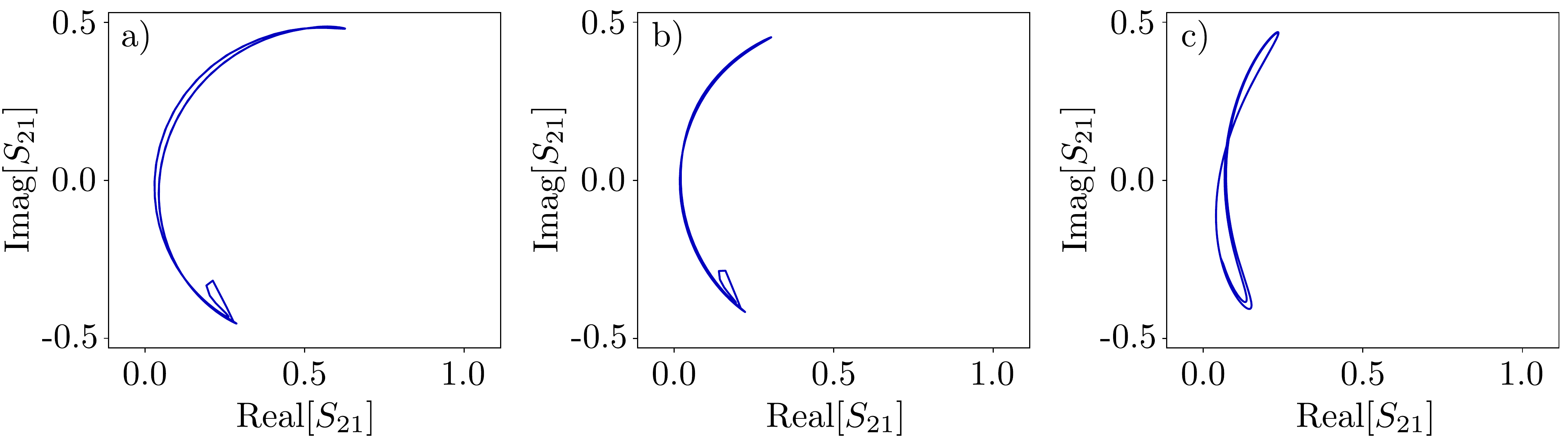}
\end{center}
\caption{Non-linear, non-equilibrium simulation results showing various ways in which the SQUID response departs from the simple theory in practical operation. a) Low power, slow flux-ramp, but with a significant flux-ramp-reset transient. b) Slow flux-ramp but with high microwave-drive power that suppresses the amplitude of the SQUID response. c) High-power, fast flux-ramp so that the resonator never has time to fully equilibrate and instead follows a ``figure-8'' trajectory along the resonance circle.}
\label{fig:umux-nonidealities}
\end{figure}

Determination of the optimal value of $P\urss{feed}$ requires analysis beyond the linear, small-signal approximations of the equations in Table \ref{tab:umux-equations}, including non-linear and non-equilibrium dynamics. As $\Phi_{\mathrm{rf}}$ grows to a significant fraction of $\Phi_{0}$, the SQUID inductance stops being effectively constant over the microwave excitation and siphons power from the microwave-probe frequency into higher harmonics. This behavior can be analyzed in expansion\cite{wegner2022analytical}, but to include all complicating phenomena generally requires simulation.

Figure~\ref{fig:umux-nonidealities} shows the results of simulations that include different sets of complicating factors, and in which we find the optimal microwave flux amplitude to be 0.25~$\Phi_{0}$ to 0.4~$\Phi_{0}$, depending on design parameters. For an example set of design parameters ($\eta = 1$, $f_{0} = 6$ GHz, $I\urss{c} = 5$ $\mu$A, $\lambda = 1/3$), this yields a tone power on the feedline of -73~dBm to -70~dBm, or approximately 120~dB above the 1 Hz noise power of the HEMT, and an arc of approximately $\pm 1$ radians.

This gives a noise spectrum in $\theta$, the angle on the resonance arc (Figure~\ref{fig:umux-s21vsflux}) of:
\begin{equation}
S_{\theta} = \left(\frac{1 - Q/Q_{i}}{2}\right)^{-2} \frac{1}{2} \frac{k_{\mathrm{B}}T\urss{N}}{P_{\mathrm{feed}}} \approx 2\frac{k_{\mathrm{B}}T\urss{N}}{P_{\mathrm{feed}}}
\end{equation}
where the first term gives the radius of the resonance circle in terms of the total quality factor $Q$ and internal quality factor $Q_{i}$. The approximation assumes $Q_{i} \gg Q$, but for narrow resonance applications ($<$\SI{100}{\kilo\hertz}) $Q_{i}$ may approach the total $Q$, which increases this noise contribution.

We can then assume a sinusoidal SQUID response, as shown in Figure~\ref{fig:umux-frdm}, with an amplitude of $\theta_{\mathrm{max}} \approx 1$ and apply the flux-ramp modulation penalty to get:

\begin{equation}
S_{\Phi}^{\mathrm{HEMT}} = S_{\theta} \left(\frac{\Phi_{0}}{2\pi \theta_{\mathrm{max}}}\right)^{2} \frac{2}{\alpha} \approx \frac{k_{\mathrm{B}}T\urss{N}}{P_{\mathrm{feed}}} \frac{2}{{\pi}^{2}} {\Phi_{0}}^{2}
\end{equation}
where in the final approximation we have assumed a 2\phinot\ flux ramp and have discarded the first oscillation for $\alpha = \frac{1}{2}$.  This implies a HEMT contribution to readout noise of approximately 0.5~\uphinotperrthz. Although this theoretical limit is rarely achieved, total readout noise of (1.0 to 2.0) \uphinotperrthz\ is common and sufficient for most low-temperature detector applications (see \blue{chapter by Gottardi \& Smith in this book}).

Two-level-system (TLS) noise arises from the coupling of the microwave fields in the resonator to two-level systems in the dielectrics. This produces a fractional-frequency noise that depends on temperature, internal power, material, and geometry\cite{gao2008experimental}. The power spectrum of this noise has a characteristic $1/\sqrt{f}$ frequency dependence, which the flux-ramp modulation of Section \ref{sec:umuxfluxrampmodulation} helps to avoid.

The TLS noise is not stationary over the period of the SQUID response, but we may bound its contribution by considering the equivalent flux noise at the steepest slope of the SQUID response, where:

\begin{equation}
\frac{\mathrm{d}\theta}{\mathrm{d}\Phi} = \theta_{\mathrm{max}}\frac{2\pi}{\Phi_{0}} \,\,\,\,\,\,\,\,\,\, \mathrm{and} \,\,\,\,\,\,\,\,\,\, \frac{\mathrm{d}\theta}{\mathrm{d}f} = \frac{4}{\Delta f\urss{BW}} = 4 Q / f_{0}
\end{equation}
so that (again assuming $\theta_{\mathrm{max}} \approx 1$) the implied contribution to flux noise is:
\begin{equation}
S_{\Phi}^{\mathrm{TLS}} = \frac{S_{\delta f_{0}}}{{f_{0}}^{2}} \left(\frac{2Q}{\pi}\right)^{2},
\end{equation}
which scales with $Q$ or inversely with the resonance bandwidth.

The number of factors determining the TLS noise makes it difficult to make a general statement about its level, but measured fractional-frequency noise for the geometry, materials, and internal power of a typical \umux\ device is less than $S_{f_{0}}/{f_{0}}^{2} \approx 10^{-19}$ Hz$^{-1}$ at 1 kHz. For example, in a 2~MHz resonator at 6~GHz, modulated at 1 MHz, we therefore expect a TLS noise contribution of less than 0.1~\uphinotperrthz. For multiplexers optimized for typical X-ray applications, with 1~MHz to 10~MHz resonator bandwidths, this contribution tends to be substantially less than the contribution of the amplifier chain.

Finally, any system for synthesizing and digitizing a large number of microwave tones adds its own noise: digital-quantization noise, clipping noise, and a pseudo-white-noise of intermodulation products due to the imperfect linearity of the DACs and ADCs. This is usually described by a dynamic range between the tone powers and the digital-noise floor. A careful analysis of this noise\cite{bennett2014integration} is beyond the scope of this chapter, except to say that multiplexing factors of $\sim$1000 push the limits of existing digitizers, which requires multiple digitizers to cover the full band with sufficient dynamic range, but that digitizer capabilities are advancing rapidly and can be expected to continue to both improve in capability and fall in cost.

\subsubsection{\umux\ crosstalk}\label{sec:umuxcrosstalk}

There are three main mechanisms of crosstalk in the microwave-SQUID multiplexer\cite{mates2019crosstalk}: Lorentzian-tail crosstalk, coupled-harmonic-oscillator crosstalk, and broadband-non-linearity crosstalk. Each constrains the multiplexer design in a different way.

Lorentzian-tail crosstalk arises from the quadratic falloff of the Lorentzian resonance shape, which allows one resonance to affect the transmission of a probe tone at the frequency of a neighboring resonance. For inductive coupling:
\begin{equation}
\chi\urss{Lorentz} \approx \frac{|S_{21}|^{2}}{2} \left(\frac{\Delta f\urss{BW}}{f_{2} - f_{1}}\right)^{2}
\end{equation}
where $\chi$ gives the fractional crosstalk between the flux-ramp-modulated signals.

This mechanism dictates the minimum frequency-spacing between resonances as a multiple of their bandwidths and therefore the ultimate bandwidth efficiency. For part-per-thousand crosstalk between frequency neighbors, a target frequency spacing of (7 to 10) $\Delta f\urss{BW}$ is typically necessary.

Coupled-harmonic-oscillator crosstalk arises when resonances are both electromagnetically coupled and close in frequency, such that the eigen modes of the system are actually linear combinations of the uncoupled-resonance modes:
\begin{equation}
\chi\urss{CHO} \approx \left(\frac{4\bar{f}^{2}}{f_{2} - f_{1}} \frac{M\urss{x}}{Z_{0}} \right)^{2}
\end{equation}
where $M\urss{x}$ is the mutual inductance between resonance terminations. Other mechanisms of electromagnetic coupling add similar terms.

This mechanism motivates a design feature seen in all multiplexers that utilize superconducting resonators, which is the placement of frequency-adjacent resonators far apart in physical space and of physically-adjacent resonators far apart in frequency space\cite{noroozian2012crosstalk}.

Broadband-non-linearity crosstalk occurs within the broadband microwave components, such as the HEMT amplifier and room-temperature mixers, whose slight non-linearity allows a 3rd-order mixing process to transfer a fraction $\chi\urss{nonlin}$ of the modulation sidebands of one carrier tone to another:
\begin{equation}
\chi\urss{nonlin} \approx 4 \frac{P_{\mathrm{feed}}|S_{21}|^{2}}{P\urss{IP3}}
\end{equation}
where $P\urss{IP3}$ is the 3rd-order intercept point (here, referred to the input of the HEMT), which is the standard measure of cubic non-linearity in amplifiers and mixers.

While this crosstalk is typically less than that of the other mechanisms, it is all-into-all and therefore remains of serious design concern. To reduce this crosstalk we must use broadband components with high $P\urss{IP3}$ or reduce the microwave power per tone at the cost of readout noise.

\subsubsection{\umux\ optimization for X-ray applications}\label{sec:umuxsummary}

For each application, the following optimization should be performed:

\begin{enumerate}
    \item Design the input coupling $M_{\mathrm{in}}$ such that the readout noise will not significantly degrade the detector signal.
    \item Calculate the flux-ramp rate necessary to accommodate the signal slew rate, which determines the resonance bandwidth.
    \item Space the resonance frequencies by a multiple of their bandwidths that is sufficient to meet the required crosstalk limits.
\end{enumerate}

For a 2\phinot\ flux-ramp and $10 \,\, \Delta f\urss{BW}$ spacing for part-per-thousand crosstalk, this implies a multiplexing factor of:
\begin{equation}
N \approx \frac{\Delta f\urss{BW-tot}}{80} \frac{\Phi_{0}}{ M_{\mathrm{in}} \left|\frac{\mathrm{d}I}{\mathrm{d}t}\right|\urss{max}}
\end{equation}
which is the number of detectors that can be read out using one HEMT, a pair of coaxial cables, and a flux-ramp line.

As an abstract example, a signal with a maximum slew-rate of 0.05~\phinot/\SI{}{\micro\second} would require a flux-ramp rate of \SI{100}{\kilo\hertz}, a resonance bandwidth of \SI{400}{\kilo\hertz}, and achieve a multiplexing factor of $\sim$1,000 in a \SI{4}{\giga\hertz} to \SI{8}{\giga\hertz} HEMT. This is more than an order of magnitude higher than can be provided by conventional multiplexing technologies ($N \sim 40$ by TDM and FDM).

Traditional X-ray applications with relatively high count-rates are generally well-matched to the practical range of \umux\ designs. As the count rate falls substantially below $\sim$\SI{100}{\hertz}, it becomes difficult to match the \umux\ bandwidth (to less than \SI{100}{\kilo\hertz}) and resonance frequency placement (to better than \SI{1}{\mega\hertz}). Reliable \umux\ designs have been proven for a range of bandwidths from \SI{100}{\kilo\hertz} to \SI{30}{\mega\hertz}.

\subsubsection{Example \umux\ systems}\label{sec:umuxsystems}

The first fielded application of \umux\ readout for TES microcalorimetry was the SLEDGEHAMMER instrument at Los Alamos National Laboratory\cite{mates2017simultaneous}, in which 128 gamma-ray microcalorimeters were read out in 1~GHz of bandwidth in the initial demonstration. The system is used to perform high-resolution gamma-ray spectroscopy of samples of nuclear materials. To appropriately sample the microcalorimeter pulses, the resonators were designed to have bandwidths of \SI{300}{\kilo\hertz} at a spacing of \SI{3}{\mega\hertz}. The typical readout noise is approximately \SI{30}{\pico\ampere\per\sqrt\hertz}, a factor of $\sim$5 below the noise of the TES. The pixel count has since been expanded to 256.

A large X-ray spectrometer is presently being assembled at NIST as part of the TOMCAT tomographic imaging system\cite{szypryt2021design}, which has demonstrated readout of $\sim$1000 detectors and will expand to 3000 detectors in the near future. To match the $\sim$\SI{1}{\ampere\per\second} slew rate of the detectors, the resonators have bandwidths of \SI{1}{\mega\hertz} on a \SI{7.5}{\mega\hertz} spacing. The detector assembly utilizes multiple modular ``microsnouts'' (Figure \ref{fig:umux-examples}), with 248 detectors per microsnout in 2~GHz of bandwidth. 

Most relevant to the topic of cryogenic X-ray imaging spectrometers for astronomy is that \umux\ readout has been proposed for the Lynx LXM\cite{bandler19} and the planning documents provide a useful guide to expected readout capability. The proposal calls for readout of $\sim$100,000 pixels via a combination of thermal and electrical multiplexing, so that the main array would require 4,000~\umux\ channels with \SI{1.4}{\mega\hertz} resonator bandwidths and 400 resonators per HEMT amplifier. Other sub-arrays have different readout requirements, which are laid out by Bandler\cite{bandler19} and Bennett\cite{bennett2019microwave}. While Lynx is not yet a funded mission, technology advancement is ongoing to increase the readiness level of the readout and to simplify its application to future large-scale cryogenic X-ray astronomy instruments.

%% file: summary_2022Aug.tex
The operation of large-scale arrays of cryogenic X-ray microcalorimeters in a satellite environment requires the minimization of heat loads and cryogenic complexity of the readout wiring. There are three main multiplexing technologies for accomplishing this: TDM, MHz-FDM, and \umux. All of these technologies are currently deployed to multiple ground-based instruments and planned for use in additional future satellite and ground-based instruments, as shown in Table \ref{tab:muxprojlist}.

\begin{table}[htp]
\caption{An overview of missions that have flown (upper section) or will fly/employ (lower section) cryogenic X-ray spectrometers.}
\def\arraystretch{1.5}
\begin{center}
\begin{tabular}{l|ccccc} \hline
Mission		    &	Launch		&	Detector &	Total \# of & Readout	& \# of sensors \\ 
                &  date        &          &  sensors     & type     & per readout \\ \hline
XQC \cite{mccammon02}           &	2002    & semiconductor &	18 		& 	JFET 	&  1	\\
Hitomi/SXS \cite{mitsuda14_SXS} &	2016	& semiconductor &	36		&	JFET	&	1	\\
Micro-X \cite{adams20_microx}	&	2018, 2022	&	TES 	&	128		&	TDM		&  16	\\ \hline
XRISM/Resolve \cite{ishisaki18}	&	2023	& semiconductor &	36		&	JFET	&	1   \\ 
Athena/X-IFU \cite{xifu18}	    &	mid-2030	&	TES			&	$\sim$2,500	&	TDM	&	33  \\
BabyIAXO$^\ast$ \cite{babyIAXO20}      &   2025       & MMC   &  TBD  & \umux & NA \\ 
HUBS$^\ast$ \cite{hubs}		    	&	$\sim$2030 & TES &	$\sim$3,500	&	TDM or FDM	&	TBD \\
LEM$^\ast$			                	&	mid-2030s	&	TES		&	$\sim$14,000 &	TDM	&	60	\\
Lynx$^\ast$ \cite{bandler19,stevenson19}   &  early 2040s &  TES or MMC & $\sim$100,000 & \umux & TBD \\
Super-DIOS$^\ast$ \cite{sdios20}		&	early 2040s	&	TES	&	$\sim$30,000 & \umux & $\sim$400\\ \hline
\multicolumn{4}{l}{$\ast$: Missions in the conceptual or proposing phase.}\\
\end{tabular}
\end{center}
\label{tab:muxprojlist}
\end{table}

In the near future, work on TDM will focus on the achievement of technology-readiness level (TRL)-6 and beyond for Athena X-IFU.  TRL-6 involves verification of the needed performance in a high-fidelity prototype that is tested in a relevant environment.  The X-IFU Engineering Model (EM) is intended to provide such verification of all X-IFU systems. Aspects of TDM to be tested in the X-IFU EM will include a fully differential wiring architecture, X-IFU's long cable harnesses and how they impact the open-loop bandwidth needed to operate with 160~ns row times, and the yield of the 50~mK TDM components.  Beyond X-IFU, TDM may hybridized to \umux\ as discussed further below.

Future work on MHz-FDM will target increasing the multiplexing factor per column, via narrowing the detector bandwidth without losing performance, increasing the upper limit of the usable bandwidth, 
and stabilizing the TES performance under high-frequency ac bias (see also \blue{Gottardi \& Smith in this book} for the state-of-art single-pixel performance under ac bias). It will also require improvements to the firmware and electronics to realize a lower-mass and simpler system. A multiplexing factor of $\sim$50 to 60 seems feasible improvement in the near future\citep{vaccaro22_MUX}. In parallel, the linearity of the energy-gain scale will be demonstrated in a full spectrometer under multi-column readout.

Future work on \umux\ will continue to advance its technological readiness to match TDM and MHz-FDM. It will also attempt to improve fabrication capability to reliably achieve $Q_{i}$ values above 200,000 and to define resonance frequencies more accurately than $\pm$0.02\%, as well as improve consistency of the fabrication quality and yield. Finally, it will require a microwave packaging solution (e.g. bump-bonded unit cells that confine the microwave fields) that allow it to be readily integrated into a TES focal plane without risking degradation of microwave performance.

For space and ground projects, TDM and FDM are planned for arrays of up to about 10$^4$ pixels, while \umux\ is planned for arrays of more than 10$^4$ pixels (or for smaller arrays of very fast detectors). However, while \umux\ provides multiple gigahertz of output bandwidth, because of the current limitations on quality factor and frequency placement it is difficult to optimize for very large numbers ($>$2,000) of low-bandwidth input signals. The bandwidth-utilization efficiency of \umux\ with resonator bandwidths of greater than $\sim$\SI{300}{\kilo\hertz} is comparable to that of TDM and MHz-FDM, but it cannot maintain that efficiency for slower input signals that require bandwidths narrower than can currently be fabricated.

For this reason several groups have begun working on hybrid multiplexing solutions\cite{schuster2022flux,yu20}, in which TDM, CDM, or MHz-FDM are used as a front end to combine multiple slow input signals into a single \umux\ ``pixel.'' These hybrid solutions could potentially enable multiplexing factors in the tens of thousands, which will probably be necessary for instruments measuring faint astronomical X-ray objects, in which the optimization will tend toward large (megapixel-scale) arrays of slow TESs to tile the focal plane.

\begin{figure}
\begin{tabular}{c}
\begin{minipage}{.9\hsize}
\begin{center}
  \includegraphics[width=\hsize]{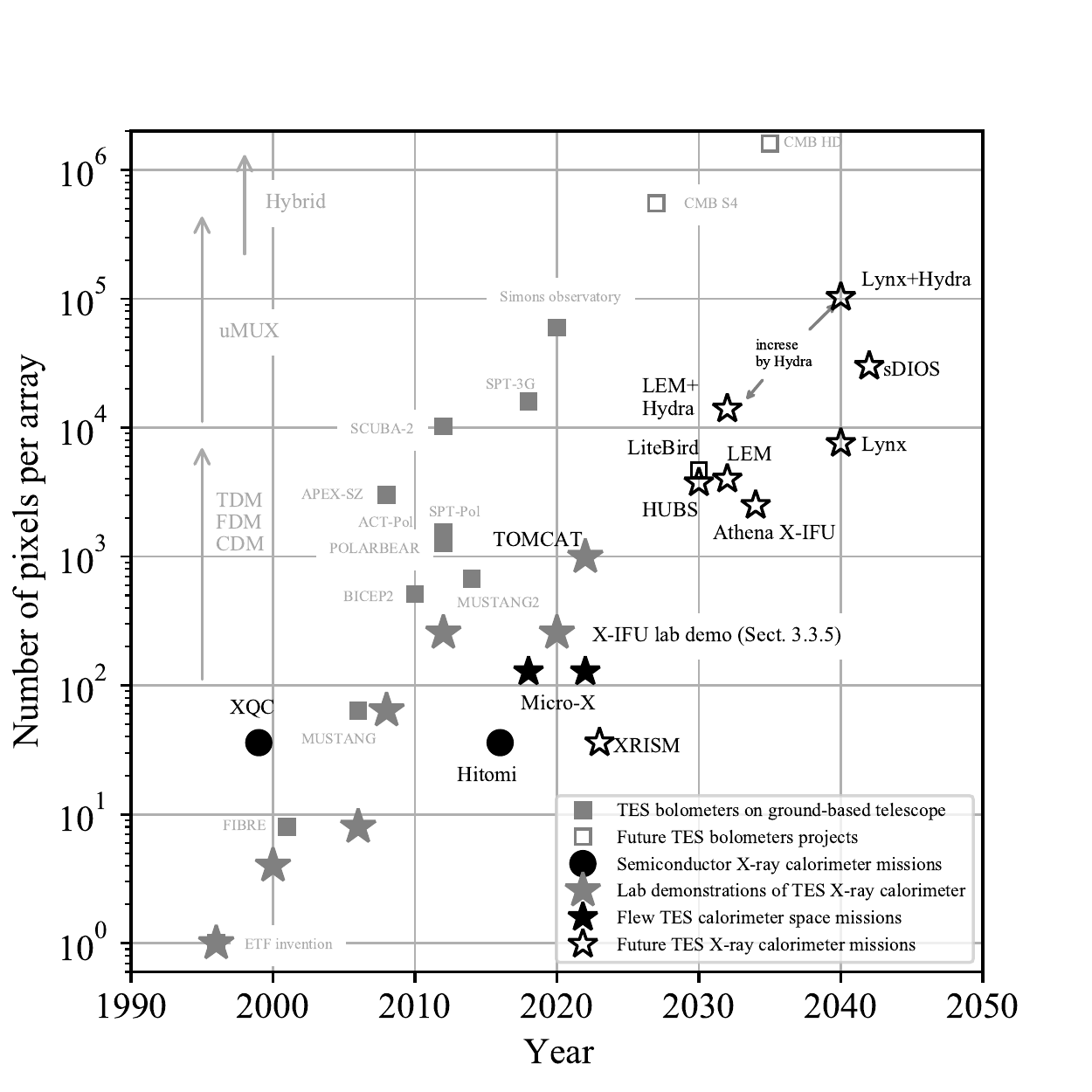}
\end{center}
\end{minipage}
\end{tabular}
\caption{
Number of pixels per microcalorimeter or microbolometer array vs.\ time. In 1995, the modern scheme for TES bias and readout (voltage bias to provide negative electrothermal feedback and current readout via a SQUID ammeter) was proposed\cite{irwin95}, which led quickly to the use of TESs in real instruments. Filled markers indicate deployed instrumentation, while open markers indicate future projects.  Circles and stars represent arrays semiconductor calorimeters and TES microcalorimeters, respectively, while squares represent arrays of sub-millimeter or microwave TES bolometers.  Grey markers represent terrestrial experiments (bolometers for astronomy and X-ray microcalorimeters for laboratory science), while black markers represent space-borne missions.  Two of the future X-ray missions show the effective number of TES absorbers in the ``Hydra'' configuration\cite{smith20}: for LEM and Lynx\cite{bandler19}, the number of Hydra pixels per TES is 4 and 5 respectively (not all TESs are configured as Hydras). TES bolometers require much less bandwidth than do microcalorimeters, and thus have higher multiplexing factors and larger arrays. For both space and ground projects, TDM and FDM are used up to $\sim10^4$ pixels, while \umux\ will be used for larger arrays.
References for projects (not perfect list):
FIBRE\cite{FIBRE}, 
APEX-SZ\citep{apex_sz},
MUSTANG\citep{mustang},
BICEP2\citep{bicep2}, 
POLARBEAR2\citep{polarbear,polarbear2},
MUSTANG2\citep{mustang2}, 
ACT-Pol\citep{act-pol},
Adv-ACTPol\citep{adv-actpol},
SPT-Pol\citep{spt-pol},
SPT-3G\cite{spt-3g},
SCUBA-2\citep{scuba2}, 
Simons Observatory\cite{so},
CMB S4\citep{cmb-s4}, 
LiteBird\cite{lb},
CMB HD\cite{cmb-hd}
TOMCAT\citep{szypryt_tomography,szypryt_1kpix_operation}
}
\label{fig:readout_summary}
\end{figure}

The development time for new technologies to be space ready is about 20 to 30 years, as seen in the gap between terrestrial and space projects in Fig.~\ref{fig:readout_summary}. Beyond X-ray astronomy, these developments will also have terrestrial X-ray spectroscopic applications in basic physics and materials analysis.

%% file: main.bbl
\begin{thebibliography}{176}
\providecommand{\natexlab}[1]{#1}
\providecommand{\url}[1]{{#1}}
\providecommand{\urlprefix}{URL }
\expandafter\ifx\csname urlstyle\endcsname\relax
  \providecommand{\doi}[1]{DOI~\discretionary{}{}{}#1}\else
  \providecommand{\doi}{DOI~\discretionary{}{}{}\begingroup
  \urlstyle{rm}\Url}\fi
\providecommand{\eprint}[2][]{\url{#2}}

\bibitem[{{Abazajian}(2022)}]{cmb-s4}
{Abazajian} Ko (2022) {CMB-S4: Forecasting Constraints on Primordial
  Gravitational Waves}. \apj 926(1):54, \doi{10.3847/1538-4357/ac1596},
  \eprint{2008.12619}

\bibitem[{{Abeln} et~al.(2021)}]{babyIAXO20}
{Abeln} A, et~al. (2021) {Conceptual Design of BabyIAXO, the intermediate stage
  towards the International Axion Observatory}. Journal of High Energy Physics
  2021(5):1--80

\bibitem[{{Adams} and others.(2020)}]{adams20_microx}
{Adams} JS, et~al. (2020) {First Operation of TES Microcalorimeters in Space
  with the Micro-X Sounding Rocket}. Journal of Low Temperature Physics
  199(3-4):1062--1071, \doi{10.1007/s10909-019-02293-5}, \eprint{1908.09689}

\bibitem[{Adams et~al.(2021)}]{microx}
Adams JS, et~al. (2021) First operation of transition-edge sensors in space
  with the {Micro-X} sounding rocket. In: Holland AD, Beletic J (eds) X-Ray,
  Optical, and Infrared Detectors for Astronomy IX, International Society for
  Optics and Photonics, SPIE, vol 11454, pp 195 -- 203,
  \doi{10.1117/12.2562645}

\bibitem[{{Ade} et~al.(2019)}]{so}
{Ade} P, et~al. (2019) {The Simons Observatory: science goals and forecasts}.
  \jcap 2019(2):056, \doi{10.1088/1475-7516/2019/02/056}, \eprint{1808.07445}

\bibitem[{Ahrens et~al.(2019)Ahrens, Wegner, Paluch, Fleischmann, Enss, and
  Kempf}]{ahrens2019superconducting}
Ahrens F, Wegner M, Paluch P, Fleischmann A, Enss C, Kempf S (2019)
  Superconducting ghz resonators for microwave squid multiplexing of metallic
  magnetic calorimeters. Verhandlungen der Deutschen Physikalischen
  Gesellschaft

\bibitem[{{Akamatsu} et~al.(2014)}]{akamatsu14_ACBias}
{Akamatsu} H, et~al. (2014) {Performance of TES X-ray Microcalorimeters with AC
  Bias Read-Out at MHz Frequencies}. Journal of Low Temperature Physics
  176:591--596, \doi{10.1007/s10909-014-1130-8}

\bibitem[{{Akamatsu} et~al.(2020)}]{akamatsu20}
{Akamatsu} H, et~al. (2020) {Progress in the Development of Frequency-Domain
  Multiplexing for the X-ray Integral Field Unit on Board the Athena Mission}.
  Journal of Low Temperature Physics 199(3-4):737--744,
  \doi{10.1007/s10909-020-02351-3}

\bibitem[{{Akamatsu} et~al.(2021)}]{akamatsu21}
{Akamatsu} H, et~al. (2021) {Demonstration of MHz frequency domain multiplexing
  readout of 37 transition edge sensors for high-resolution x-ray imaging
  spectrometers}. Applied Physics Letters 119(18):182601,
  \doi{10.1063/5.0066240}, \eprint{2111.01797}

\bibitem[{{Anderson} and {Rowell}(1963)}]{anderson63}
{Anderson} PW, {Rowell} JM (1963) {Probable Observation of the Josephson
  Superconducting Tunneling Effect}. \prl 10(6):230--232,
  \doi{10.1103/PhysRevLett.10.230}

\bibitem[{{Austermann}(2012)}]{spt-pol}
{Austermann} JEo (2012) {SPTpol: an instrument for CMB polarization
  measurements with the South Pole Telescope}. In: {Holland} WS, {Zmuidzinas} J
  (eds) Millimeter, Submillimeter, and Far-Infrared Detectors and
  Instrumentation for Astronomy VI, Society of Photo-Optical Instrumentation
  Engineers (SPIE) Conference Series, vol 8452, p 84521E,
  \doi{10.1117/12.927286}, \eprint{1210.4970}

\bibitem[{{Bandler} et~al.(2019)}]{bandler19}
{Bandler} SR, et~al. (2019) {Lynx x-ray microcalorimeter}. Journal of
  Astronomical Telescopes, Instruments, and Systems 5:021017,
  \doi{10.1117/1.JATIS.5.2.021017}

\bibitem[{{Barret} et~al.(2018)}]{xifu18}
{Barret} D, et~al. (2018) {The ATHENA X-ray Integral Field Unit (X-IFU)}. In:
  \procspie, Society of Photo-Optical Instrumentation Engineers (SPIE)
  Conference Series, vol 10699, p 106991G, \doi{10.1117/12.2312409}

\bibitem[{Battistelli et~al.(2008)}]{MCE}
Battistelli ES, et~al. (2008) Functional description of read-out electronics
  for time-domain multiplexed bolometers for millimeter and sub-millimeter
  astronomy. Journal of Low Temperature Physics 151(3):908--914

\bibitem[{{Benford}(2002)}]{FIBRE}
{Benford} DJo (2002) {First astronomical use of multiplexed transition edge
  bolometers}. In: {Porter} FS, {McCammon} D, {Galeazzi} M, {Stahle} CK (eds)
  Low Temperature Detectors, American Institute of Physics Conference Series,
  vol 605, pp 589--592, \doi{10.1063/1.1457715}

\bibitem[{Bennett et~al.(2019)Bennett, Mates, Bandler
  et~al.}]{bennett2019microwave}
Bennett DA, Mates JA, Bandler SR, et~al. (2019) Microwave squid multiplexing
  for the lynx x-ray microcalorimeter. Journal of Astronomical Telescopes,
  Instruments, and Systems 5(2):021007

\bibitem[{Bennett et~al.(2012)}]{bennett_gamma}
Bennett DA, et~al. (2012) A high resolution gamma-ray spectrometer based on
  superconducting microcalorimeters. Review of Scientific Instruments
  83(9):093113

\bibitem[{Bennett et~al.(2014)}]{bennett2014integration}
Bennett DA, et~al. (2014) Integration of tes microcalorimeters with microwave
  squid multiplexed readout. IEEE Transactions on Applied Superconductivity
  25(3):1--5

\bibitem[{{Bergen} et~al.(2016)}]{bergen16}
{Bergen} A, et~al. (2016) {Design and validation of a large-format transition
  edge sensor array magnetic shielding system for space application}. Review of
  Scientific Instruments 87(10):105109, \doi{10.1063/1.4962157}

\bibitem[{Beyer and Drung(2008)}]{beyer_drung_2008}
Beyer J, Drung D (2008) A squid multiplexer with
  superconducting-to-normalconducting switches. Superconductor Science and
  Technology 21(10):105022

\bibitem[{{Bruijn} et~al.(2014)}]{bruijn14}
{Bruijn} MP, et~al. (2014) {Tailoring the High-Q LC Filter Arrays for Readout
  of Kilo-Pixel TES Arrays in the SPICA-SAFARI Instrument}. Journal of Low
  Temperature Physics 176(3-4):421--425, \doi{10.1007/s10909-013-1003-6}

\bibitem[{{Bruijn} et~al.(2018)}]{bruijn18}
{Bruijn} MP, et~al. (2018) {LC Filters for FDM Readout of the X-IFU TES
  Calorimeter Instrument on Athena}. Journal of Low Temperature Physics
  193(5-6):661--667, \doi{10.1007/s10909-018-1951-y}

\bibitem[{{Cash}(1979)}]{cash79}
{Cash} W (1979) {Parameter estimation in astronomy through application of the
  likelihood ratio}. \apj 228:939--947, \doi{10.1086/156922}

\bibitem[{{Chervenak} et~al.(1999)}]{chervenak99}
{Chervenak} JA, et~al. (1999) {Superconducting multiplexer for arrays of
  transition edge sensors}. Applied Physics Letters 74(26):4043,
  \doi{10.1063/1.123255}

\bibitem[{Chervenak et~al.(1999)}]{chervenak1999}
Chervenak JA, et~al. (1999) Superconducting multiplexer for arrays of
  transition edge sensors. Applied Physics Letters 74(26):4043--4045

\bibitem[{Chesca et~al.(2004)Chesca, Kleiner, and Koelle}]{squid}
Chesca B, Kleiner R, Koelle D (2004) SQUID Theory, John Wiley \& Sons, Ltd,
  chap~2, pp 29--92. \doi{https://doi.org/10.1002/3527603646.ch2}

\bibitem[{{Cohen} and {Givler}(1972)}]{cohen72}
{Cohen} D, {Givler} E (1972) {Magnetomyography: magnetic fields around the
  human body produced by skeletal muscles}. Applied Physics Letters
  21(3):114--116, \doi{10.1063/1.1654294}

\bibitem[{{Cui} et~al.(2020)}]{hubs}
{Cui} W, et~al. (2020) {HUBS: a dedicated hot circumgalactic medium explorer}.
  In: Society of Photo-Optical Instrumentation Engineers (SPIE) Conference
  Series, Society of Photo-Optical Instrumentation Engineers (SPIE) Conference
  Series, vol 11444, p 114442S, \doi{10.1117/12.2560871}

\bibitem[{{Cunningham} et~al.(2002)}]{cunningham02}
{Cunningham} MF, et~al. (2002) {High-resolution operation of
  frequency-multiplexed transition-edge photon sensors}. Applied Physics
  Letters 81(1):159, \doi{10.1063/1.1489486}

\bibitem[{D'Andrea et~al.(2017)}]{dandrea17}
D'Andrea M, et~al. (2017) {The Cryogenic AntiCoincidence detector for ATHENA
  X-IFU: a scientific assessment of the observational capabilities in the hard
  X-ray band}. Experimental Astronomy 44(3):359--370,
  \doi{10.1007/s10686-017-9543-4}

\bibitem[{D'Andrea et~al.(2020)}]{dandrea20}
D'Andrea M, et~al. (2020) {The Demonstration Model of the ATHENA X-IFU
  Cryogenic AntiCoincidence Detector}. Journal of Low Temperature Physics
  199(1-2):65--72, \doi{10.1007/s10909-019-02300-9}

\bibitem[{D'Andrea et~al.(2021)}]{dandrea21}
D'Andrea M, et~al. (2021) {Single Pixel Performance of a 32 {\texttimes} 32
  Ti/Au TES Array With Broadband X-Ray Spectra}. IEEE Transactions on Applied
  Superconductivity 31(5):3065303, \doi{10.1109/TASC.2021.3065303}

\bibitem[{Dawson et~al.(2019)}]{Dawson_TLS}
Dawson CS, et~al. (2019) Two-level switches for advanced time-division
  multiplexing. IEEE Transactions on Applied Superconductivity 29(5):2500205

\bibitem[{{de Haan} et~al.(2012){de Haan}, {Smecher}, and {Dobbs}}]{dehaan12}
{de Haan} T, {Smecher} G, {Dobbs} M (2012) {Improved performance of TES
  bolometers using digital feedback}. In: {Holland} WS, {Zmuidzinas} J (eds)
  Millimeter, Submillimeter, and Far-Infrared Detectors and Instrumentation for
  Astronomy VI, Society of Photo-Optical Instrumentation Engineers (SPIE)
  Conference Series, vol 8452, p 84520E, \doi{10.1117/12.925658}

\bibitem[{{de Haan} et~al.(2020)}]{dehaan19}
{de Haan} T, et~al. (2020) {Recent Advances in Frequency-Multiplexed TES
  Readout: Vastly Reduced Parasitics and an Increase in Multiplexing Factor
  with Sub-Kelvin SQUIDs}. Journal of Low Temperature Physics
  199(3-4):754--761, \doi{10.1007/s10909-020-02403-8}

\bibitem[{{de Wit} et~al.(2022)}]{dewit22_SRON}
{de Wit} M, et~al. (2022) {Performance of the SRON Ti/Au Transition Edge Sensor
  X-ray Calorimeters}. arXiv e-prints arXiv:2208.12556, \eprint{2208.12556}

\bibitem[{{den Hartog} et~al.(2009)}]{BBFB}
{den Hartog} R, et~al. (2009) {Baseband Feedback for
  Frequency-Domain-Multiplexed Readout of TES X-ray Detectors}. In: {Young} B,
  {Cabrera} B, {Miller} A (eds) American Institute of Physics Conference
  Series, vol 1185, pp 261--264, \doi{10.1063/1.3292328}

\bibitem[{{den Hartog} et~al.(2018)}]{denhartog18_DAC}
{den Hartog} R, et~al. (2018) {Performance of a state-of-the-art DAC system for
  FDM readout}. In: {den Herder} JWA, {Nikzad} S, {Nakazawa} K (eds) Space
  Telescopes and Instrumentation 2018: Ultraviolet to Gamma Ray, Society of
  Photo-Optical Instrumentation Engineers (SPIE) Conference Series, vol 10699,
  p 106994Q, \doi{10.1117/12.2312793}

\bibitem[{{den Herder} et~al.(2001)}]{rgs}
{den Herder} JW, et~al. (2001) {The Reflection Grating Spectrometer on board
  XMM-Newton}. \aap 365:L7--L17, \doi{10.1051/0004-6361:20000058}

\bibitem[{{Dicker}(2006)}]{mustang}
{Dicker} SRo (2006) {A 90-GHz bolometer array for the Green Bank Telescope}.
  In: {Zmuidzinas} J, {Holland} WS, {Withington} S, {Duncan} WD (eds) Society
  of Photo-Optical Instrumentation Engineers (SPIE) Conference Series, Society
  of Photo-Optical Instrumentation Engineers (SPIE) Conference Series, vol
  6275, p 62751B, \doi{10.1117/12.672166}

\bibitem[{{Dicker}(2014)}]{mustang2}
{Dicker} SRo (2014) {MUSTANG2: a large focal plan array for the 100 meter Green
  Bank Telescope}. In: {Holland} WS, {Zmuidzinas} J (eds) Millimeter,
  Submillimeter, and Far-Infrared Detectors and Instrumentation for Astronomy
  VII, Society of Photo-Optical Instrumentation Engineers (SPIE) Conference
  Series, vol 9153, p 91530J, \doi{10.1117/12.2056455}

\bibitem[{{Dobbs} et~al.(2008){Dobbs}, {Bissonnette}, and {Spieler}}]{dobbs08}
{Dobbs} M, {Bissonnette} E, {Spieler} H (2008) {Digital Frequency Domain
  Multiplexer for Millimeter-Wavelength Telescopes}. IEEE Transactions on
  Nuclear Science 55(1):21--26, \doi{10.1109/TNS.2007.911601}

\bibitem[{Doriese et~al.(2006)}]{doriese_2006}
Doriese WB, et~al. (2006) Progress toward kilopixel arrays: 3.8 ev
  microcalorimeter resolution in 8-channel squid multiplexer. Nuclear
  Instruments and Methods in Physics Research Section A: Accelerators,
  Spectrometers, Detectors and Associated Equipment 559(2):808--810

\bibitem[{Doriese et~al.(2016)}]{doriese2016}
Doriese WB, et~al. (2016) Developments in time-division multiplexing of x-ray
  transition-edge sensors. Journal of low temperature physics 184(1):389--395

\bibitem[{Doriese et~al.(2017)}]{doriese2017}
Doriese WB, et~al. (2017) A practical superconducting-microcalorimeter x-ray
  spectrometer for beamline and laboratory science. Review of Scientific
  Instruments 88(5):053108

\bibitem[{Doriese et~al.(2019)}]{doriese_ASC2019}
Doriese WB, et~al. (2019) Optimization of time-and code-division-multiplexed
  readout for athena x-ifu. IEEE Transactions on Applied Superconductivity
  29(5):2500305

\bibitem[{Drung et~al.(2005)}]{drung_XXF}
Drung D, et~al. (2005) dc squid readout electronics with up to 100 mhz
  closed-loop bandwidth. IEEE Transactions on Applied Superconductivity
  15(2):777--780

\bibitem[{{Drung} et~al.(2007)}]{drung07_SQUID}
{Drung} D, et~al. (2007) {Highly Sensitive and Easy-to-Use SQUID Sensors}. IEEE
  Transactions on Applied Superconductivity 17(2):699--704,
  \doi{10.1109/TASC.2007.897403}

\bibitem[{Duh et~al.(1988)}]{duh1988ultra}
Duh KG, et~al. (1988) Ultra-low-noise cryogenic high-electron-mobility
  transistors. IEEE transactions on electron devices 35(3):249--256

\bibitem[{{Durkin} et~al.(2019)}]{durkin19}
{Durkin} M, et~al. (2019) {Demonstration of Athena X-IFU Compatible 40-Row
  Time-Division-Multiplexed Readout}. IEEE Transactions on Applied
  Superconductivity 29(5):2904472, \doi{10.1109/TASC.2019.2904472}

\bibitem[{{Durkin} et~al.(2020)}]{durkin20}
{Durkin} M, et~al. (2020) {A Predictive Control Algorithm for
  Time-Division-Multiplexed Readout of TES Microcalorimeters}. Journal of Low
  Temperature Physics 199(1-2):275--280, \doi{10.1007/s10909-020-02342-4}

\bibitem[{Durkin et~al.(2021)}]{durkin2021}
Durkin M, et~al. (2021) Mitigation of finite bandwidth effects in
  time-division-multiplexed {SQUID} readout of {TES} arrays. IEEE Transactions
  on Applied Superconductivity 31(5):1600905

\bibitem[{Fleischmann et~al.(2005)Fleischmann, Enss, and
  Seidel}]{fleischmann05}
Fleischmann A, Enss C, Seidel G (2005) Metallic Magnetic Calorimeters,
  Springer, pp 151--216

\bibitem[{Fowler et~al.(2021)}]{fowler_2021}
Fowler JW, et~al. (2021) Absolute energies and emission line shapes of the l
  x-ray transitions of lanthanide metals. Metrologia 58(1):015016

\bibitem[{Friedrich(2006)}]{friedrich2006}
Friedrich S (2006) Cryogenic x-ray detectors for synchrotron science. Journal
  of synchrotron radiation 13(2):159--171

\bibitem[{Gandilo et~al.(2016)}]{piper}
Gandilo NN, et~al. (2016) The primordial inflation polarization explorer
  (piper). In: Millimeter, Submillimeter, and Far-Infrared Detectors and
  Instrumentation for Astronomy VIII, vol 9914, pp 372--379

\bibitem[{Gao et~al.(2008)}]{gao2008experimental}
Gao J, et~al. (2008) Experimental evidence for a surface distribution of
  two-level systems in superconducting lithographed microwave resonators.
  Applied Physics Letters 92(15):152505

\bibitem[{Gonzales et~al.(2022)}]{APC_WFEE}
Gonzales M, et~al. (2022) Fully differential broadband lna with active
  impedance matching for squid readout. Journal of Low Temperature Physics p in
  review

\bibitem[{Gottardi and Nagayashi(2021)}]{gottardi21}
Gottardi L, Nagayashi K (2021) A review of x-ray microcalorimeters based on
  superconducting transition edge sensors for astrophysics and particle
  physics. Applied Sciences 11(9), \doi{10.3390/app11093793},
  \urlprefix\url{https://www.mdpi.com/2076-3417/11/9/3793}

\bibitem[{{Gottardi} et~al.(2014{\natexlab{a}})}]{gottardi14a}
{Gottardi} L, et~al. (2014{\natexlab{a}}) {Josephson effects in an alternating
  current biased transition edge sensor}. Applied Physics Letters
  105(16):162605, \doi{10.1063/1.4899065}

\bibitem[{{Gottardi} et~al.(2014{\natexlab{b}})}]{gottardi14}
{Gottardi} L, et~al. (2014{\natexlab{b}}) {Weak-Link Phenomena in AC-Biased
  Transition Edge Sensors}. Journal of Low Temperature Physics 176:279--284,
  \doi{10.1007/s10909-014-1093-9}

\bibitem[{{Gottardi} et~al.(2018)}]{gottardi18}
{Gottardi} L, et~al. (2018) {Josephson Effects in Frequency-Domain Multiplexed
  TES Microcalorimeters and Bolometers}. Journal of Low Temperature Physics
  193(3-4):209--216, \doi{10.1007/s10909-018-2006-0}

\bibitem[{{Gottardi} et~al.(2019)}]{gottardi19}
{Gottardi} L, et~al. (2019) {A six-degree-of-freedom micro-vibration acoustic
  isolator for low-temperature radiation detectors based on superconducting
  transition-edge sensors}. Review of Scientific Instruments 90(5):055107,
  \doi{10.1063/1.5088364}

\bibitem[{{Grace}(2014)}]{act-pol}
{Grace} Eo (2014) {ACTPol: on-sky performance and characterization}. In:
  {Holland} WS, {Zmuidzinas} J (eds) Millimeter, Submillimeter, and
  Far-Infrared Detectors and Instrumentation for Astronomy VII, Society of
  Photo-Optical Instrumentation Engineers (SPIE) Conference Series, vol 9153, p
  915310, \doi{10.1117/12.2057243}

\bibitem[{Harper et~al.(2018)}]{hawc_plus}
Harper DA, et~al. (2018) Hawc+, the far-infrared camera and polarimeter for
  sofia. Journal of Astronomical Instrumentation 7(04):1840008

\bibitem[{Hashimoto et~al.(2022)}]{hashimoto_kaonic}
Hashimoto T, et~al. (2022) Measurements of strong-interaction effects in
  kaonic-helium isotopes at sub-ev precision with x-ray microcalorimeters.
  Physical review letters 128(11):112503

\bibitem[{{Hazumi} et~al.(2020)}]{hazumi20}
{Hazumi} M, et~al. (2020) {LiteBIRD satellite: JAXA's new strategic L-class
  mission for all-sky surveys of cosmic microwave background polarization}. In:
  Society of Photo-Optical Instrumentation Engineers (SPIE) Conference Series,
  Society of Photo-Optical Instrumentation Engineers (SPIE) Conference Series,
  vol 11443, p 114432F, \doi{10.1117/12.2563050}

\bibitem[{{Henderson}(2016)}]{adv-actpol}
{Henderson} SWo (2016) {Advanced ACTPol Cryogenic Detector Arrays and Readout}.
  Journal of Low Temperature Physics 184(3-4):772--779

\bibitem[{{Henning}(2015)}]{spt-3g}
{Henning} Jo (2015) {SPT-3G: The third generation camera and survey for the
  South Pole Telescope}. In: American Astronomical Society Meeting Abstracts
  \#225, American Astronomical Society Meeting Abstracts, vol 225, p 220.02

\bibitem[{Hirayama et~al.(2013)}]{hirayama2013microwave}
Hirayama F, et~al. (2013) Microwave squid multiplexer for tes readout. IEEE
  transactions on applied superconductivity 23(3):2500405--2500405

\bibitem[{{Hitomi Collaboration} et~al.(2016)}]{Hitomi16}
{Hitomi Collaboration}, et~al. (2016) {The quiescent intracluster medium in the
  core of the Perseus cluster}. \nat 535(7610):117--121,
  \doi{10.1038/nature18627}

\bibitem[{Holland et~al.(2013)}]{scuba2}
Holland WS, et~al. (2013) {SCUBA}-2: the 10,000 pixel bolometer camera on the
  {James} {Clerk} {Maxwell} telescope. Monthly Notices of the Royal
  Astronomical Society 430(4):2513--2533

\bibitem[{{H{\"o}lzer} et~al.(1997)}]{holzer}
{H{\"o}lzer} G, et~al. (1997) {K{\ensuremath{\alpha}}$_{1,2}$ and
  K{\ensuremath{\beta}}$_{1,3}$ x-ray emission lines of the 3d transition
  metals}. \pra 56(6):4554--4568, \doi{10.1103/PhysRevA.56.4554}

\bibitem[{Huber et~al.(1997)}]{huber_SSA}
Huber ME, et~al. (1997) Dc squid series arrays with intracoil damping to reduce
  resonance distortions. Applied Superconductivity 5(7-12):425--429

\bibitem[{Huber et~al.(2001)}]{huber_SSA_2}
Huber ME, et~al. (2001) Dc squid series array amplifiers with 120 mhz bandwidth
  (corrected). IEEE Transactions on Applied Superconductivity 11(2):4048--4053

\bibitem[{{Inoue}(2016)}]{polarbear2}
{Inoue} Yo (2016) {POLARBEAR-2: an instrument for CMB polarization
  measurements}. In: {Holland} WS, {Zmuidzinas} J (eds) Millimeter,
  Submillimeter, and Far-Infrared Detectors and Instrumentation for Astronomy
  VIII, Society of Photo-Optical Instrumentation Engineers (SPIE) Conference
  Series, vol 9914, p 99141I, \doi{10.1117/12.2231961}, \eprint{1608.03025}

\bibitem[{{Irwin}(1995)}]{irwin95}
{Irwin} KD (1995) {An application of electrothermal feedback for high
  resolution cryogenic particle detection}. Applied Physics Letters
  66(15):1998--2000, \doi{10.1063/1.113674}

\bibitem[{{Irwin}(2009)}]{irwin09}
{Irwin} KD (2009) {Shannon Limits for Low-Temperature Detector Readout}. In:
  {Young} B, {Cabrera} B, {Miller} A (eds) The Thirteenth International
  Workshop on Low Temperature Detectors - LTD13, American Institute of Physics
  Conference Series, vol 1185, pp 229--236, \doi{10.1063/1.3292320}

\bibitem[{Irwin and Hilton(2005)}]{irwin05}
Irwin KD, Hilton GC (2005) Transition-edge sensors, Springer, pp 63--150

\bibitem[{{Irwin} and {Lehnert}(2004)}]{irwin04}
{Irwin} KD, {Lehnert} KW (2004) {Microwave SQUID multiplexer}. Applied Physics
  Letters 85(11):2107, \doi{10.1063/1.1791733}

\bibitem[{{Irwin} et~al.(2010)}]{irwin10}
{Irwin} KD, et~al. (2010) {Code-division multiplexing of superconducting
  transition-edge sensor arrays}. Superconductor Science Technology
  23(3):034004, \doi{10.1088/0953-2048/23/3/034004}

\bibitem[{Irwin et~al.(2012)}]{irwin_FAS}
Irwin KD, et~al. (2012) Advanced code-division multiplexers for superconducting
  detector arrays. Journal of Low Temperature Physics 167(5):588--594

\bibitem[{{Ishisaki} et~al.(2018)}]{ishisaki18}
{Ishisaki} Y, et~al. (2018) {Resolve Instrument on X-ray Astronomy Recovery
  Mission (XARM)}. Journal of Low Temperature Physics 193(5-6):991--995,
  \doi{10.1007/s10909-018-1913-4}

\bibitem[{{Jackson} et~al.(2016)}]{jackson16}
{Jackson} BD, et~al. (2016) {The focal plane assembly for the Athena X-ray
  Integral Field Unit instrument}. In: {den Herder} JWA, {Takahashi} T, {Bautz}
  M (eds) Space Telescopes and Instrumentation 2016: Ultraviolet to Gamma Ray,
  Society of Photo-Optical Instrumentation Engineers (SPIE) Conference Series,
  vol 9905, p 99052I, \doi{10.1117/12.2232544}

\bibitem[{{Jaklevic} et~al.(1964)}]{jakelevic64}
{Jaklevic} RC, et~al. (1964) {Quantum Interference Effects in Josephson
  Tunneling}. \prl 12(7):159--160, \doi{10.1103/PhysRevLett.12.159}

\bibitem[{{Kaastra} and {Bleeker}(2016)}]{kaastra16}
{Kaastra} JS, {Bleeker} JAM (2016) {Optimal binning of X-ray spectra and
  response matrix design}. \aap 587:A151, \doi{10.1051/0004-6361/201527395},
  \eprint{1601.05309}

\bibitem[{{Kelley} et~al.(2007)}]{xrs}
{Kelley} RL, et~al. (2007) {The Suzaku High Resolution X-Ray Spectrometer}.
  \pasj 59:77--112, \doi{10.1093/pasj/59.sp1.S77}

\bibitem[{{Kelley} et~al.(2008)}]{kelley09}
{Kelley} RL, et~al. (2008) {Ion-Implanted Silicon X-Ray Calorimeters: Present
  and Future}. Journal of Low Temperature Physics 151(1-2):375--380,
  \doi{10.1007/s10909-007-9663-8}

\bibitem[{{Kempf} et~al.(2018)}]{kemph18}
{Kempf} S, et~al. (2018) {Physics and Applications of Metallic Magnetic
  Calorimeters}. Journal of Low Temperature Physics 193(3-4):365--379,
  \doi{10.1007/s10909-018-1891-6}

\bibitem[{{Kermish}(2012)}]{polarbear}
{Kermish} Zo (2012) {The POLARBEAR experiment}. In: {Holland} WS, {Zmuidzinas}
  J (eds) Millimeter, Submillimeter, and Far-Infrared Detectors and
  Instrumentation for Astronomy VI, Society of Photo-Optical Instrumentation
  Engineers (SPIE) Conference Series, vol 8452, p 84521C,
  \doi{10.1117/12.926354}, \eprint{1210.7768}

\bibitem[{{Kimura} et~al.(2008)}]{kimura08}
{Kimura} S, et~al. (2008) {Performance Measurement of the 8-Input SQUIDs for
  TES Frequency Domain Multiplexing}. Journal of Low Temperature Physics
  151(3-4):946--951, \doi{10.1007/s10909-008-9771-0}

\bibitem[{{Kiviranta} et~al.(2018){Kiviranta}, {Gr{\"o}nberg}, and {van der
  Kuur}}]{kiviranta18}
{Kiviranta} M, {Gr{\"o}nberg} L, {van der Kuur} J (2018) {Two SQUID amplifiers
  intended to alleviate the summing node inductance problem in multiplexed
  arrays of Transition Edge Sensors}. arXiv e-prints arXiv:1810.09122

\bibitem[{{Kiviranta} et~al.(2002)}]{kiviranta02}
{Kiviranta} M, et~al. (2002) {SQUID-based readout schemes for microcalorimeter
  arrays}. In: {Porter} FS, {McCammon} D, {Galeazzi} M, {Stahle} CK (eds) Low
  Temperature Detectors, American Institute of Physics Conference Series, vol
  605, pp 295--300, \doi{10.1063/1.1457649}

\bibitem[{{Kiviranta} et~al.(2004)}]{mikko04}
{Kiviranta} M, et~al. (2004) {Design and performance of multiloop and washer
  SQUIDs intended for sub-kelvin operation}. Superconductor Science Technology
  17(5):S285--S289, \doi{10.1088/0953-2048/17/5/038}

\bibitem[{{Kiviranta} et~al.(2021)}]{kiviranta21}
{Kiviranta} M, et~al. (2021) {Two-Stage SQUID Amplifier for the Frequency
  Multiplexed Readout of the X-IFU X-Ray Camera}. IEEE Transactions on Applied
  Superconductivity 31(5):3060356, \doi{10.1109/TASC.2021.3060356}

\bibitem[{Lee et~al.(2022)}]{lee_2022}
Lee S, et~al. (2022) Generic character of charge and spin density waves in
  superconducting cuprates. Proceedings of the National Academy of Sciences
  119(15):e2119429119

\bibitem[{Lee et~al.(2019)}]{ssrl}
Lee SJ, et~al. (2019) Soft x-ray spectroscopy with transition-edge sensors at
  stanford synchrotron radiation lightsource beamline 10-1. Review of
  Scientific Instruments 90(11):113101, \doi{10.1063/1.5119155}

\bibitem[{Likharev(2022)}]{likharev2022dynamics}
Likharev KK (2022) Dynamics of Josephson junctions and circuits. Routledge

\bibitem[{Lucas et~al.(2022)}]{lucas2022indium}
Lucas TJ, et~al. (2022) Indium bump process for low-temperature detectors and
  readout. Journal of Low Temperature Physics pp 1--6

\bibitem[{Macculi et~al.(2016)}]{macculi16}
Macculi C, et~al. (2016) {The Cryogenic AntiCoincidence detector for ATHENA
  X-IFU: a program overview}. In: {den Herder} JWA, {Takahashi} T, {Bautz} M
  (eds) Space Telescopes and Instrumentation 2016: Ultraviolet to Gamma Ray,
  Society of Photo-Optical Instrumentation Engineers (SPIE) Conference Series,
  vol 9905, p 99052K, \doi{10.1117/12.2231298}

\bibitem[{Malnou et~al.(2022)}]{malnou}
Malnou M, et~al. (2022) Performance of a kinetic inductance traveling-wave
  parametric amplifier at 4 kelvin: Toward an alternative to semiconductor
  amplifiers. Physical Review Applied 17(4):044009

\bibitem[{{Mantegazzini} et~al.(2022)}]{Mantegazzini22}
{Mantegazzini} F, et~al. (2022) {Metallic magnetic calorimeter arrays for the
  first phase of the ECHo experiment}. Nuclear Instruments and Methods in
  Physics Research A 1030:166406, \doi{10.1016/j.nima.2022.166406},
  \eprint{2111.09945}

\bibitem[{{Markert} et~al.(1994)}]{markert94}
{Markert} TH, et~al. (1994) {High-Energy Transmission Grating Spectrometer for
  the Advanced X-ray Astrophysics Facility (AXAF)}. In: {Siegmund} OH,
  {Vallerga} JV (eds) EUV, X-Ray, and Gamma-Ray Instrumentation for Astronomy
  V, Society of Photo-Optical Instrumentation Engineers (SPIE) Conference
  Series, vol 2280, pp 168--180, \doi{10.1117/12.186812}

\bibitem[{Mates et~al.(2008)}]{mates2008demonstration}
Mates J, et~al. (2008) Demonstration of a multiplexer of dissipationless
  superconducting quantum interference devices. Applied Physics Letters
  92(2):023514

\bibitem[{Mates et~al.(2012)}]{mates2012flux}
Mates J, et~al. (2012) Flux-ramp modulation for squid multiplexing. Journal of
  Low Temperature Physics 167(5):707--712

\bibitem[{Mates et~al.(2017)}]{mates2017simultaneous}
Mates J, et~al. (2017) Simultaneous readout of 128 x-ray and gamma-ray
  transition-edge microcalorimeters using microwave squid multiplexing. Applied
  Physics Letters 111(6):062601

\bibitem[{Mates et~al.(2019)}]{mates2019crosstalk}
Mates J, et~al. (2019) Crosstalk in microwave squid multiplexers. Applied
  Physics Letters 115(20):202601

\bibitem[{Mates(2011)}]{mates2011microwave}
Mates JAB (2011) The microwave squid multiplexer. PhD thesis

\bibitem[{McCammon(2005)}]{mcCammon05_semiconductor}
McCammon D (2005) Semiconductor Thermistors, Springer, pp 35--62

\bibitem[{{McCammon}(2005)}]{mccammon05}
{McCammon} D (2005) {Thermal Equilibrium Calorimeters - An Introduction},
  vol~99, Springer, p~1

\bibitem[{{McCammon} et~al.(2002)}]{mccammon02}
{McCammon} D, et~al. (2002) {A High Spectral Resolution Observation of the Soft
  X-Ray Diffuse Background with Thermal Detectors}. \apj 576(1):188--203,
  \doi{10.1086/341727}

\bibitem[{{Mehl}(2008)}]{apex_sz}
{Mehl} Jo (2008) {TES Bolometer Array for the APEX-SZ Camera}. Journal of Low
  Temperature Physics 151(3-4):697--702, \doi{10.1007/s10909-008-9738-1}

\bibitem[{Miaja-Avila et~al.(2016)}]{miaja_2016}
Miaja-Avila L, et~al. (2016) Ultrafast time-resolved hard x-ray emission
  spectroscopy on a tabletop. Phys Rev X 6:031047

\bibitem[{{Mitsuda} et~al.(1999)}]{mitsuda99}
{Mitsuda} K, et~al. (1999) {Multi-pixel readout of transition-edge sensors
  using a multi-input SQUID}. Nuclear Instruments and Methods in Physics
  Research A 436(1-2):252--255, \doi{10.1016/S0168-9002(99)00630-0}

\bibitem[{{Mitsuda} et~al.(2007)}]{mitsuda07}
{Mitsuda} K, et~al. (2007) {The X-Ray Observatory Suzaku}. \pasj 59:1--7,
  \doi{10.1093/pasj/59.sp1.S1}

\bibitem[{{Mitsuda} et~al.(2014)}]{mitsuda14_SXS}
{Mitsuda} K, et~al. (2014) {Soft x-ray spectrometer (SXS): the high-resolution
  cryogenic spectrometer onboard ASTRO-H}. In: {Takahashi} T, {den Herder} JWA,
  {Bautz} M (eds) Space Telescopes and Instrumentation 2014: Ultraviolet to
  Gamma Ray, Society of Photo-Optical Instrumentation Engineers (SPIE)
  Conference Series, vol 9144, p 91442A, \doi{10.1117/12.2057199}

\bibitem[{{Montgomery} et~al.(2022)}]{montgomery22}
{Montgomery} J, et~al. (2022) {Performance and characterization of the SPT-3G
  digital frequency-domain multiplexed readout system using an improved noise
  and crosstalk model}. Journal of Astronomical Telescopes, Instruments, and
  Systems 8(1):014001, \doi{10.1117/1.JATIS.8.1.014001}, \eprint{2103.16017}

\bibitem[{{Morgan} et~al.(2016)}]{morgan16}
{Morgan} KM, et~al. (2016) {Code-division-multiplexed readout of large arrays
  of TES microcalorimeters}. Applied Physics Letters 109(11):112604,
  \doi{10.1063/1.4962636}

\bibitem[{{Moseley} et~al.(1984){Moseley}, {Mather}, and {McCammon}}]{mmm}
{Moseley} SH, {Mather} JC, {McCammon} D (1984) {Thermal detectors as x-ray
  spectrometers}. Journal of Applied Physics 56(5):1257--1262,
  \doi{10.1063/1.334129}

\bibitem[{{Nagayoshi} et~al.(2020)}]{nagayoshi20}
{Nagayoshi} K, et~al. (2020) {Development of a Ti/Au TES Microcalorimeter Array
  as a Backup Sensor for the Athena/X-IFU Instrument}. Journal of Low
  Temperature Physics 199(3-4):943--948, \doi{10.1007/s10909-019-02282-8}

\bibitem[{{Nakashima} et~al.(2020)}]{nakashima20}
{Nakashima} Y, et~al. (2020) {Low-noise microwave SQUID multiplexed readout of
  38 x-ray transition-edge sensor microcalorimeters}. Applied Physics Letters
  117(12):122601, \doi{10.1063/5.0016333}

\bibitem[{Noroozian et~al.(2012)}]{noroozian2012crosstalk}
Noroozian O, et~al. (2012) Crosstalk reduction for superconducting microwave
  resonator arrays. IEEE Transactions on Microwave Theory and Techniques
  60(5):1235--1243

\bibitem[{Okumura et~al.(2021)}]{okumura_muonic}
Okumura T, et~al. (2021) Deexcitation dynamics of muonic atoms revealed by
  high-precision spectroscopy of electronic k x rays. Physical Review Letters
  127(5):053001

\bibitem[{O'Neil et~al.(2017)}]{oneil_2017}
O'Neil GC, et~al. (2017) Ultrafast time-resolved x-ray absorption spectroscopy
  of ferrioxalate photolysis with a laser plasma x-ray source and
  microcalorimeter array. The Journal of Physical Chemistry Letters
  8(5):1099--1104

\bibitem[{{BICEP2Collaboration}and others(2014)}]{bicep2}
{BICEP2Collaboration}and others (2014) {Detection of B-Mode Polarization at
  Degree Angular Scales by BICEP2}. \prl 112(24):241101,
  \doi{10.1103/PhysRevLett.112.241101}, \eprint{1403.3985}

\bibitem[{{LiteBIRDCollaboration}and others(2022)}]{lb}
{LiteBIRDCollaboration}and others (2022) {Probing Cosmic Inflation with the
  LiteBIRD Cosmic Microwave Background Polarization Survey}. arXiv e-prints
  arXiv:2202.02773, \eprint{2202.02773}

\bibitem[{Palosaari et~al.(2016)}]{palosaari_2016}
Palosaari MRJ, et~al. (2016) Broadband ultrahigh-resolution spectroscopy of
  particle-induced x rays: extending the limits of nondestructive analysis.
  Physical Review Applied 6(2):024002

\bibitem[{{Porter} et~al.(2009)}]{porter09}
{Porter} FS, et~al. (2009) {The Astro-H Soft X-ray Spectrometer (SXS)}. In:
  {Young} B, {Cabrera} B, {Miller} A (eds) The Thirteenth International
  Workshop on Low Temperature Detectors - LTD13, American Institute of Physics
  Conference Series, vol 1185, pp 91--94, \doi{10.1063/1.3292564}

\bibitem[{{Porter} et~al.(2010)}]{porter10}
{Porter} FS, et~al. (2010) {The detector subsystem for the SXS instrument on
  the ASTRO-H Observatory}. In: {Arnaud} M, {Murray} SS, {Takahashi} T (eds)
  Space Telescopes and Instrumentation 2010: Ultraviolet to Gamma Ray, Society
  of Photo-Optical Instrumentation Engineers (SPIE) Conference Series, vol
  7732, p 77323J, \doi{10.1117/12.857888}

\bibitem[{{Ravera} et~al.(2014)}]{xifu14}
{Ravera} L, et~al. (2014) {The X-ray Integral Field Unit (X-IFU) for Athena}.
  In: Space Telescopes and Instrumentation 2014: Ultraviolet to Gamma Ray,
  \procspie, vol 9144, p 91442L, \doi{10.1117/12.2055884}

\bibitem[{Reintsema et~al.(2003)}]{reintsema_2003}
Reintsema CD, et~al. (2003) Prototype system for superconducting quantum
  interference device multiplexing of large-format transition-edge sensor
  arrays. Review of Scientific Instruments 74(10):4500--4508

\bibitem[{Reintsema et~al.(2009)}]{reintsema_2009}
Reintsema CD, et~al. (2009) Electronics for a next-generation squid-based
  time-domain multiplexing system. In: AIP Conference Proceedings, vol 1185, pp
  237--240

\bibitem[{{Roelfsema} et~al.(2012)}]{roelfsema12}
{Roelfsema} P, et~al. (2012) {The SAFARI imaging spectrometer for the SPICA
  space observatory}. In: {Clampin} MC, {Fazio} GG, {MacEwen} HA, {Oschmann} J
  Jacobus~M (eds) Space Telescopes and Instrumentation 2012: Optical, Infrared,
  and Millimeter Wave, Society of Photo-Optical Instrumentation Engineers
  (SPIE) Conference Series, vol 8442, p 84420R, \doi{10.1117/12.927010}

\bibitem[{{Sadleir} et~al.(2010)}]{sadleir10}
{Sadleir} JE, et~al. (2010) {Longitudinal Proximity Effects in Superconducting
  Transition-Edge Sensors}. \prl 104(4):047003,
  \doi{10.1103/PhysRevLett.104.047003}

\bibitem[{{Sadleir} et~al.(2011)}]{sadleir11}
{Sadleir} JE, et~al. (2011) {Proximity effects and nonequilibrium
  superconductivity in transition-edge sensors}. \prb 84(18):184502,
  \doi{10.1103/PhysRevB.84.184502}

\bibitem[{Sakai et~al.(2022)}]{kazu_COTS}
Sakai K, et~al. (2022) Developments of laboratory-based transition-edge sensor
  readout electronics using commercial-off-the-shelf modules. Journal of Low
  Temperature Physics p in review

\bibitem[{{Sato} et~al.(2020)}]{sdios20}
{Sato} K, et~al. (2020) {Super DIOS mission for exploring ``dark baryon''}. In:
  Society of Photo-Optical Instrumentation Engineers (SPIE) Conference Series,
  Society of Photo-Optical Instrumentation Engineers (SPIE) Conference Series,
  vol 11444, p 114445O, \doi{10.1117/12.2561681}

\bibitem[{Schuster et~al.(2022)}]{schuster2022flux}
Schuster C, et~al. (2022) Flux ramp modulation based hybrid microwave squid
  multiplexer. Applied Physics Letters 120(16):162601

\bibitem[{{Sehgal}(2019)}]{cmb-hd}
{Sehgal} No (2019) {CMB-HD: An Ultra-Deep, High-Resolution Millimeter-Wave
  Survey Over Half the Sky}. In: Bulletin of the American Astronomical Society,
  vol~51, p~6, \eprint{1906.10134}

\bibitem[{{Smith} et~al.(2012)}]{smith12}
{Smith} SJ, et~al. (2012) {Small Pitch Transition-Edge Sensors with Broadband
  High Spectral Resolution for Solar Physics}. Journal of Low Temperature
  Physics 167(3-4):168--175, \doi{10.1007/s10909-012-0574-y}

\bibitem[{{Smith} et~al.(2020)}]{smith20}
{Smith} SJ, et~al. (2020) {Toward 100,000-Pixel Microcalorimeter Arrays Using
  Multi-absorber Transition-Edge Sensors}. Journal of Low Temperature Physics
  199(1-2):330--338, \doi{10.1007/s10909-020-02362-0}

\bibitem[{Smith et~al.(2021)}]{smith2021}
Smith SJ, et~al. (2021) Performance of a broad-band, high-resolution,
  transition-edge sensor spectrometer for x-ray astrophysics. IEEE Transactions
  on Applied Superconductivity 31(5):2100806

\bibitem[{Stevenson et~al.(2019)}]{stevenson19}
Stevenson TR, et~al. (2019) {Magnetic calorimeter option for the Lynx x-ray
  microcalorimeter}. Journal of Astronomical Telescopes, Instruments, and
  Systems 5(2):1 -- 9, \doi{10.1117/1.JATIS.5.2.021009}

\bibitem[{Swetz et~al.(2011)}]{act_swetz}
Swetz DS, et~al. (2011) Overview of the {Atacama} {Cosmology} {Telescope}:
  receiver, instrumentation, and telescope systems. The Astrophysical Journal
  Supplement Series 194(2):41, \doi{10.1088/0067-0049/194/2/41}

\bibitem[{Szypryt et~al.(2021{\natexlab{a}})Szypryt, Bennett
  et~al.}]{szypryt2021design}
Szypryt P, Bennett DA, et~al. (2021{\natexlab{a}}) Design of a 3000-pixel
  transition-edge sensor x-ray spectrometer for microcircuit tomography. IEEE
  Transactions on Applied Superconductivity 31(5):1--5

\bibitem[{Szypryt et~al.(2019)}]{szypryt_EBIT}
Szypryt P, et~al. (2019) A transition-edge sensor-based x-ray spectrometer for
  the study of highly charged ions at the national institute of standards and
  technology electron beam ion trap. Review of Scientific Instruments
  90(12):123107

\bibitem[{Szypryt et~al.(2021{\natexlab{b}})}]{szypryt_tomography}
Szypryt P, et~al. (2021{\natexlab{b}}) Design of a 3000-pixel transition-edge
  sensor x-ray spectrometer for microcircuit tomography. IEEE Transactions on
  Applied Superconductivity 31(5):2100405

\bibitem[{Szypryt et~al.(2022)}]{szypryt_1kpix_operation}
Szypryt P, et~al. (2022) A tabletop x-ray tomography instrument for
  nanometer-scale imaging: demonstration of the 1,000-element transition-edge
  sensor subarray. IEEE Transactions on Applied Superconductivity (in
  preparation)

\bibitem[{{Takahashi} et~al.(2012)}]{takahashi12}
{Takahashi} T, et~al. (2012) {The ASTRO-H X-ray Observatory}. In: Society of
  Photo-Optical Instrumentation Engineers (SPIE) Conference Series, Society of
  Photo-Optical Instrumentation Engineers (SPIE) Conference Series, vol 8443,
  \doi{10.1117/12.926190}

\bibitem[{{Takei} et~al.(2009)}]{takei09_FDM}
{Takei} Y, et~al. (2009) {SQUID multiplexing using baseband feedback for space
  application of transition-edge sensor microcalorimeters}. Superconductor
  Science Technology 22(11):114008, \doi{10.1088/0953-2048/22/11/114008}

\bibitem[{{Takei} et~al.(2018)}]{takei18}
{Takei} Y, et~al. (2018) {Vibration isolation system for cryocoolers of soft
  x-ray spectrometer on-board ASTRO-H (Hitomi)}. Journal of Astronomical
  Telescopes, Instruments, and Systems 4:011216,
  \doi{10.1117/1.JATIS.4.1.011216}

\bibitem[{{Taralli} et~al.(2020)}]{taralli20}
{Taralli} E, et~al. (2020) {Characterization of High Aspect-Ratio TiAu TES
  X-ray Microcalorimeter Array Under AC Bias}. Journal of Low Temperature
  Physics 199(1-2):80--87, \doi{10.1007/s10909-019-02254-y}

\bibitem[{Taralli et~al.(2021)}]{taralli21}
Taralli E, et~al. (2021) {Ti/Au TES 32 {\texttimes} 32 Pixel Array: Uniformity,
  Thermal Crosstalk and Performance at Different X-Ray Energies}. IEEE
  Transactions on Applied Superconductivity 31(5):3061022,
  \doi{10.1109/TASC.2021.3061022}

\bibitem[{{Tashiro} et~al.(2018)}]{xrism18}
{Tashiro} M, et~al. (2018) {Concept of the X-ray Astronomy Recovery Mission}.
  In: \procspie, Society of Photo-Optical Instrumentation Engineers (SPIE)
  Conference Series, vol 10699, p 1069922, \doi{10.1117/12.2309455}

\bibitem[{{The Lynx Team}(2018)}]{lynx}
{The Lynx Team} (2018) {The Lynx Mission Concept Study Interim Report}. arXiv
  e-prints arXiv:1809.09642

\bibitem[{Uhlig et~al.(2013)}]{uhlig_2013}
Uhlig J, et~al. (2013) Table-top ultrafast x-ray microcalorimeter spectrometry
  for molecular structure. Phys Rev Lett 110:138302

\bibitem[{{Ullom} and {Bennett}(2015)}]{ullom15}
{Ullom} JN, {Bennett} DA (2015) {Review of superconducting transition-edge
  sensors for x-ray and gamma-ray spectroscopy}. Superconductor Science
  Technology 28(8):084003, \doi{10.1088/0953-2048/28/8/084003}

\bibitem[{{Ullom} et~al.(2003)}]{ullom03}
{Ullom} JN, et~al. (2003) {A frequency-domain read-out technique for large
  microcalorimeter arrays demonstrated using high-resolution
  {\ensuremath{\gamma}}-ray sensors}. IEEE Transactions on Applied
  Superconductivity 13(2):643--648, \doi{10.1109/TASC.2003.813981}

\bibitem[{{Vaccaro} et~al.(2021)}]{vaccaro21_FSA}
{Vaccaro} D, et~al. (2021) {Frequency Shift Algorithm: Application to a
  Frequency-Domain Multiplexing Readout of X-ray Transition-Edge Sensor
  Microcalorimeters}. arXiv e-prints arXiv:2102.06092

\bibitem[{{Vaccaro} et~al.(2022{\natexlab{a}})}]{vaccaro22_MUX}
{Vaccaro} D, et~al. (2022{\natexlab{a}}) {Frequency domain multiplexing readout
  for large arrays of transition-edge sensors}. arXiv e-prints
  arXiv:2208.12604, \eprint{2208.12604}

\bibitem[{{Vaccaro} et~al.(2022{\natexlab{b}})}]{vaccaro22_sus}
{Vaccaro} D, et~al. (2022{\natexlab{b}}) {Susceptibility study of TES
  micro-calorimeters for X-ray spectroscopy under FDM readout}. arXiv e-prints
  arXiv:2208.10875, \eprint{2208.10875}

\bibitem[{{Vaccaro} et~al.(2022{\natexlab{c}})}]{vaccaro22}
{Vaccaro} D, et~al. (2022{\natexlab{c}}) {Thermal Crosstalk of X-Ray
  Transition-Edge Sensor Micro-Calorimeters Under Frequency Domain Multiplexing
  Readout}. IEEE Transactions on Applied Superconductivity 32(1):3128710,
  \doi{10.1109/TASC.2021.3128710}

\bibitem[{{van der Hulst} et~al.(2021)}]{FSA}
{van der Hulst} P, et~al. (2021) {Frequency shift algorithm: Design of a
  baseband phase locked loop for frequency-domain multiplexing readout of x-ray
  transition-edge sensor microcalorimeters}. Review of Scientific Instruments
  92(7):073101, \doi{10.1063/5.0044968}

\bibitem[{{van der Kuur} et~al.(2004)}]{vanderkuur04}
{van der Kuur} J, et~al. (2004) {Implementation of frequency domain
  multiplexing in imaging arrays of microcalorimeters}. Nuclear Instruments and
  Methods in Physics Research A 520(1-3):551--554,
  \doi{10.1016/j.nima.2003.11.312}

\bibitem[{{van der Kuur} et~al.(2016)}]{vanderkuur16}
{van der Kuur} J, et~al. (2016) {Optimising the multiplex factor of the
  frequency domain multiplexed readout of the TES-based microcalorimeter
  imaging array for the X-IFU instrument on the Athena x-ray observatory}. In:
  {den Herder} JWA, {Takahashi} T, {Bautz} M (eds) Space Telescopes and
  Instrumentation 2016: Ultraviolet to Gamma Ray, Society of Photo-Optical
  Instrumentation Engineers (SPIE) Conference Series, vol 9905, p 99055R,
  \doi{10.1117/12.2232830}

\bibitem[{{van Weers}(2013)}]{vanweers13}
{van Weers} HJ (2013) {Niobium flex cable for low temperature high density
  interconnects}. Cryogenics 55:1--4, \doi{10.1016/j.cryogenics.2012.10.006}

\bibitem[{Walsh(1923)}]{walsh}
Walsh JL (1923) A closed set of normal orthogonal functions. American Journal
  of Mathematics 45(1):5--24

\bibitem[{{Wang} et~al.(2020)}]{wang20}
{Wang} Q, et~al. (2020) {Noise Measurements of a Low-Noise Amplifier in the FDM
  Readout System for SAFARI}. Journal of Low Temperature Physics
  199(3-4):817--823, \doi{10.1007/s10909-019-02328-x}

\bibitem[{Wegner et~al.(2022)Wegner, Enss, and Kempf}]{wegner2022analytical}
Wegner M, Enss C, Kempf S (2022) Analytical model of the readout power and
  squid hysteresis parameter dependence of the resonator characteristics of
  microwave squid multiplexers. Superconductor Science and Technology
  35(7):075011

\bibitem[{{Wegner} et~al.(2018)}]{wegner18}
{Wegner} M, et~al. (2018) {Microwave SQUID Multiplexing of Metallic Magnetic
  Calorimeters: Status of Multiplexer Performance and Room-Temperature Readout
  Electronics Development}. Journal of Low Temperature Physics
  193(3-4):462--475, \doi{10.1007/s10909-018-1878-3}

\bibitem[{Welty and Martinis(1991)}]{welty}
Welty RP, Martinis JM (1991) A series array of dc squids. IEEE Transactions on
  Magnetics 27(2):2924--2926

\bibitem[{Yamada et~al.(2021)}]{spring8}
Yamada S, et~al. (2021) Broadband high-energy resolution hard x-ray
  spectroscopy using transition edge sensors at spring-8. Review of Scientific
  Instruments 92(1):013103

\bibitem[{{Yoon} et~al.(2001)}]{yoon01}
{Yoon} J, et~al. (2001) {Single superconducting quantum interference device
  multiplexer for arrays of low-temperature sensors}. Applied Physics Letters
  78(3):371, \doi{10.1063/1.1338963}

\bibitem[{{Yoon} et~al.(2018)}]{yoon18}
{Yoon} W, et~al. (2018) {Toward Large Field-of-View High-Resolution X-ray
  Imaging Spectrometers: Microwave Multiplexed Readout of 28 TES
  Microcalorimeters}. Journal of Low Temperature Physics 193(3-4):258--266,
  \doi{10.1007/s10909-018-1917-0}

\bibitem[{{Yu} et~al.(2020)}]{yu20}
{Yu} C, et~al. (2020) {An impedance-modulated code-division microwave SQUID
  multiplexer}. Engineering Research Express 2(1):015011,
  \doi{10.1088/2631-8695/ab68a4}

\bibitem[{Zappe(1977)}]{zappe}
Zappe H (1977) Josephson quantum interference computer devices. IEEE
  Transactions on Magnetics 13(1):41--47

\end{thebibliography}
